\documentclass[11pt,a4paper]{article}
\pdfoutput=1
\usepackage{jheppub}
\usepackage{latexsym,amsfonts,amsmath,amssymb}
\usepackage{bbm}
\usepackage{empheq}
\usepackage{graphicx}
\usepackage{color}
\usepackage{caption}
\usepackage{subcaption}
\usepackage[normalem]{ulem}
\usepackage{comment}
\usepackage{url}
\usepackage{slashed}
\usepackage{tabu}
\usepackage{multirow}
\usepackage{extarrows}

\usepackage{mathtools}

\usepackage{amsxtra,graphics,epsfig,bm,tikz,xfrac,lscape}
\usetikzlibrary{decorations.pathmorphing}
\usetikzlibrary{decorations.markings}
\usetikzlibrary{arrows, decorations.markings, calc, fadings, decorations.pathreplacing, patterns, decorations.pathmorphing, positioning}

\usetikzlibrary{positioning,shapes}
\usetikzlibrary{chains}
\usetikzlibrary{arrows,fit,decorations.pathreplacing}
\tikzstyle{every picture}+=[remember picture]
\tikzstyle{na} = [baseline=-.5ex]

\newcommand\lag{\mathcal{L}}

\newcommand{\CK}{\mathcal{K}}

\newcommand{\CV}{\mathcal{V}}

\newcommand\cB{{\cal B}}

\newcommand\cD{{\cal D}}

\newcommand\cH{{\cal H}}

\newcommand\cL{{\cal L}}
\newcommand\cM{{\cal M}}
\newcommand\cN{{\cal N}}
\newcommand\cO{{\cal O}}

\newcommand\cV{{\cal V}}
\newcommand\cZ{{\cal Z}}

\renewcommand{\Re}{{\rm Re}}
\newcommand{\Tr}{\mbox{Tr}}
\newcommand{\re}{{\rm e}}

\def\R{{L}}

\def\Z{\mathbb{Z}}

\newcommand{\half}{\frac{1}{2}}

\newcommand{\nn}{\nonumber}

\def\i{\mathrm{i}}

\def\beqa{\begin{eqnarray}}
\def\eeqa{\end{eqnarray}}
\def\be{\begin{equation}}
\def\ee{\end{equation}}
\def\bse{\begin{subequations}}
\def\ese{\end{subequations}}

\newcommand{\bem}{\begin{pmatrix}}
\newcommand{\eem}{\end{pmatrix}}

\renewcommand{\=}{\;  = \;}
\def\+{\, + \,}

\def\wt{\widetilde}

\def\bar{\overline}
\def\ads2{AdS$_2$}
\def\ss2{S$^2$}
\def\vm{\text{v.m.}}
\def\cm{\text{c.m.}}

\def\rt2{\sqrt{2}}
\def\gym{g_{\scriptscriptstyle\text{{YM}}}}

\newcommand\sbeqns[1] {\begin{subequations}#1\end{subequations}}
\newcommand\gathr[1] {\begin{gather}#1\end{gather}}
\newcommand\pmat[1]{\begin{pmatrix}	#1 \end{pmatrix}}

\newcommand\qeq{Q_{eq}}

\renewcommand{\Re}{\mbox{Re}}

\def\s{\sigma}

\def\a{\alpha}

\def\k{\kappa}
\def\eps{\epsilon}

\def\l{{\lambda}}

\def\ve{\varepsilon}

\def\r{ r}

\def\S{{\sigma}}
\def\P{{\rho}}

\newcommand\quater{{\textstyle{\frac{1}{4}}}}

\newcommand{\ba}{\begin{array}}
\newcommand{\ea}{\end{array}}

\title{
Supersymmetric localization: \\
$\cN=(2,2)$ theories on S$^2$ and AdS$_2$
}

\author{Alfredo Gonz\' alez Lezcano$^{a}$} 
\author{Imtak Jeon$^{a,b}$} 
\author{and Augniva Ray$^a$}

\affiliation{$^a$ Asia Pacific Center for Theoretical Physics, Postech, Pohang 37673, Korea }
  \affiliation{$^b$ Department of Physics, Postech, Pohang 37673, Korea  }

\emailAdd{alfredo.gonzalez@apctp.org}
\emailAdd{imtakjeon@gmail.com}
\emailAdd{augniva.ray@apctp.org}

\abstract{
Application of the supersymmetric localization method to theories on 
anti-de Sitter spacetime  has received recent interest, yet still remains as a challenging problem.  In this paper, we focus on (global) Euclidean \ads2, on which we  consider
an Abelian $\cN=(2,2)$ theory and implement localization computation to obtain the exact partition function.  For comparison, we also revisit the theory on \ss2 and perform a parallel computation. We refine the notion of equivariant supersymmetry and use appropriate functional integration measure. For \ads2 we choose a supersymmetric boundary condition which is compatible with the principle of variation. 
 To evaluate the 1-loop determinant about the localization saddle, we use index theory and fixed point formula, where we pay attention to the effect of zero modes and their superpartners. The existence of fermionic superpartner of 1-form boundary zero modes is proven.  Obtaining the 1-loop determinant requires expansion of the index that presents an ambiguity, which we resolve using boundary condition. The resulting partition function reveals an  overall dependence on the size of the background manifold, \ads2 as well as~\ss2, as a sum of two types of contributions: a local one from local conformal anomaly through the index computation and a global one coming from zero modes. This  overall size dependence matches with the perturbative 1-loop evaluation using heat kernel method.

}

\begin{document}

\maketitle

%%%%%%%%%%%%%%%%%%%%%%%%%%%%%%%%%%%%%%%%%%%%%%%%%%%%%%%%%%
%\section{Introduction and summary}
\section{Introduction}
\label{sec:introduction}
Supersymmetric localization \cite{Duistermaat:1982vw, berline1982classes, Atiyah:1984px, Witten:1988ze, Witten:1988xj, Nekrasov:2002qd, Pestun:2007rz} is a powerful method for exact computation of supersymmetric observables like the partition function. While it has been successfully applied for many supersymmetric field theories on various compact manifolds (see  \cite{Pestun:2016zxk} and references therein), its application to the theories on non-compact manifold, such as anti-de Sitter space, has received  relatively recent attention, in particular, see \cite{Gupta:2015gga,Murthy:2015yfa} for AdS$_2 \times$S$^2$, \cite{David:2016onq, David:2018pex,David:2019ocd, Cabo-Bizet:2017jsl, Pittelli:2018rpl} for AdS$_2 \times$S$^1$, \cite{Assel:2016pgi} for~AdS$_3$ and even more generally, \cite{Aharony:2015hix} for  AdS$_p \times $S$ ^q$ . The main motivation  for such application would be to provide an exact test of the AdS/CFT correspondence through the exact computation not only for the conformal field theory but also for the supergravity theory on anti-de Sitter space. 
There has been a considerable collective effort to pursue this direction 
\cite{Banerjee:2009af,Dabholkar:2010uh,Dabholkar:2011ec, Gupta:2012cy, Dabholkar:2014wpa, LopesCardoso:2022hvc, Hristov:2021zai, deWit:2018dix,Jeon:2018kec,Iliesiu:2022kny,Gupta:2021roy,Ciceri:2023mjl}.

%\sout{Through the efforts of many contributors, } 
The most concrete and successful example of the application of localization to supergravities so far consists  of computing the entropy of  $1/8$ BPS black hole with asymptotically flat spacetime in type II supergravity  and showing that the degeneracy of states of such black holes is given by integer numbers \cite{Dabholkar:2011ec, Iliesiu:2022kny}. Although this precise result proves to be a stringent test for AdS/CFT correspondence, localization computation on anti-de Sitter space remains both computationally challenging and conceptually less understood \cite{Sen:2023dps} and we need to gain a better understanding of it. This will allow us to consider more generic systems such as less supersymmetric theories or gauged supergravities and eventually collect evidence in favor of the AdS/CFT correspondence and exploit it to explore quantum aspects of gravity.

  Our aim here is to  deepen our understanding of the localization method applied to theories defined on non-compact spaces with boundary, specifically the case of anti-de Sitter space. 
 For this purpose, we focus on a simplified setup which is two-dimensional Euclidean anti-de Sitter space in global frame. This is also motivated by the fact that \ads2 is the common geometric factor that any asymptotically flat extremal black hole contains in its near horizon. 
 On this background, we take an explicit toy example, which is $R$-symmetric $\cN = (2,2)$ Abelian theory and study its partition function using localization method.  For a comparative study we also revisit the evaluation of the partition function of the same theory defined on \ss2 which was already  well studied in the literature \cite{Benini:2012ui,  Doroud:2012xw, Alekseev:2022zao,Park:2016dpb}. By comparing the two parallel computations, we gain better understanding on the localization on \ads2.

For the exact 1-loop computation within the localization technique, we utilize an equivariant index 
\cite{Atiyah:1973ad,Atiyah:1974obx}, which we compute through  Atiyah-Bott fixed point formula \cite{atiyah1966lefschetz,10.2307/1970694,10.2307/1970721}  (see \cite{Pestun:2007rz, Hama:2010av} for application of the index method in supersymmetric localization).
 Although we lack a mathematically rigorous proof for the Atiyah-Bott fixed point formula in non-compact spaces we will assume that it can be applied to \ads2 and we support this assumption with a direct calculation of de-Rham cohomology that we present in  appendix
~\ref{app:Atiyah}.  Our motivation to use the index method is that, in general, it beautifully simplifies the computation,  eventually circumventing potentially prohibitive technical difficulties when analyzing non-compact spaces.  For example, in our case the computation of the determinant of kinetic operators turns into computation of an equivariant index with respect to a $U(1)$ isometry generated by  equivariant supersymmetry $\qeq^2$, and it further reduces down to evaluations only at the fixed points of the $U(1)$ isometry. Some of the complications we can avoid with this method are the following: if we want to deal with any non-trivial background rather than a pure anti-de Sitter space, such as black hole geometry, finding the spectrum itself would be demanding. 
In addition, since the anti-de Sitter space has continuous spectrum,  cancellation between determinant over bosonic and fermionic spectrum may be subtle. Bearing those generic situations and potential complications in mind, we want to use the index method and systematize its implementation in non-compact spaces. 

One direction that we pay special attention to in the localization computation is to obtain the dependence of the partition function on the size of background manifold. In particular, the overall scale dependence  itself is a good observable as it is  protected under renormalization and thus is useful as a test of the AdS/CFT correspondence. Further, we take this quantity to test the validity of our localization method, for which we compare the result with the perturbative 1-loop computation using heat kernel method.  From general analysis of the heat kernel method, it is known that the overall scaling dependence is captured by two contributions \cite{Vassilevich:2003xt}: One is local contribution obtained through the heat kernel expansion, which is local quantity as the heat kernel is evaluated over the complete spectrum in functional space or equivalently, it is obtained through local scalar quantities describing the geometry such as curvature. Here the coefficient of zeroth order in the expansion, called the Seeley-DeWitt coefficient, leads to the conformal anomaly of the theory.  The other is a global contribution obtained through the zero mode effect. 
As the evaluation of 1-loop does not involve its zero modes, one adds the zero modes effect into the heat kernel to make it local, and to compensate this fictitious addition, we subtract the zero modes and treat their effect to the partition function separately. The integration measure of the zero mode may add extra effect.   Explicit example for those two contributions can be found in the study of logarithmic correction to black hole entropy~\cite{ Sen:2012kpz, Sen:2012cj} or free energy~\cite{Bhattacharyya:2012ye}  and many  further works since then.

From the localization computation, we show a similar structure of the overall size dependence obtained as the sum of local and global contributions. The local contribution is captured by the computation of the index that is evaluated over the complete spectrum in the functional space. This reproduces the known anomaly result for theory on \ss2 \cite{Silverstein:1995re, Hori:2006dk, Hori:2013ika} and also obtains the anomaly for the theory on \ads2. We confirm both of them  by computing the Seeley-DeWitt coefficient of heat kernel.  As for the zero modes on the localization saddle, there are boundary zero modes from 1-form field (also from their fermionic superpartner) for the \ads2 case, and one bulk zero mode from a scalar field as a mode along usual localization saddle both for \ss2 and \ads2 cases. For \ss2 case, one also needs to take into account the fact that zero modes of ghost and anti-ghost are eliminated in 1-loop. Thus for the index computation we fictitiously add them and compensate this addition by separately treating those zero mode contributions. All those effects are compared with the heat kernel computation, where we show that the contribution of the zero modes agrees with the one in the localization computation.

The treatment of the zero modes in the context of  localization is subtle and it was prescribed in the context of localization on \ads2$\times$\ss2 \cite{Jeon:2018kec}, where the fermionic superpartner of the boundary zero modes, which were also zero modes of the canonical $\qeq$-exact deformation, are promoted to be non-zero modes by adding corresponding $\qeq$-exact action. We apply the same prescription in this paper. 
For this application, we first give an improved interpretation for the equivariant supercharge whose algebra includes large gauge transformation with all the non-normalizable parameters.
Next, we prove that the fermionic superpartner of the boundary 1-form zero modes exist in the normalizable spectrum of fermion, which was assumed in \cite{Jeon:2018kec}. 
A specific feature of \ads2  that is different from \ads2$\times$\ss2  case is that the fermionic superpartner of the boundary 1-form zero modes are also zero modes of  the kinetic term of physical action, which would make the partition function vanish like the gravitino zero modes would do in the quantum entropy function \cite{Banerjee:2011jp}.  We regard this addition of the $\qeq$-action for those fermionic zero mode as a regulator to obtain finite contributions to the partition function.  For the bulk zero mode along the localization saddle, it turns out that this mode does not contribute to the scaling dependence of partition function both for \ss2 and \ads2 case as it is completely factored out from the index computation. For the case of \ss2,  we do not need any prescription to capture the zero modes of ghost, but we only need to keep track of where and how those zero modes affect the scaling dependence. As a result, we find their contribution to the scaling dependence in addition to the known conformal anomaly contribution \cite{Silverstein:1995re, Hori:2006dk, Hori:2013ika}. 

Since the classical action of the theories we consider does not depend on the scaling of the size of background, all of the size dependence comes through its dependence in the functional integration measure. To determine the functional integration measure, we use  Fujikawa’s method together with supersymmetry. Fujikawa’s ultra-local argument ensures the invariance of the measure under BRST transformation associated to the diffeomorphism~\cite{Fujikawa:1984qk}, and this method was applied to obtain the \ss2 partition function~\cite{Hori:2013ika}. As diffeomorphism does not fully determine the measure for the complex field and its conjugate field, we fix this ambiguity by using supersymmetry such that all the bosonic measure and fermionic measure are related by the same supersymmetry. In fact, the supercharge that maps the bosonsic and fermionic measure is the rescaled one by the length scale of the background, and this rescaled supercharge enters in the computation of the equivariant index. This is how index computation shows the size dependence of partition function as was prescribed in \cite{Gupta:2015gga}.  We realise this idea in our example by explicitly listing the functional integration measure of all the fields including ghost field (and  also ghost of ghost multiplet for \ss2 case).

The main difference between \ads2 and \ss2 is that \ads2 is a non-compact space endowed with its asymptotic boundary while \ss2 has no boundary.  
Therefore, for the case of \ads2, we have to set an asymptotic boundary condition on each field in order to define a theory. We use two criteria to set boundary conditions, which are based on the variational principle and supersymmetry. 
The variational principle relies on what boundary terms we have in the theory. Furthermore, these boundary terms should be chosen such that the total action is supersymmetric  for our purpose of studying a supersymmetric theory. In this paper, however, we do not explore all possible choices of supersymmetric boundary terms. Instead, given the supersymmetric boundary terms that we find, we show that imposing normalizable boundary condition for all bosonic fields and letting supersymmetry transformation to determine the boundary condition of fermionic fields is compatible with the variational principle.\footnote{The resulting boundary condition of fermion says that its asymptotic behavior will be different from the behavior of normalizable modes given in \cite{Camporesi:1992tm}. Therefore, the boundary condition for supersymmetric localization in section \ref{sec:susyloc} and for heat kernel computation in section \ref{sec:HK} are different. This issue will be addressed in the paper \cite{GonzalezLezcano:2023uar} that is currently in preparation. } Under these conditions, we will have to turn off the Fayet-Iliopoulos (FI) parameter and parameter for topological term for \ads2 which are generically allowed for~\ss2.

To read the 1-loop determinant from  the index, we need to expand the index with respect to the equivariant parameter. Here, the way of expansion presents an ambiguity depending on whether to expand in terms of the parameter or its inverse. 
In our specific case for the chiral multiplet, we will have four possibilities.
This would yield to four different possible results for the 1-loop contribution of the chiral multiplet, which is an issue that has been already encountered in the context of localization in non-compact spaces, in particular for AdS$_3$ in \cite{Assel:2016pgi}. In fact, this ambiguity is a generic feature of transversally elliptic operators, and it is also present in the case of compact spaces. In that context it has been given a resolution \cite{Hama:2011ea, Alday:2013lba, Closset:2013sxa} that relays on the compactness of the space.  We classify the possible expansions according to  whether they admit the presence of zero modes in the result. Our boundary conditions will single out one of these options unambiguously. 

We eventually arrive at our final results that we summarize as follows:
\begin{itemize}

\item In \eqref{s2result} we display the exact result of partition function of the theory on \ss2.  Our new contribution here is that we obtain the  scale dependence of the partition function using the method of index for localization. This in particular includes the zero mode contribution in addition to the known conformal anomaly contribution.  This result  calibrates our method to move on towards evaluating the partition function on \ads2.

\item After the choice of boundary conditions that is compatible with supersymmetry and variational principle, we carry out the localization computation on \ads2. This yields the result of full partition function of the theory as we present  in \eqref{Ads2result}.
The result  also captures the conformal anomaly as well as the zero mode contribution to overall scaling dependence of the partition function.  
 To determine the scale independent part of the partition function, we shed light on the role of boundary conditions in eliminating the ambiguities associated to expanding the index of certain  differential operator. Our result is comparable to the hemisphere partition function with certain boundary condition \cite{Hori:2013ika, Honda:2013uca, Sugishita:2013jca}.
\item The overall scale dependence  in the result consisting of local and global contribution is confirmed further by perturbative 1-loop computation using heat kernel method both for \ads2 and \ss2 . 
\end{itemize}

The rest of this paper is organized as follows. In section \ref{sec2}, we set up the  $\cN=(2,2)$ theories on the background \ss2 as well as \ads2 and collect important facts that will be needed for computing the partition function. 
%We give further clarification of the equivariant supercharge and its algebra which was interpreted in~\cite{deWit:2018dix,Jeon:2018kec}, write down supersymmetric action with boundary terms,  define the functional integration measure of the theories, find the on-shell saddle, and organize the cohomological variables. 
We devote section \ref{bdycondition}  to motivate and present our choice of boundary conditions (viz., normalizable boundary conditions for bosonic fields and we let the asymptotic behavior of fermions to be dictated by supersymmetry). 
%To this end we perform the asymptotic analysis that fixes the theory on \ads2. Furthermore, we explicitly check the consistency of our boundary conditions with the variational principle. 
In section~\ref{sec:susyloc}, we compute the partition function using supersymmetric localization and the index method. 
%We first find the localization solutions in section~\ref{subsec:locsad}, which are classified in cohomological variables, and we write down the induced measure of them.  In the following subsection  \ref{subsec:IT}, we apply index theory for the 1-loop determinant, where we analyze how to deal with the zero modes both for \ss2 and \ads2 cases. In particular for \ads2, we also treat the fermionic superpartner of 1-form boundary zero modes. After showing the transversal ellipticity of the operator for the index, we apply fixed point formula to compute the index.  Using appropriate  expansion of the index and regularization, we obtain the 1-loop determinant and then the result of partition function  \ref{subsec:1loop}. 
In section \ref{sec:HK}, we support our localization result by  computing perturbative 1-loop around the on-shell saddle using heat kernel method. We finalize in section \ref{sec:conclusions} with the conclusions. Several technical aspects have been collected in a series of appendices. During our calculations we use 2-component Dirac spinor notation for fermions. Details about the specific representation of the gamma matrices and our conventions for fermion multiplication are shown in appendix \ref{gammaconvention}. 
%We refer the reader to footnote \ref{footnote1} for a precise relation between our notation and that of \cite{Closset:2014pda, Benini:2012ui} for ease of correspondence.
Appendix \ref{sec:appKilling} presents the Killing spinors for both \ss2 and \ads2. In appendix \ref{app:Atiyah} we discuss  the Atiyah-Bott fixed point formula and we apply it on the de-Rham cohomology on \ss2 and \ads2. Appendix \ref{subsubsec:Reg} sketches a proof of the Zeta function regularization that we use to evaluate the 1-loop determinant. In appendix~\ref{basisfunctions } we collect the basis functions for scalar, vector and fermions on both backgrounds. Some derivations in the heat kernel method are presented in appendix \ref{appendixheatkernel}.

%%%%%%%%%%%%%%%%%%%%%%%%%%%%%%%%%%%%%%%%%%%%%%%%%%%%%%%%%%%%%%%%%%%%%%%%%%%%%%%%%%%%%%%%%%%%%%%%%%%%%%%%%%%%%%%%%%%%

\section{$\cN= (2\,,2)$ theory on \ss2 and \ads2} \label{sec2}

In this section, we discuss the  $R$-symmetric $\cN=(2,2)$ theory on S$^2$ and AdS$_2$ presenting details as would be required for us in future sections.
For our purposes it will be sufficient to focus on a theory containing an Abelian vector multiplet and a chiral multiplet. We shall follow the prescription developed in \cite{Closset:2014pda, Festuccia:2011ws} for the construction of such theories in general 2-dimensional manifolds. 
 In subsection \ref{background}, we introduce the gravity multiplet and the Killing spinor equations, and then present the main properties of the supersymmetric backgrounds that are relevant to our work. We then move on to subsection \ref{mattercontents}, where we discuss in detail the theories on those backgrounds. Given our field content, we present how equivariant supersymmetry acts. In particular its action on the ghost field is refined for the case of \ads2.
 We also present the action including appropriate boundary terms for the case of \ads2. Subsection \ref{Sec:Measure} is devoted to define the integration measure and the reality properties we impose on the fields. In subsection \ref{subsec:classicalsaddles} we present the classical saddles. These saddles will be used in section \ref{sec:HK} to do computations using the method of heat kernels. Finally, we end in subsection \ref{subsec:cohomVar} by introducing cohomological variables, which are a reorganization of the physical variables into a representations of the equivariant supersymmetry $\qeq$.

%%%%%%%%%%%%%%%%%%%%%%%%%%%%%%%%%%%%%%%%%%%%%%%
%%%%%%%%%%%%%%%%%%%%%%%%%%%%%%%%%%%%%%%%%%%%%%%

%%%%%%%%%%%%%%%%%%%%%%%%%%%%%%%%%
%%%%%%%%%%%%%%%%%%%%%%%%%%%%%%%%%

\subsection{Supersymmetric background }\label{background}

 To look at the supersymmetric backgrounds we start by considering the gravity multiplet, 
 \be \nonumber \{h_{\mu \nu}, A^{R}_{\mu}, C_{\mu}, \bar{C}_{\mu}, \psi_{\mu }, \bar{\psi}_{\mu }\}\, ,\ee where $h_{\mu \nu}$ is the graviton,  $A^{R}_{\mu}$ is the $R$-symmetry field,  the complex gauge fields $C_{\mu}$ and $\bar{C}_{\mu}$ are the graviphotons and $\psi_{\mu }$ and  $\bar{\psi}_{\mu }$ are the gravitini coupled to supersymmetric current. 
 The graviphotons have dual field strengths given in terms of $H$ and $G$ defined in the following way:
\begin{align}
H+ \i G & = - \i \epsilon^{\mu \nu} \partial_{\mu}C_{\nu} , \hspace{3mm} H - \i G = - \i \epsilon^{\mu \nu} \partial_{\mu}\bar{C}_{\nu}.
\end{align}
where $\epsilon^{\mu \nu}$ is the Levi-Civita tensor. 

The Killing spinor equations governing the $\cN =(2,2)$ supersymmetry on a curved 2-dimensional manifold can be obtained by setting to zero the supersymmetric variation of the gravitini, which yields:
\beqa\label{KSeq}
&&D_\mu \epsilon= -\half H \gamma_\mu \epsilon - \i \half  G \gamma_\mu \gamma_3 \epsilon\,,
\\
&&D_\mu \bar\epsilon = -\half H \gamma_\mu \bar\epsilon + \i \half  G \gamma_\mu \gamma_3 \bar\epsilon\,,\label{KSeq2}
\eeqa
where $D_\mu \epsilon =( \partial_\mu + \omega_\mu + \i A^R_\mu)\epsilon $,   $D_\mu \bar\epsilon =( \partial_\mu + \omega_\mu - \i A^R_\mu)\bar\epsilon $, with $\omega_{\mu}$ the spin connection and $A_{\mu}^{R}$ the R-symmetry background field. The Killing spinors, $( \epsilon\,, \bar\epsilon )$, have R-charge~$(-1\,, +1)$ and mass dimension~$(-\half\,, -\half)$ respectively.

For later purposes, it will be convenient to note that the auxiliary fields $H $ and $G$ are related to the Ricci curvature as
\be\label{scalarcurvature}
R = -2 (H^2 +G^2)\,,
\ee
which can be obtained from the integrability condition 
\be
\left[D_\mu\,, D_\nu \right]\epsilon = \frac{1}{4}R_{\mu\nu }{}^{ab}\gamma_{ab}\epsilon\,,
\ee
 and  the Killing spinor equation \eqref{KSeq} assuming that the $R$-charge background connection is flat.

\subsection*{The S$^2$ and AdS$_2$ background}

All the previous data considerably simplifies for the cases of 
 S$^2$ and AdS$_2$ that we will ultimately study. In the following, we collect some relevant information about S$^2$ and AdS$_2$ that will serve to fix terminology and the notation that we employ along the paper. The metrics have the form
\sbeqns{\gathr{ 
		\text{\ss2: \quad}ds^2 =L^2 (d\psi^2 + \sin^2\psi ~d\theta^2)\,, \qquad 0 \le \psi <\pi\,,~  0\le \theta <2 \pi\,, \label{spheremetric} \\
		\text{\ads2: \quad}ds^2 =\R^2 (d\eta^2 + \sinh^2\eta ~d\theta^2)\,, \qquad 0 \le \eta <\infty\,, ~ 0\le \theta <2 \pi\,,\label{adsmetric}
	}
}
where $L$ is a length scale associated to the size of each manifold.
Given these background metrics, obtaining the supersymmetric backgrounds by solving the Killing spinor equations \eqref{KSeq} and \eqref{KSeq2} requires us to set the background fields $A^R_\mu$, $H$ and $G$  to be: $ H= \mp \i L^{-1}\,,~A^{R}_\mu = G=0\,,~~~~\text{or   }~~  G= \mp \i L^{-1}\,,~A^{R}_\mu = H=0$ on S$^2$ and $H= \mp L^{-1}\,,~A^{R}_\mu = G=0\,,~~~~\text{or   }~~ G= \mp L^{-1}\,,~A^{R}_\mu = H=0   $ on AdS$_2$. For later convenience, we would like the two Killing spinors $\epsilon$ and $\bar{\epsilon}$ to be a symplectic Majorana pair, which is guaranteed by the following choice
\sbeqns{\gathr{ 
		\text{\ss2: \quad} H= -  \frac{\i}{L}\,,\quad A^{R}_\mu = G=0\,,\label{auxiliaryS2} \\
\text{\ads2: \quad} G= -\frac{1}{ L}\,,\quad A^{R}_\mu = H=0    \,, \label{auxiliaryads2}
	}
}

\subsection*{Killing spinors on S$^2$} 

For S$^2$ we have that the  Killing spinor equations acquire the form
\be
D_\mu \epsilon =    \frac{\i}{2L} \gamma_\mu \epsilon\,,\qquad D_\mu \bar\epsilon =    \frac{\i}{2L} \gamma_\mu \bar\epsilon\, \label{eq:KSS21}.%\,,~~~~D_\mu \ve^2_{\textrm{S$^2$}} = - s^2 \i \frac{1}{2}\tau_3 \tau_\mu \ve^2_{\textrm{S$^2$}} 
\ee
Here, the covariant derivative is $D_\mu = \partial_\mu +\omega_\mu$ and the non-trivial component of the spin connection is $\omega_\theta = -\half \cos\psi \, \gamma_{12}$. 
 The general solutions to \eqref{eq:KSS21} are presented 
in appendix \ref{S2KS}. However we single out the solution
\be\label{KSonS2}
\epsilon=\sqrt{L}{\rm e}^{\i\theta/2}\begin{pmatrix}  \cos\frac{\psi}{2}\\ \i  \sin \frac{\psi}{2} \end{pmatrix}\,,\qquad\bar{\epsilon}=\sqrt{L}{\rm e}^{-\i\theta/2}\begin{pmatrix}   -\i \sin\frac{\psi}{2}\\ - \cos \frac{\psi}{2} \end{pmatrix}\,,
\ee
for the purpose of the localization. They form a symplectic Majorana pair as
\be\label{realityKS}
\epsilon^\dagger = \i \bar{\epsilon}^T C\,,\quad\quad \bar{\epsilon}^\dagger = -\i \epsilon^T C\,,
\ee
and thus have the following orthogonality property
\be
\epsilon^\dagger \bar\epsilon =0\,,\qquad\bar\epsilon^\dagger \epsilon =0\,.
\ee
More detailed properties of this choice of Killing spinors are used in section \ref{S2KS}.
In terms of $\bar\epsilon$ and $\epsilon$ it is possible to define the following bispinors
\beqa \label{bispinorsS2}
\bar{\epsilon}\gamma_A \epsilon & = & L \left(0\,, -\i \sin\psi\,,-\i \cos \psi\,, -\i  \right)\,, \\ \nonumber
\epsilon \gamma_A \epsilon & =&L  {\rm e}^{\i \theta} \left(- \i \,, \cos \psi \, , - \sin \psi \, , 0 \right)\, , \\ \nonumber
\bar{\epsilon} \gamma_A \bar{\epsilon} & = & L {\rm e}^{- \i \theta} \left(\i \,, \cos \psi \, , - \sin \psi \, , 0\right),
\eeqa
where we have denoted $\gamma_A \equiv (\gamma_1\,,\gamma_2\,,\gamma_3\,,1)$.
In particular, the corresponding Killing vector is  associated to the following $U(1)$ rotation of S$^2$,
\be
\xi \,\equiv\, \i \bar{\epsilon}\gamma^\mu \epsilon \partial_\mu \= \partial_\theta\,.
\ee

\subsection*{Killing spinors on AdS$_2$} In this case we have the following Killing spinor equations
\be
D_\mu \epsilon =    \frac{\i }{2L} \gamma_\mu\gamma_3 \epsilon\,,\qquad D_\mu \bar\epsilon =   -  \frac{\i}{2L} \gamma_\mu \gamma_3 \bar\epsilon\,.
\ee
Here, the covariant derivative is $D_\mu = \partial_\mu +\omega_\mu$ and the non-trivial component of the spin connection is $\omega_\theta = -\half \cosh\eta \, \gamma_{12}$. 
Among the general solutions presented in appendix \ref{AdS3KS}, we  select
\be\label{KSonAdS2}
\epsilon=\sqrt{L}{\rm e}^{\i\theta/2}\Biggl(\ba{c}  \cosh\frac{\eta}{2}\\ \i  \sinh \frac{\eta}{2} \ea\Biggr)\,,\qquad \bar{\epsilon}=\sqrt{L} {\rm e}^{-\i\theta/2}\Biggl( \ba{c} -\i \sinh\frac{\eta}{2}\\ - \cosh \frac{\eta}{2}\ea \Biggr)\,,
\ee
for the purpose of the localization. They from a symplectic Majorana pair as
\be\label{realityKS2}
\epsilon^\dagger = \i \bar{\epsilon}^T C  \,, \quad \quad \bar{\epsilon}^\dagger =-\i \epsilon^T C \,,
\ee
and thus have the following orthogonality property
\be
\epsilon^\dagger \bar\epsilon =0\,,\qquad\bar\epsilon^\dagger \epsilon =0\,.
\ee
More detailed properties of this choice of Killing spinors are used in section \ref{AdS3KS}. The bispinors between $\bar\epsilon$ and $\epsilon$ are given by
\beqa \label{bispinorsAdS2}
\bar{\epsilon}\gamma_A \epsilon & = & L\left(0\,, -\i \sinh\eta\,, -\i \,,-\i \cosh \eta \right)\,, \\ \nonumber
\epsilon \gamma_A \epsilon & = &L {\rm e}^{i \theta}\left( - \i \cosh \eta\, , 1 \, , - \sinh \eta \, ,0 \right)\, ,\\ \nonumber
\bar{\epsilon} \gamma_A \bar{\epsilon} & = & L {\rm e}^{-i \theta}\left(  \i \cosh \eta\, , 1 \, , - \sinh \eta \, ,0 \right) \, .
\eeqa
In particular, the corresponding Killing vector is associated to $U(1)$ rotation of AdS$_2$,
\be
\xi \,\equiv\, \i \bar{\epsilon}\gamma^\mu \epsilon \partial_\mu \=  \partial_\theta\,.
\ee

\subsection{Equivariant supersymmetry, multiplets and action}\label{mattercontents}
On the \ss2 or \ads2 backgrounds described above, we consider $\cN= (2,2)$ Abelian vector multiplet including ghost multiplet (and ghost of ghost multiplet for \ss2), and chiral multiplet with gauge charge.  

The vector multiplet consists of a vector, two real scalars, two Dirac spinors and an auxiliary scalar, 
\be
\mbox{Vector}~ :\quad \{ A_\mu \,, \S \,,\P  \,,\lambda\,,\bar\lambda\,,\hat{D}\}\,,
\ee 
whose $R$-charge  assignment is $(0,0,0,-1, 1,0)$ and the mass dimensions are (0\,,1\,,1\,, $\frac{3}{2}$\,,$\frac{3}{2}$\,,2).
For systematic treatment of gauge fixing using BRST quantization, we include the ghost multiplet to the vector multiplet, consisting of ghost, anti-ghost and auxiliary field, 
\be
\mbox{Ghost}~ :\quad \{ c \,, \Bar{c} \,,b \}\,,
\ee
Here, the Grassmann odd scalars $c$ and $\Bar{c}$ are ghost and anti-ghost fields and Grassmann even scalar $b$ is auxiliary field, whose $R$-charges are all zero and the mass dimensions are $(0\,,2\,,2)$.
By adding this ghost multiplet the vector multiplet now has $6+6$ bosonic and fermionic degrees of freedom respectively.
In the case of our theory on~\ss2, there are zero modes of the ghost fields which we need to freeze out. To this end we further include the ghost of ghost multiplet,
\be\label{GGmultiplet}
\mbox{Ghost of ghost}~ :\quad  \{\Lambda_0\,, b_0\,, c_0\,, \Bar{\Lambda}_0\,,\Bar{c}_0\}\,.
\ee
Here $\Lambda_0\,,\Bar{\Lambda}_0\,, b_0$ are Grassmann even and $\Bar{c}_0\,, c_0$ are Grassmann odd variables, where  we in particular call $\Lambda_0$ as the `ghost of ghost'. Their mass dimensions are given as $(0\,,2\,,2\,,4\,,4)$. These variables are not defined as local fields but just defined as certain modes: in fact they are constant modes as the zero modes of ghosts for \ss2 case are constants.  The role of each variable will be clear later in this subsection.  

The chiral multiplet consists of two complex scalars, two Dirac spinors and two auxiliary fields,
 \be
\mbox{Chiral}~ :\quad  \{\phi\,, \bar{\phi}\,, \psi\,, \bar{\psi}\,, F\,, \bar{F}\}\,,
 \ee
whose  $R$-charge assignment is $(\r \,, - \r \,, \r -1 \,, -\r +1\, , \r-2\, , - r+2 )$, the gauge charges are $(1\,, -1\,, 1\,,-1\,, 1\,, -1)$, and the mass dimensions are ($0$\,,$0$\,,$\frac{1}{2}$\,,$\frac{1}{2}$\,,$1$\,,$1$). \footnote{
As a $\cN=(2,2)$ multiplet, its Weyl scaling dimension is ($\frac{r}{2}$\,,$\frac{r}{2}$\,,$\frac{r+1}{2}$\,,$\frac{r+1}{2}$\,,$\frac{r+2}{2}$\,,$\frac{r+2}{2}$). }

Note that in Euclidean space, in contrast to the Lorentzian space, the barred field and unbarred field are not related by complex conjugation.   For bosonic fields, we will impose appropriate reality conditions  to ensure the kinetic term to be positive. For the barred and unbarred Dirac fermions, we will treat them to be independent, so the fermion degrees of freedom are formally doubled \cite{Festuccia:2011ws}. As the formal doubling doesn't mean the number of path integral measures have doubled, we still have $4+4$ bosonic and fermionic physical degree of freedom both for vector and chiral multiplet.  More details  on the reality condition will be discussed subsection \ref{Sec:Measure}.

\subsection*{Equivariant supersymmetry }
Let us denote the supercharge $Q$ as
\be
Q\;\equiv\; Q_\epsilon + Q_{\bar{\epsilon}}\,,
\ee
where $Q_\epsilon$ and $Q_{\Bar{\epsilon}}$ are supersymmetry variation with the Grassmann even Killing spinors $\epsilon $ and $\Bar{\epsilon}\,$, i.e. 
$
Q_\epsilon \;\equiv\; \delta(\epsilon)$ and $   Q_{\bar{\epsilon}}\;\equiv\; \delta({\bar{\epsilon}})\,.
$
Let us further define equivariant supercharge $\qeq$ by combining the supercharge with BRST charge for the $U(1)$ gauge symmetry as
\be\label{qeqDef}
\qeq \;\equiv \;Q + Q_{\text{brst}}\,.
\ee
Here, $\qeq$ stems from the BRST charge of  supergravities \cite{deWit:2018dix, Jeon:2018kec}, where all the spacetime symmetries including supersymmetries are local gauge symmetries.  In fact the Killing spinors generating the rigid supersymmetry $Q$ are  the background values of the ghost fields associated to the local supersymmetries.  As a consistent supergravity theory should be invariant under BRST symmetry, our gauge theory should also be invariant under the $\qeq$ symmetry.   

For the vector multiplet fields,
the  transformation rule of the equivariant supercharge is given by 
\be\ba{lll}
\label{eq:deltaA}
\qeq A_\mu &=& -\i \half (\bar\epsilon \gamma_\mu \lambda + \epsilon \gamma_\mu \bar\lambda) + \partial_\mu c \\
\qeq \S  &=& -\half  (\bar\epsilon \lambda - \epsilon  \bar\lambda)\\
\qeq \P  &=& -\i \half  (\bar\epsilon \gamma_3 \lambda + \epsilon \gamma_3 \bar\lambda)\\
\qeq \lambda &=& \i \gamma_3 \epsilon F_{12} -\hat{D} \epsilon - \i \gamma^\mu \epsilon \,\partial_\mu \S  - \gamma_3 \gamma^\mu \epsilon \,\partial_\mu \P  + (\i H - G\gamma_3)\epsilon \,\S  +(H \gamma_3 +\i G) \epsilon\,\P  

\\
&=& \i \gamma_3 \epsilon F_{12} -\hat{D} \epsilon - \i \gamma^\mu D_\mu (\epsilon \S ) - \gamma_3 \gamma^\mu D_\mu (\epsilon \P  )
\\
\qeq \bar\lambda &=& \i \gamma_3 \bar\epsilon F_{12} +\hat{D} \bar\epsilon +
 \i \gamma^\mu \bar\epsilon  \,\partial_\mu  \S  - \gamma_3 \gamma^\mu \bar\epsilon \,\partial_\mu   \P  -(\i H +G\gamma_3)\bar\epsilon  \,\S  + (H \gamma_3 -\i G) \bar\epsilon  \,\P  
 \\
 &=& \i \gamma_3 \bar\epsilon F_{12} +\hat{D} \bar\epsilon +
 \i \gamma^\mu D_\mu( \bar\epsilon  \S ) - \gamma_3 \gamma^\mu D_\mu ( \bar\epsilon  \P  )
 \\
\qeq \hat{D}&=& - \i\half  \bar\epsilon \gamma^\mu  D_\mu  \lambda + \i \half \epsilon \gamma^\mu  D_\mu  \bar\lambda   - \i\half  \bar\epsilon( H+\i G\gamma_3) \lambda + \i \half  \epsilon( H -\i G\gamma_3) \bar\lambda  

\\
&=& - \i\half  D_\mu (\bar\epsilon \gamma^\mu \lambda) + \i \half D_\mu (\epsilon \gamma^\mu  \bar\lambda  )
\,,
\ea\ee
where $F_{12}= \half \epsilon^{\mu\nu}F_{\mu\nu}$.\footnote{\label{footnote1}To compare with the notation in \cite{Benini:2012ui}, we redefine
$
\S  = -\sigma
$ and $\P  =\eta$. 
 To compare with \cite{Closset:2014pda}, we choose the charge conjugation matrix $C= -\i \tau_2$ instead of $C = \tau_2$, set the gamma matrix representation $\gamma_a =(-\tau_1\,,-\tau_2)$, and do the following redefinition of the supersymmetry parameters (on the left hand side we write our parameters and on the right hand side those of \cite{Closset:2014pda}):
\begin{equation*}\ba{l}
\epsilon_\pm \rightarrow -\sqrt{2}\tilde\zeta_\mp\,,~~~~\bar\epsilon_\pm \rightarrow \sqrt{2}\zeta_\mp\,,
\ea\end{equation*}
the vector multiplet fields
\begin{equation*}\ba{ll}
\S  \rightarrow \half (\sigma+\tilde{\sigma})\,,~~~~&\P  \rightarrow \frac{1}{2\i}(\sigma-\tilde{\sigma})\,,
\\
\lambda_\pm \rightarrow \sqrt{2} \tilde\lambda_\mp\,,~~~~&\bar\lambda_\pm \rightarrow - \sqrt{2}\lambda_\mp\,,
\\
\hat{D}\rightarrow -\i D\,,
\ea\end{equation*}
the chiral multiplet fields
\begin{equation*}\ba{lll}
\phi \rightarrow \phi \,,~~~~~&\bar \phi \rightarrow \tilde\phi\,,
~~~~~&F\rightarrow F \,,~~~\bar{F}\rightarrow - \tilde{F}\,,\\
\psi_\pm \rightarrow \psi_{\mp}\,,~~~&\bar\psi_\pm \rightarrow \tilde\psi_{\mp}\,,\\
\ea\end{equation*}
}

To complete the vector multiplet under $\qeq$, we include  the ghost multiplet fields which transform as
\be\label{qeqGhost}
\qeq c\= - \Lambda + \Lambda_0\,,\qquad
\qeq \Bar{c}\= b\,,\qquad
\qeq b \= \xi^\mu  \partial_\mu  \Bar{c}\,.
\ee
Here, we have used $\Lambda$ defined as
\be\label{LambdaParameter}
\Lambda \; \equiv \; -\xi^\mu A_\mu -\bar\epsilon \epsilon \S  - \i \bar\epsilon \gamma_3 \epsilon\, \P  \,,
\ee
and one can check using the Fierz identity that it is invariant under $Q$ and thus 
\be\label{QeqLambda}
\qeq \Lambda \= -\xi^\mu \partial_\mu c\,.
\ee
In the first equation of \eqref{qeqGhost}, the $\Lambda_0$ is introduced  in order to  make the equality valid by eliminating the modes of $\Lambda$ in \eqref{LambdaParameter} that would not match with any modes in $\qeq c$ on the left-hand side of the equality. For the case of \ads2, the ghost field~$c$ is a normalizable scalar in order not to gauge away the large gauge symmetry of~$U(1)$, and thus its superpartner $\qeq c$ should also be normalizable. However, the quantity $\Lambda$~\eqref{LambdaParameter}  can include non-normalizable modes due to the asymptotically diverging behavior of bi-spinor, $\Bar{\epsilon}\epsilon = -\i \cosh\eta$ as in \eqref{bispinorsAdS2}, as well as existence of non-normalizable modes in $\xi^\mu A_\mu $. Therefore, we remove such non-normalizable modes of $\Lambda$ by introducing $\Lambda_0$ in  \eqref{qeqGhost}. For the case of \ss2, the ghost field $c$ does not include a constant mode as it is zero mode, otherwise it would spuriously eliminate unphysical degrees of freedom that were not part of the spectrum to begin with, since constant modes produce a trivial pure gauge transformation. Therefore, we remove the constant mode of $\Lambda$ by introducing $\Lambda_0$ in \eqref{qeqGhost}. 
In summary the variable $\Lambda_0$ is identified as \footnote{In \cite{deWit:2018dix,Jeon:2018kec}, the $\Lambda_0$ was interpreted as background value of $\Lambda$ and it makes the algebra closed in equivariant way. However, we would like to point out that $\Lambda_0$ is not restricted only to the background value of $\Lambda$ but should include all the modes of $\Lambda$ that would not match with $\qeq c$, otherwise the equality in \eqref{qeqGhost}  would not be valid.  Moreover, since $\Lambda_0$ is a singlet under $\qeq$, as explained after \eqref{L0ads2}, the algebra is still equivariantly closed as can later be seen in \eqref{EqAlgebra}.}
\beqa
&\mbox{for \ss2:}& \qquad \Lambda_0\= \mbox{constant mode of } \Lambda\,,\label{L0s2} \\
&\mbox{for \ads2:}& \qquad \Lambda_0\= \mbox{non-normalizable modes of 
} \Lambda\,.\label{L0ads2}
\eeqa
 We note that the $\Lambda_0$ is singlet under $\qeq$ transformation,
 i.e., $
 \qeq \Lambda_0 =0.
 $ 
 This is because~$\qeq \Lambda_0$  would give constant or non-normalizable part of  $ \xi^\mu \partial_\mu c $ according to~\eqref{QeqLambda}. However, the ghost field $c$ doesn't include the constant or non-normalizable modes in it.

For the case of \ss2, we can deal with the ghost field $c$ within a complete Hilbert space by including its constant zero mode in it. It is done by promoting $\Lambda_0$ to a $\Lambda$ independent  constant called `ghost of ghost':
\be\label{a0ghostofghost}
\Lambda_0 \;\rightarrow \; \mbox{ ghost of ghost}\,.
\ee
 Then,  the first equality \eqref{qeqGhost} is valid because on the left hand side we include zero mode of ghost and on the right hand side we introduce new constant $\Lambda_0$. The new constant field $\Lambda_0$  is accompanied by the constant multiplet \eqref{GGmultiplet} which have transformation rule as 
\be\ba{lll}\label{qeqGGmultiplet}
\qeq \Lambda_0 \= 0\,,\qquad\qquad &\qeq \bar{\Lambda}_0\= \bar{c}_0\,, \qquad\qquad &\qeq b_0\= c_0\,,
\\
 &\qeq \bar{c}_0 \= 0 \,,\qquad\qquad  &\qeq c_0 \=0\,.
\ea\ee
  We will see in \eqref{actionghost} that integration over $\Bar{\Lambda}_0$ identifies the~$\Lambda_0$ with the constant mode of $\Lambda$ as was originally introduced as in \eqref{L0s2}. Also, the constant zero mode of ghost field is removed via the on-shell condition of ghost of ghost multiplet. 
  
For the chiral multiplet fields, the transformation rule for the equivariant supercharge is given by
\be\ba{l} \label{deltachiral}
 \qeq \phi = \bar\epsilon \psi+ \i c \phi\,\\
\qeq \bar\phi = \epsilon \bar\psi - \i c\bar\phi \\
\qeq \psi = \i \gamma^\mu \epsilon D_\mu \phi  
-\i \epsilon (\S +\frac{\r}{2} H)\phi   +\gamma_3 \epsilon (\P +\frac{\r}{2}G)\phi +\bar\epsilon F
+ \i c \psi \\
\qeq \bar\psi = \i \gamma^\mu \bar\epsilon D_\mu \bar\phi  
-\i \bar\epsilon ( \S +\frac{\r}{2} H)\bar\phi   -\gamma_3 \bar\epsilon ( \P +\frac{\r}{2}G)\bar\phi +\epsilon \bar F - \i c \Bar{\psi}
\\
\qeq F =    \i \epsilon\gamma^\mu D_\mu \psi  +\i  \epsilon\psi (\S +\frac{\r}{2}H)+\epsilon\gamma_3 \psi ( \P +\frac{\r}{2}G)   -\i \epsilon\lambda \phi + \i c F
\\
\qeq \bar F =    \i \bar\epsilon\gamma^\mu D_\mu \bar\psi  +\i  \bar\epsilon\bar\psi (  \S +\frac{\r}{2}H )-\bar\epsilon\gamma_3 \bar\psi ( \P +\frac{\r}{2}G)     +\i \bar\epsilon \bar\lambda \,\bar\phi- \i c \Bar{F} \,.
\ea\ee
Here, the covariant derivative on each field is summarized as 
\begin{equation}
    D_\mu \= \partial_\mu +\omega_\mu -\i  G_{U(1)}  A_\mu\,, 
\end{equation}
where $\omega_\mu$ is spin connection and $G_{U(1)}$ is gauge charge. 

\subsection*{The  algebra}
If we consider only supercharge $Q$ without BRST charge, then its  algebra closes off-shell with field dependent symmetry parameter as follows,
\be
\label{susyalgebra}
Q^2 = \cL_\xi + %\text{Gauge}
\delta_{\text{gauge}} (\Lambda) + \delta_{R}(\Lambda_R )\,,
\ee
where $\cL_\xi$ is Lie derivative along the Killing vector $\xi^\mu \equiv \i \Bar{\epsilon}\gamma^\mu \epsilon$, $\Lambda$ is the  field dependent parameter defined in \eqref{LambdaParameter}  and the $\Lambda_R$ is $R$-symmetry parameter given by
\be\ba{l}
\Lambda_R \= -\frac{1}{4} ((D_\mu \bar\epsilon)  \gamma^\mu \epsilon  - \bar\epsilon  \gamma^\mu D_\mu \epsilon ) \=  -\frac{1}{2}\left(H \bar{\epsilon} \epsilon +  \i G \bar{\epsilon} \gamma_3 \epsilon\right)\,.
\ea
\ee
For the particular choice of the Killing spinors, viz., \eqref{KSonS2} for \ss2 or \eqref{KSonAdS2} for \ads2, the symmetry parameters in the supersymmetry algebra in \eqref{susyalgebra} are given by
\beqa
&&\textrm{S}^2:\qquad \xi^\mu\partial_\mu = \partial_\theta  %=\left(0\,,1\right)
\,,\qquad \Lambda= -  A_{\theta} +\i L \S -  \cos\psi\, L \P  \,, \qquad \Lambda_R= \frac{1}{2}\, ,   \label{eq:LS2}
\\
&&\textrm{AdS}_2:\qquad \xi^\mu \partial_\mu = \partial_\theta%\left(0\,,  1\right)
\,,  \qquad \Lambda = - A_\theta + \i \cosh\eta\, L\S  - L\P  \,, \qquad \Lambda_R = \frac{1}{2}\, \,.
\label{eq:LAdS2}
\eeqa
We note that the Killing vector generates compact $U(1)$ isometry of \ss2 or \ads2, where in particular it preserves the boundary of \ads2.
 We also note that the gauge symmetry generated with the parameter $\Lambda$ includes not only the `canonical' gauge symmetry but also large gauge transformation. 

If we consider the equivariant supercharge $\qeq$  as defined in \eqref{qeqDef}, which is indeed what we use for the supersymmetric localization in section \ref{sec:susyloc}, then its algebra closes only with ``large gauge" transformation parameter as follows,
\be\label{EqAlgebra}
\qeq^2 \= \cL_\xi + %\text{Gauge}
\delta_{\text{gauge}} (\Lambda_0) + \delta_{R}(\Lambda_R )\,.
\ee
We note that, due to the transformation rule of the ghost \eqref{qeqGhost}, the gauge symmetry parameter $\Lambda$ in \eqref{susyalgebra} is replaced by the $\Lambda_0$ that is identified in \eqref{L0s2} or \eqref{L0ads2} for \ss2 or \ads2 respectively. This transformation indeed generates only the large gauge transformation since it is not gauged away by the ghost field. 

%%%%%%%%%%%%%%%%%%%%%%%%%%%%%%%%%%%%%%%%%%

%%%%%%%%%%%%%%%%%%%%%%%%%%%%%%%%%%%%%%%%%%%%%%%%%%%%%%%%%%%%%%%%%%%%%%
\subsection*{Supersymmetric action} 
The total action of the theory that we consider is of the form: 
\begin{align}
S_{\text{tot}} & = S_{\text{v.m.}} + S_{\text{ghost}}+ S_{\text{FI}}+ S_{\text{top.}} + S_{\text{c.m.}} %+ S_{W+\bar{W}}
\,. \label{eq:Ltot}
\end{align}

For the vector multiplet, the supersymmetric action  consists of bulk Lagrangian and  total derivative terms which gives boundary terms as
\be\label{Actionvector}
S_{\text{v.m.}}=\frac{1}{\gym^2} \int {\textrm{d}^2 x} \sqrt{g}\left[ \cL^{\text{bulk}}_{\text{v.m.}} + D_{\mu}V^\mu_{\text{v.m.}}\right]\,,
\ee
 where $\gym$ is the super-renormalizable gauge coupling.
In the action \eqref{Actionvector}, the bulk Lagrangian is given by  
\begin{align} 
\begin{split}
%S_{\text{v.m.}}
\cL^{\text{bulk}}_{\text{v.m.}}
& =%\frac{1}{\gym^2} \int {\textrm{d}^2 x} \sqrt{g}\,\Bigl(
\frac{1}{2} \Big(F_{12} + \i (G \S  - H \P  )\Big)^2+ \frac{1}{2}\partial_{\mu }\S  \partial^{\mu}\S  +  \frac{1}{2}\partial_{\mu }\P  \partial^{\mu}\P  \label{eq:LYM}\\
& 
\quad +\frac{1}{2} \Big(\hat{D} - \i \left( H\S  + G \P  \right)\Big)^2+ \frac{\i}{2} \bar{\lambda} \gamma^{\mu}D_{\mu}\lambda
\\
&= 
\frac{1}{2} \Big(F_{12} + \i (G \S  - H \P  )\Big)^2+ \frac{1}{2}\partial_{\mu }\S  \partial^{\mu}\S  +  \frac{1}{2}\partial_{\mu }\P  \partial^{\mu}\P   +\frac{1}{2} D^2+ \frac{\i}{2} \bar{\lambda} \gamma^{\mu}D_{\mu}\lambda\,,
\end{split}
\end{align}
where for the second line, we have used field redefinition of the auxiliary field for convenience
\beqa\label{Dredef}
D&\;\equiv \; &   \hat{D}-\i \left( H\S  + G \P  \right)\,.
\eeqa
For \ads2 case ($H= 0\,, G= -\frac{1}{L}$), it is essentially shifting  the path integration contour of $\hat{D}$ by $\frac{\i}{L} \P $ in the complex plane. Then imposing reality condition such that $D$ real   makes the $D^2$ term in the action manifestly positive. \footnote{ Similarly, one can think of redefining the $F_{12}$ to absorb the $-\frac{\i}{L} \S$ and make the term, $(F_{12} -\frac{\i}{\R} \S)^2$ in the action on \ads2, manifestly positive. However, this field redefinition may not be possible because $\S$ is not dual of a 2-form  and thus finding the corresponding redefinition of 1-form field $A_\mu$ is not possible. 
Therefore, treating the term $(F_{12} -\frac{\i}{\R} \S)^2$ causes some modes having wrong sign of the action (see subsection \ref{ads2vm}).  To treat this wrong sign,  we implicitly use analytic continuation of the path integral in the same spirit of \cite{Sen:2012kpz}.  
}

The total derivative term in the action 
\eqref{Actionvector} is to make the action supersymmetric without ignoring the boundary term when taking the supersymmetry variation. It is determined by the fact that the vector multiplet action is $\qeq$-exact as
\beqa
\textrm{S$^2$}:&& \frac{1}{4 \i \bar \epsilon \epsilon} \qeq \left(\overline{\left( Q \lambda \right)}~  \lambda + \overline {\left(Q \bar {\lambda}\right)}~  \bar {\lambda}\right)\= \cL_{\text{v.m.}}+ D_\mu V^\mu\,,
\\
\textrm{AdS$_2$:}&&-\frac{1}{4\i \Bar{\epsilon}\gamma_3\epsilon}\qeq\left(\overline{\left( Q \lambda \right)}~  \lambda + \overline {\left(Q \bar {\lambda}\right)}~  \bar {\lambda}\right)\= \cL^{\text{bulk}}_{\text{v.m.}} + D_\mu V^\mu\,,
\eeqa
with
\beqa
\textrm{S$^2$}:&&  V^\mu \=\frac{1}{4\i\Bar{\epsilon}\epsilon} \left[ \dfrac{\i}{2} \left( \bar \epsilon \gamma^{\mu} \bar \epsilon ~ \lambda \lambda +  \epsilon \gamma^{\mu}  \epsilon ~ \bar \lambda \bar \lambda \right) + \i  \bar \lambda \gamma^\mu \lambda \right]\, ,
\\
\textrm{AdS$_2$:}&&V^\mu \=\frac{1}{4\i\Bar{\epsilon}\gamma_3 \epsilon} \left[- \dfrac{1}{2} \epsilon^{\mu \nu} \left( \bar \epsilon \gamma_\nu \bar \epsilon ~ \lambda \lambda +  \epsilon \gamma_\nu  \epsilon ~ \bar \lambda \bar \lambda \right) + \i \bar \lambda \gamma^\mu \lambda \right]\, . \label{eq:bdryLambda}
\eeqa
Here,  we denote  the \textit{bar} operation on the bracket, $({Q \lambda})$ and $({Q \bar{\lambda}})$,  
as the following exchange
\beqa
\textrm{S$^2$}:&& \overline{\left(Q \lambda\right)}  \equiv Q \lambda \Big{|}_{\epsilon \rightarrow \i \bar \epsilon}\, , \hspace{5mm} \overline{\left(Q \bar{\lambda}\right)}  \equiv Q \bar \lambda \Big{|}_{\bar \epsilon \rightarrow - \i \epsilon}\,,
\\
\textrm{AdS$_2$:}&&\overline{\left(Q \lambda\right)}  \equiv Q \lambda \Big{|}_{\epsilon \rightarrow \i \gamma_3 \bar \epsilon}\, , \hspace{5mm} \overline{\left(Q \bar{\lambda}\right)}  \equiv Q \bar \lambda \Big{|}_{\bar \epsilon \rightarrow \i \gamma_3 \epsilon}\, .
\eeqa
Note that for \ss2 case, taking \textit{bar} operation is equivalent to taking dagger operation.

The supersymmetric action for the ghost multiplet is given by $\qeq$-exact form \cite{Pestun:2007rz}
\be\label{actionghost}
S_{\text{ghost}}\=\frac{1}{\gym^2} \qeq \int {\textrm{d}^2 x} \sqrt{g}\, \i \Big(\bar{c}\nabla_\mu A^\mu + \Bar{c}b_0 + c \Bar{\Lambda}_0 \Big)\,.
\ee
The last two terms are absent for \ads2 case. For \ss2 case, using the transformation rules~\eqref{eq:deltaA},~\eqref{qeqGhost} and~\eqref{qeqGGmultiplet}, we can expand the action \eqref{actionghost} as
\be\label{actionghostExpand}
S_{\text{ghost}}=\frac{1}{\gym^2}\! \int {\textrm{d}^2 x}\! \sqrt{g}\, \i \Big( b\nabla_\mu A^\mu - \bar{c}\nabla^2 c -\bar{c}\nabla_\mu \lambda^\mu + b b_0 - \Bar{c}c_0- c \Bar{c}_0 -(\Lambda\! -\! \Lambda_0) \Bar{\Lambda}_0\Big)\,.
\ee
The first two terms are standard ghost multiplet action, giving gauge fixing condition as well as the Faddeev Popov determinant. The third term is irrelevant  because $\bar{c}$ can be connected only to $c$ but there are no vertices in those extra terms containing $c$. From the fourth to sixth term,  the zero modes of $\{c\,,\Bar{c}\,,b\}$ are eliminated by the integration over $\{b_0 \,, c_0\,,\Bar{c}_0\}$. From the last term, integration over $\bar{\Lambda}_0$ provides the identification of $\Lambda_0$ and the constant modes of $\Lambda$.\footnote{This identification happens because the integration over $\bar \Lambda_0$ gives a Delta function. Note, that since $\Lambda - \Lambda_0$ is generically complex valued, we need to use a generalized notion of Delta function with complex argument. We can think of such Delta function as the following limit $\delta(z) = \lim_{s\to \infty} \int_{- \infty}^{\infty} d p \,{\rm e}^{- \i \, p\, z - \frac{1}{2 \, s}p^2}$ with $z \in \mathbb{C}$.}

Since we are dealing with an Abelian vector multiplet, then a corresponding FI term and topological term can be added.  They need to be treated separately for \ss2 and \ads2 case since \ads2 has boundary and requires additional boundary terms. For \ss2, we have supersymmetric FI term and topological term given by
\begin{align}
S^{\text{S}^2}_{\text{FI}} + S^{\text{S}^2}_{\text{top.}} & =- \i \xi \int {\textrm{d}^2 x} \sqrt{g}  \hat{D}  + \i \frac{\vartheta}{2 \pi} \int {\textrm{d}^2 x}\sqrt{g} \,F_{12}\,,
\end{align}
and each terms is supersymmetric as it is. For \ads2, let us set the boundary at a large value of fixed $\eta $. Then, we have supersymmetric FI term consisting of bulk, boundary and counter term as
\beqa
S^{\text{AdS}_2}_{FI}&\=&S^{\text{bulk}}_{FI} +S^{\text{bdry}}_{FI}+S^{\text{c.t.}}_{FI}\,,
\eeqa
where
\beqa\label{FIActions}
S^{\text{bulk}}_{FI}&\=&- \i \xi \int {\textrm{d}^2 x} \sqrt{g}   \hat{D} \; \equiv \; - \i \xi \int {\textrm{d}^2 x} \sqrt{g}   \Bigl(D - \frac{\i}{L}\P\Bigr) \,,\nn
\\
S^{\text{bdry}}_{FI}&\=&-
 {\xi} \int {\rm{d} \theta} \sqrt{\gamma} \frac{1}{L}  \big( \i \Bar{\epsilon}\epsilon\, A_2   +  \Bar{\epsilon}\gamma_2 \epsilon\, \S \bigr)\,,
  \\
 S^{\text{c.t.}}_{FI}&\=& - {\xi} \int {\rm{d} \theta} \sqrt{\gamma} \frac{1}{L}  \Lambda \; \equiv \; {\xi} \int {\rm{d} \theta} \sqrt{\gamma} \frac{1}{L} (\xi^\mu A_\mu +\bar\epsilon \epsilon \S  + \i \bar\epsilon \gamma_3 \epsilon\, \P  )  \,,\nn
\eeqa
where the $\gamma$ in the boundary action in \eqref{FIActions} is induced metric on the boundary of \ads2. The supersymmetry variation of the bulk action  generates boundary term and it is exactly canceled by the variation of the boundary action in \eqref{FIActions}, which can be easily checked using the projection property of the Killing spinors in~\eqref{projectionKSads2}.  
The counter term action \eqref{FIActions} is added in order to cancel the divergence of the bulk and boundary action associated to the infinite volume of \ads2, in the spirit of holographic renormalization \cite{Skenderis:2002wp}. Due to the transformation property of $\Lambda$ as in \eqref{QeqLambda}, each term  of the counter term action is supersymmetric as it is. 

 We may also make supersymmetric topological term for \ads2 by including boundary term as follows,
\beqa\label{topAction}
S^{\text{AdS}_2}_{\text{top.}}&\=&    \i \frac{\vartheta}{2 \pi} \int {\textrm{d}^2 x}\sqrt{g} \,F_{12} + \i \frac{\vartheta}{2\pi}  \int {\rm{d} \theta} (\Bar{\epsilon}\epsilon  \S + \i\Bar{\epsilon}\gamma_3 \epsilon  \P  )
\\
&\=& \i \frac{\vartheta}{2\pi}  \int {\rm{d} \theta} \left(A_\theta +\Bar{\epsilon}\epsilon  \S + \i\Bar{\epsilon}\gamma_3 \epsilon  \P  \right)\,.
\nn
\eeqa
Then, this action \eqref{topAction} is invariant under equivariant supersymmetry since the last expression  is  nothing but integration of the symmetry parameter $\Lambda$ \eqref{LambdaParameter}. Also,  this action is  finite at on-shell saddle. However, we will see in section \ref{bdycondition} that this term is not in favor of variational principle. Therefore we will turn off the topological term, $\vartheta=0$.  
\\

For the chiral  multiplet with $R$-charge $r$, the supersymmetric action  consists of bulk Lagrangian and  total derivative terms which gives boundary terms as
\be\label{Actionchiral}
S_{\text{c.m.}}=\frac{1}{\gym^2} \int {\textrm{d}^2 x} \sqrt{g}\Big( \cL^{\text{bulk}}_{\text{c.m.}} + D_{\mu}V^\mu_{\text{c.m.}}\Big)\,,
\ee
The bulk Lagrangian in \eqref{Actionchiral} is given by
\begin{eqnarray} \label{mattermultiplet}
\mathcal{L}_{\text{c.m.}}^{\text{bulk}} & \= &D_{\mu}\bar{\phi}D^{\mu}\phi  
+ M_\phi^2 \bar{\phi}\phi +\bar{F}F  %\quad\\&&
- \i \bar{\psi} \gamma^{\mu} D_{\mu} \psi  +  \bar{\psi} M_\psi \psi      - \i \bar{\psi}\lambda \phi - \i \bar{\phi}\bar{\lambda}\psi \,,
\end{eqnarray}
where the mass square of the scalar field and mass of the fermion are
\beqa\label{MassScalar}
M_\phi^2 &\=& \left(\S+ \frac{\r}{2}H \right)^2 +\left(\P +\frac{\r}{2}G \right)^2 + \frac{r}{4}R +\i \hat{D}\nn
\\
&\=&\left(\S+ \frac{\r}{2}H \right)\left(\S+ \frac{\r-2}{2}H  \right) +\left(\P +\frac{\r}{2}G \right)\left(\P +\frac{\r-2}{2}G \right) + \i {D}\,,
\\ \label{MassFermion}
M_\psi &\=& -\i\left(\S+ \frac{\r}{2}H \right)  - \left(\P +\frac{\r}{2}G \right) \gamma_3\,.
\eeqa
where for \eqref{MassScalar} we have used the relation between curvature and $G$ and $H$~\eqref{scalarcurvature}, and the redefinition of auxiliary field, $\hat{D}\equiv \i (H\S  + G \P  ) +{D} $ in \eqref{Dredef}. 
Note that the scalar and fermion masses have holomorphic dependence on $\S + \frac{r}{2}H$ and~$\P +\frac{\r}{2}G$. 
For \ss2, since the mass square of the scalar is given by
\be
M^2_{\phi,\text{S}^2} \=\frac{r}{4L^2}(2-r) + \S^2 + \P^2 +\frac{\i}{L}\S (1-r) +\i D\,,
\ee
the  requirement that real part of the mass square be positive imposes restriction on the range of $r$, viz., $0\leq r \leq 2$. On the other hand, the mass square of scalar for \ads2 is given by
\be
M^2_{\phi,\text{AdS}_2}
\= \Bigl(\P -\frac{r}{2L}+\frac{1}{2L}\Bigr)^2 -\frac{1}{4L^2}+ \S^2+\i D\,,
\ee
and the real part of the mass square is always greater than its BF bound, $M^2_{\text{AdS}_2} \geq -\frac{1}{4L^2}$  \cite{Breitenlohner:1982jf,Breitenlohner:1982bm}. Therefore, there is no restriction on the range of $R$-charge $\r$.

The total derivative term in \eqref{Actionchiral} is determined by the fact that the chiral multiplet action is $\qeq$-exact as we discuss in the following:

For \ss2,
\be
\frac{1}{2\i \Bar{\epsilon}\epsilon} \qeq \left[ (  Q_{\bar{\epsilon}} - Q_\epsilon)\left(\i  \bar{\psi}  \psi - \bigl(2 \S + (r-1)H\bigr)\bar{\phi}  \phi  \right)\right] \=  \mathcal{L}_{\text{c.m.}}^{\text{bulk}} + D_\mu V_{\text{c.m.}}^\mu\,,
\ee
where
\begin{eqnarray}
 V_{\text{c.m.}}^{\mu}|_{\text{boson}} &=& \frac{1}{\i \Bar{\epsilon}\epsilon} \Bigl[\frac{\i}{2}\epsilon^{\mu \nu}   (\i \bar \epsilon \gamma_3 \epsilon)    \left(\bar \phi D_\nu \phi -  \phi D_\nu \bar\phi \right) - {\bar \epsilon \gamma_3 \gamma^\mu \epsilon }\,  \rho \bar \phi \phi \Bigr]\, , 
 \\
 V_{\text{c.m.}}^\mu|_{\text{fermion}} &=& 
-\frac{1}{2} \frac{1}{\i\bar{\epsilon}\epsilon} \Bigl[(\bar\epsilon \gamma^\mu \bar \psi ) ~ \epsilon  \psi + (\epsilon \bar \psi ) \bar \epsilon  \gamma^\mu  \psi \Bigr]\,,
\end{eqnarray}

and for \ads2, 
\be 
\frac{1}{2\i \Bar{\epsilon}\gamma_3\epsilon}\qeq \left[ (Q_{\epsilon} - Q_{\bar{\epsilon}})\left( \i \bar{\psi} \gamma_3 \psi - \i \Bigl(2 \P + (r-1)G \Bigr)\bar{\phi}  \phi  \right)\right] \= \mathcal{L}_{\text{c.m.}}^{\text{bulk}} + D_\mu  V_{\text{c.m.}}^\mu\,,
\ee 
where 
\beqa
V_{\text{c.m.}}^\mu \Bigl|_{\text{boson}}&\=&\frac{1}{\i\bar{\epsilon}\gamma_3\epsilon}\Bigl[\frac{\i}{2}\epsilon^{\mu \nu} ({\i\bar{\epsilon}\epsilon})  \left(\bar{\phi} D_\nu \phi - \phi D_\nu\bar{\phi} \right)  -   \bar{\epsilon} \gamma_3 \gamma^\mu \epsilon \, \S  \bar{\phi}\phi \Bigr] \, ,  \label{eq:bdryPhi}
\\
V_{\text{c.m.}}^\mu\Bigl|_{\text{fermion}} &=&
\frac{1}{2} \frac{1}{\i \bar{\epsilon}\gamma_3\epsilon} \Bigl[(\bar\epsilon \gamma^\mu \bar \psi ) ~ \epsilon \gamma_3 \psi - (\epsilon \bar \psi ) \bar \epsilon \gamma_3 \gamma^\mu  \psi \Bigr]\,.   \label{eq:bdryPsi}
\eeqa

%%%%%%%%%%%%%%%%%%%%%%%%%%%%%%%%%%%%%%%%%%%%%%%%%%%%%%%%%%%%%%%%%%%%%%%%%%%%%%%%%%%%%%
%%%%%%%%%%%%%%%%%%%%%%%%%%%%%%%%%%%%%%%%%%%%%%%

\subsection{Functional integration measure}\label{Sec:Measure}
 For the functional integration of Euclidean theory to be well defined,  the real part of the bosonic action needs to be positive, which requires us to impose appropriate reality conditions on each fields. 
\subsection*{Reality condition}
For the fluctuations of bosonic fields in the vector multiplet, the natural reality condition we choose is  
\be\ba{llll}
A_\mu^\ast = A_\mu \,,\quad\quad & \S ^\ast= \S  \,,\quad\quad &\P  ^\ast =\P  \,,\quad\quad& D^\ast = D\,. \label{eq:ReVec}
\ea\ee
As was discussed in \eqref{Dredef}, we shift the contour of $\hat{D}$ such that the redefined auxiliary field is $D$ real. 
If we keep this reality condition and take the supersymmetry variation given in~\eqref{eq:deltaA}, then we never end up with any consistent reality condition on two gaugini $\lambda$ and $\bar{\lambda}$. Therefore, we are forced to give up compatibility between supersymmetry transformation and reality condition. i.e.,  the supersymmetry variation $\qeq$ and complex conjugation do not commute. For example, even if we start with the real vector field $A_\mu$, its variation can not be real. Furthermore, we are also forced to treat the two gaugini $\lambda$ and $\bar{\lambda}$ independent, which `formally' doubles the fermionic degree of freedoms, as a standard treatment of fermions in Euclidean theory \cite{Festuccia:2011ws}. It is `formal' doubling as we do not double the number of path integration measure as discussed before. 

We could have tried to impose the reality condition on the gaugini in the same way as \eqref{realityKS} on the Killing spinors , i.e.,
$
\lambda^\dagger = \i \bar\lambda^T C\,$ and $\bar\lambda^\dagger = -\i \lambda^T C\,$. Then we could impose the following reality condition on bosonic fields while guaranteeing the compatibility between supersymmetry and the reality condition:
$
A_\mu^\ast = A_\mu \,, \S ^\ast= -\S  \,,\P  ^\ast =\P  \,, D^\ast = -D\,.
$
However, the imaginary nature of $\S $  and $D$ makes the kinetic term of $\S$ and the $D$-term in \eqref{eq:LYM} to have the wrong sign and the path integral ill-defined. Therefore we give up the supersymmetry compatible reality condition and follow the condition \eqref{eq:ReVec}. 

In the same way for the chiral multiplet, in order to make the path integral well-defined,  we choose the reality condition for bosonic variables as 
\be
\phi^\ast = \bar{\phi}\,,\quad\quad F^\ast = \bar{F}\,, \label{eq:ReChiral}
\ee
and let the fermions $\psi$ and $\Bar{\psi}$ be two independent Dirac spinors. Again, we could have considered the reality condition
compatible with supersymmetry transformation rule given in~\eqref{deltachiral},  which is 
$
\phi^\ast = \bar{\phi}\,, F^\ast =- \bar{F}\,, \psi^\dagger = \i \bar{\psi}^T\! C\,,\,\bar\psi^\dagger = -\i {\psi}^T C\,
$. This condition the $F \bar F$ term to have wrong sign and therefore the path integral ill-defined. 
Therefore, we  give up the supersymmetry compatible reality condition also for the chiral multiplet and follow the condition \eqref{eq:ReChiral}.

\subsection*{The measure}  
From supergravity point of view, the functional integration measure should be determined by invariance under BRST transformation associated to all the local symmetry of supergravity. In our case, the BRST invariance practically turns into two guides: one is Fujikawa's prescription for diffeomorphism invariance \cite{Fujikawa:1984qk}, and the other is invariance under the equivariant supersymmetry defined in \eqref{qeqDef}.  Consequently, the integration measure for a generic field $\varphi(x)$ is dictated in terms of integration over actual integration variable $\wt{\varphi}(x)$ as
\be\label{scalarmeasure}
\cD\varphi \= \prod_x {\rm d}\wt{ \varphi}(x) \,,
\ee
and we determine the actual integration variable $\wt{\varphi}$.

In \cite{Fujikawa:1984qk}, the integration measure for theory with gravitational coupling  was obtained by imposing  BRST invariance associated to diffeomorphism. For example, the invariant measure of scalar field $\phi$ is  determined  by the actual integration variable $\wt{\phi}$ as
\be\label{scalarmeasure0}
\wt{\phi}(x) \;\sim\; g^{1/4} \phi(x)\,
\ee
 up to overall constant, where  determinant of metric $g$ is involved and it makes the measure invariant under BRST transformation of diffeomorphism.   This  was understood as imposing the following condition, 
 \be\label{Ultralocal}
\text{constant} \= \int \cD \varphi \,\, {\rm e}^{-|| \varphi||^2}\,,
\ee
and accordingly we  determine the measures. Here the square of norm is dictated by the kinetic term in the action \eqref{eq:LYM}, \eqref{actionghostExpand} and \eqref{mattermultiplet}. For example, the square of norm for scalar $\S$ in vector multiplet and scalar $\phi$ in chiral multiplet are respectively given by  
\be\label{Norm2}
||\S ||^2 \=\frac{1}{\gym^2} \int {\textrm{d}^2 x} \sqrt{g}~\S^2\,,\qquad \qquad ||\phi ||^2 \= \int {\textrm{d}^2 x} \sqrt{g}\,\bar{\phi} \phi\,.
 \ee
Since the square of norm is diffeomorphism invariant and so is the condition \eqref{Ultralocal}, the resulting measure becomes diffeomorphism invariant. 

The above condition \eqref{Ultralocal} does not fully determine the  measure. For complex scalars, $\phi$ and $\Bar{\phi}$, the definition of norm  given in \eqref{Norm2} does not fix the individual measure $\cD \phi $ and $\cD\Bar{\phi}$, but just fix $\cD \phi \cD\Bar{\phi}$ together. Same ambiguity happens also for Dirac spinors and ghost anti-ghost field. We fix this ambiguity by the relation between bosonic and fermionic measures  under  the  equivariant supersymmetry.

We summarize actual integration variables that dictate the integration measures as in \eqref{scalarmeasure} for all fields  as follows.  From the determinant of metric appearing in the measure such as in \eqref{scalarmeasure}, we will only keep size factor for convenience, i.e., $g^{1/4}\sim L$. Also, in order to make the measure dimensionless, we have also inserted the reference length scale~$L_0$ appropriately: For vector multiplet, 
%%%%%%%%%%%%%%%%%%%%%%%%%%%%%%%%%%%%%%%%%%%%%%%%%%%% 
%%%%%%%%%%%%%%%%%%%%%%%%%%%%%%%%%%%%%%%%%%%%%%%%%%%%%%%%%%%%%%%%%%%%%%%%%%%%%%%%%%%%%%%%%%%%%%%%%%%%%%%%%%%%%%%%%%
\begin{eqnarray}\label{IntVariableVM}
\left\{ \wt{A}_a\,, \wt{\S}\,, \wt{\P}\,, \wt{\lambda}\,, \wt{\overline{\lambda}}\,,\wt{D}\right\}\equiv
\left\{ \frac{A_a L}{\gym L_0}\,, \frac{\S L}{\gym L_0}\,, \frac{\P L}{\gym L_0}\,, \frac{\lambda L}{\gym \sqrt{L_0}} \,, \frac{\overline{\lambda} L}{\gym \sqrt{L_0}} \,,\frac{D L}{\gym}\right\}\,,
 \end{eqnarray}
 for ghost multiplet, 
 \beqa
 \left\{\wt{c}\,, \wt{\Bar{c}}\,, \wt{b}\right\} \equiv \left\{
 \frac{c \sqrt{L}}{\gym L_0^{3/2}}\,, \frac{\Bar{c}L^{3/2}}{\gym \sqrt{L_0}}\,, \frac{bL}{\gym}\right\}\,,
 \eeqa
 for the ghost of ghost multiplet,
 \beqa
 \left\{\wt{\Lambda_0}\,, \wt{b_0}\,, \wt{c_0}\,,\wt{\bar{\Lambda}_0}\,,\wt{\Bar{c}_0}\right\} \equiv \left\{\frac{\Lambda_0}{\gym L_0}\,, \frac{b_0 L}{\gym}\,, \frac{c_0\sqrt{L L_0}}{\gym}\,,\frac{\bar{\Lambda}_0 L^2 L_0}{\gym}\,,\frac{\Bar{c}_0(L L_0)^{3/2}}{\gym}\right\}\,,
 \eeqa
and for chiral multiplet, 
\begin{eqnarray}\label{IntVariableCM}
 \left\{\wt{\phi}\,,\wt{\overline{\phi}}\,, \wt{\psi}\,, \wt{\overline{\psi}}\,,\wt{F}\,, \wt{\overline{F}}
\right\}\equiv \left\{\frac{\phi L}{L_0}\,,\frac{\overline\phi L}{L_0}\,,\frac{\psi L}{ \sqrt{L_0}} \,, \frac{\overline{\psi} L}{ \sqrt{L_0}}\,, F L\,, \overline{F}L \right\}\,.
\end{eqnarray}

We find that the bosonic and fermionic integration variables given above  are mapped by the rescaled equivariant supercharge,
\begin{eqnarray}\label{tildeQeq}
 \wt{Q}_{eq} \;\equiv\; \sqrt{\frac{L_0}{L}}\qeq\,,
\end{eqnarray}
and the equivariant algebra from $\wt{Q}_{eq}$ is closed in terms of the above variables as
\begin{eqnarray}\label{tildeQAlgebra}
 \wt{Q}_{eq}^2 \= \frac{L_0}{L} \left(\cL_\xi + \delta_{\text{gauge}}(\Lambda_0)+ \delta_{R} (\Lambda_R) \right)\,.
\end{eqnarray}
The mapping of the boson and fermion   by the \eqref{tildeQeq} will be more manifest when we organize new variables called ``cohomological variable'' in section \ref{subsec:cohomVar}. 
 Since the bosonic and fermionic integration variables are mapped by \eqref{tildeQeq}, the invariance of the full integration measure under the supercharge $\qeq$ in guaranteed. 

Note that the total dependence on the size of the background $L$ in the integration variables  in from~\eqref{IntVariableVM} to~\eqref{IntVariableCM} seem to completely cancel each other in between bosonic  and fermionic ones. However, since the functional integration measure consist of infinite product of modes, it is not guaranteed that such cancellation  occurs.  In fact, we will see in section \ref{sec:susyloc} that partition functions have non-trivial $L$ dependence, which accounts for the anomaly associated to the scaling symmetry on theory on \ss2   \cite{Gomis:2015yaa} as well as on \ads2\,.  In section \ref{sec:susyloc}, we will compute equivariant index with respect to rescaled operator $\wt{Q}_{eq}^2$. Since this operator acquires the $L$ dependence as in \eqref{tildeQAlgebra}, this rescaling of the supercharge is the source for $L$ dependence of the 1-loop in the index computation as pointed out in \cite{Gupta:2015gga}.

%%%%%%%%%%%%%%%%

\subsection{Classical saddles}\label{subsec:classicalsaddles}
Classical equations of motion for bosonic fields of vector and chiral multiplet are given by
\be\label{eqofmotion}
\ba{lll}
0&\=& D - \i \gym^2 \big(\xi - \Bar{\phi}\phi\big) \,,
\\
0&\=&  \i H  \bigl(F_{12} + \i (G \S  - H \P  )\bigr) + D_\mu D^\mu \rho  - \gym^2 \bigl(  2\P +(r-1)G \bigr)\Bar{\phi}\phi\,,
\\
0&\=&  \i G  \bigl(F_{12} + \i (G \S  - H \P  )\bigr) - D_\mu D^\mu \S  + \gym^2 \bigl(  2\S +(r-1)H \bigr)\Bar{\phi}\phi\,,
\\
0&\=& -\frac{1}{2} \epsilon^{\mu\nu} \partial_\nu \bigl(F_{12} + \i (G \S  - H \P  )\bigr) + \i\gym^2   \bigl( \Bar{\phi}D^\mu \phi - \phi D^\mu \Bar{\phi} \bigr)\,,
\\
0&=&(-D^\mu D_\mu + M_\phi^2\,)\,\phi\,,
\\
0&=&(-D^\mu D_\mu + M_\phi^2\, )\,\bar\phi\,,
\ea\ee
and these are solved by two branches of solutions depending on the value of FI parameter~$\xi$. 

When $\xi=0$, the above equations of motion are solved by 
\be \label{classicalsaddles} \ba{l}
\S \= \S_0\,,\qquad \P_0 \= \P_0\,,\qquad F_{12}= -\i (G \S_0 -H \P_0) \,,
\\
\Bar{\phi}\= \phi \= D\=0\,,
\ea\ee
where $\S_0$ and $\P_0$ are real constants. 

For the case of \ss2 $(G=0 \,,H=-\i/L)$, two constants $\S_0$ and $\P_0$. While $\S_0$ is being integrated, the constant $\P_0$ is quantized to integer $\mathbf{m}$ with appropriate unit\footnote{We may set the units of magnetic flux in ``tilde'' variables in terms of $\gym L_0$, i.e. $\wt{m} = \frac{1}{2 \pi} \int \wt{F}_{12}= \frac{m}{\gym L_0}$ with $m \in \mathbb{Z}$.   } 
as the magnetic flux on S$^2$, $\int_{\textrm{S$^2$}} F_{12}= 2\pi \mathbf{m}$. Therefore, the  non-vanishing configurations of solution on \ss2  are
\beqa \label{GNOs2}
&&\S= \S_0\,,\quad\quad \P  =  \frac{ \mathbf{m}}{2 L}\,,\quad \quad  F_{12} = \frac{\mathbf{m}}{2 L^2}\, \quad \Rightarrow \quad A =-\frac{\mathbf{m}}{2 } \left(\cos \psi \mp 1\right)d\theta \,, 
\eeqa
where  we present the solution of gauge field $A_\theta$ in  two patches up to gauge transformation.  In one patch, the solution is regular at the north pole, where~$\psi =0$, and in the other patch, the solution is regular at the south pole, where~$\psi = \pi $.

For the case of \ads2 $(G=-1/L \,,H=0)$ neither $\S_0$ nor $\P_0$  parameterize a moduli space as they are not normalizable modes and remain fixed as constant numbers. Therefore, we write the non-vanishing configurations of solution on \ads2 up to gauge transformation  as
\beqa\label{solutionads2}
&&\P \= \P_0\,,\quad\quad \qquad\S=\S_0\,,\\ \nn
&&F_{12}\= \i \frac{\S_0}{L} \quad \Rightarrow \quad A \= \i\, L \S_0 \,(\cosh\eta -1) d \theta + \sum_{\ell \neq 0, \,\ell \in \mathbb{Z}}\alpha^{(\ell)}_{\text{bdry}} A^{(\ell)}_{\text{bdry}}  \, \,.
\eeqa
We note here that unlike the case of \ss2, there are additional modes called ``boundary zero modes'',~$A^{(\ell)}_{\text{bdry}}$, given in terms of pure gauge mode, yet with non-normalizable parameters~$\Lambda^{(\ell)}_{\text{bdry}}$ \cite{Camporesi:1992tm}. Explicitly, 
\beqa \label{nonnormalizablescalarsads2}
A_{\text{bdry}}^{(\ell)}=d\Lambda^{(\ell)}_{\text{bdry}}\, ,\qquad  \Lambda^{(\ell)}_{\text{bdry}} = \dfrac{1}{\sqrt{2\pi |\ell|}} \Biggl(\dfrac{\sinh \eta }{1+\cosh \eta }\Biggr)^{|\ell|} {\rm e}^{\i \ell \theta} \, , \qquad \ell = \pm 1\,,\pm 2\,,\, \cdots  \,.
\eeqa 
As the name suggest,  they survive at the boundary of \ads2 and they have trivial field strength being the zero modes of the theory. However, those modes cannot be gauged away and should be taken into account for physical contribution \cite{Sen:2012kpz}. The regularized number of boundary zero modes is known as $n_{\text{zm}}^{\text{bdry}}= -1$. (See appendix \ref{appendix:AdS2} for the counting.)
We again note that the solution of gauge field $A$ in \eqref{solutionads2} is regular at the center of~\ads2, where~$\eta=0$. Later, we will set $\S_0=0$ for our action to have well-defined variational principle which we will see in section \ref{bdycondition}.

When $\xi \neq 0$, the equations of motion \eqref{eqofmotion} are solved by another branch of solutions,
\beqa\label{Higgsbranch}
&&\phi \=\sqrt{\xi}{\rm e}^{\i (\a_0 +  \a(x)) } \,, \qquad \Bar{\phi}\= \sqrt{\xi}{\rm e}^{-\i (\a_0 +\a(x)) }\,,\qquad \\
&&A_\mu \= \partial_\mu \a(x)\,,\qquad \P \= \half (1-r)G\,,\qquad \S\=\half (1-r)H\,,\qquad D \= 0\,.
\eeqa
For \ss2, the constant phase $\a_0$ parameterize the moduli, and $\a(x)$ can be gauged away. 
For \ads2, if we use normalizable condition for the scalar $\phi$ and $\Bar{\phi}$, $\a_0$ cannot be a moduli but is instead a fixed value. Moreover, only normalizable $\a(x)$ is allowed as a solution, and it can be gauged away.  
If we allow non-normalizable boundary condition for $\phi$ and $\Bar{\phi}$, then constant $\a_0$ is a moduli and non-normalizable $\a(x)$ is allowed as a classical solution. In this case $\a(x)$ cannot be gauged away and is identified with the boundary zero modes as $\a(x) = \sum_{\ell}\a^{(\ell)}_{\text{bdry}} \Lambda^{(\ell)}_{\text{bdry}}(x)$ up to gauge transformation.

In this paper, we will focus on the case $\xi =0$, and normalizable boundary condition for the scalars $\phi$ and $\Bar{\phi}$.

\subsection{Cohomological variables} \label{subsec:cohomVar}

For applying  supersymmetric localization, it is convenient to reorganize the degrees of freedom of our theory into a certain representation of $\qeq$ which we call ``cohomological variables". The cohomological variables consist of $\qeq$-cohomology complex $(\Phi\,, \qeq \Phi\,, \Psi\,, \qeq \Psi)$ and possibly a singlet of $\qeq$. 
Here, we call $\Phi$ the elementary boson, $\Psi$ the elementary fermion, and $\qeq \Phi$ and $\qeq \Psi$ are  their  superpartners. They naturally form a cohomology complex with respect to the equivariant supercharge $\qeq$.  
There are two main benefits of using this representation. One is that the structure of supersymmetry among  the variables is manifest. Therefore, imposing boundary conditions respecting supersymmetry becomes straightforward. The other benefit is that we can evaluate the 1-loop partition function systematically using index  theory, which we will do in section \ref{sec:susyloc}. We devote this section to find cohomological variables for our theory and appropriately define the integration measure in terms of them.

The organization of variables into cohomological form is implemented via a change of variables  which must be non-singular in order to take all degrees of freedom correctly into account. The algorithm that we will follow to implement such change of variables is as follows 
\begin{itemize}
\item[1)] We choose a twisting of all the fermions of the theory by combining the Killing spinor and fermions to make them have the same spin structure as bosonic variables of the theory. Then we check the invertibility of this change of variables.  
\item[2)] We start with a given bosonic component $\phi_{\mathfrak{R}}$ in some representation $\mathfrak{R}$ of the gauge group, and consider its variation $\qeq\phi_{\mathfrak{R}}$ which is certainly in the same representation.  This $\qeq\phi_{\mathfrak{R}}$ may be a combination of the twisted fermionic variables and other bosonic fields with coefficient consisting of bilinears of the Killing spinors. 
 
\item[3)] To decide if $\phi_{\mathfrak{R}}$  is elementary ( i.e., it is part of $\Phi$ ) and  $ \qeq \phi_{\mathfrak{R}} \in \qeq\Phi$, we need to verify if there is a twisted fermion $\psi_{\mathfrak{R}}$ within the expression of $\qeq\phi_{\mathfrak{R}}$ such that it does not contain derivatives on it and that it has coefficient  that is regular everywhere to guarantee the invertibility of the change of variables. If this happens, then we classify $\phi_{\mathfrak{R}} \in \Phi$ and $\psi_{\mathfrak{R}} \not\in \Psi$.

\item[3)] Proceed similarly with a fermionic variables in  $\Psi$.

\item[4)] Continue the process until all variables in the $\qeq$-complex are classified. In case of failure to classify, then we restart from step $1$ and repeat the process with another choice of variables.
\item[5)] The $\qeq$-singlet is identified with the constant (or non-normalizable)  part of the field dependent parameter for the local symmetries in the supersymmetry algebra.  It is singlet since $\qeq$ acting on this mode would give constant  (or non-normalizable) part of ghost  field for the corresponding symmetry and this is absent, as already explained in~\eqref{L0s2} and~\eqref{L0ads2}.
\end{itemize}

Let us now use the above algorithm to find the elementary fields in the case of Abelian vector multiplet and chiral multiplet.

\subsection*{The vector multiplet}  \label{subsec:VMtwisted}
Let us consider  the  full vector multiplet including ghost multiplet  which contains the ghost of ghost multiplet for the case of \ss2
\beqa\label{offshellVM}
{V}^{\text{v.m.}}& = &\left(A_{a}, \S , \P  , \lambda, \bar{\lambda}, D\right) \oplus \left(c ,\bar{c}, b\,;\, \Lambda_0\,, b_0\,,c_0\,,\Bar{\Lambda}_0\,, \Bar{c}_0\right)\,,
\eeqa
where we remind that for \ads2 case $\Lambda_0$ is not an independent variable and $b_0\,,c_0\,,\Bar{\Lambda}_0$ and $\Bar{c}_0$ are absent.

For the gaugini, we first propose the following twisted variables:
\begin{align}
\begin{split}\label{twistinglambda}
\lambda_A & = - \frac{\i}{2} \left(\bar{\epsilon} \gamma_A \lambda + \epsilon \gamma_A \bar{\lambda}\right) \,,  \quad A =1,2,3,4 \,,\\
\end{split}
\end{align}
where by $\gamma_4$ we mean $\gamma_4 =-\i \gamma_1 \gamma_2 \gamma_3 =1$. This change of variables has invertible as the inverse relation is
\begin{align}
\begin{split}
\lambda  = - \frac{1}{\i \bar{\epsilon} \epsilon} \left( \gamma^A \epsilon \lambda_A\right) \label{eq:LambdaTW}\,,\quad 
\bar{\lambda}  = \frac{1}{\i \bar{\epsilon} \epsilon} \left( \gamma^A \bar\epsilon \lambda_A\right)\,.
\end{split}
\end{align}
The Jacobian of the transformation is non singular since it is given by \footnote{In fact, for AdS$_2$ case, this Jacobian is singular at spatial infinity since the bispinor in the Jacobian is $\bar{\epsilon}\epsilon  = -\i \cosh\eta$ and it diverges as $\eta \rightarrow \infty $.   However, if we consider  the entire change of variable from original variables \eqref{offshellVM} to the cohomological variables \eqref{eq:VQarray}, the Jacobian is non-singular as shown in~\eqref{JacobialFull}.  The singularity of the Jacobian  at the boundary of AdS$_2$ in the \eqref{Jtwisting} may represent the fact that the asymptotic boundary condition for cohomological variables and the original variables are different as we will discuss in section~\ref{BCsaddle}. }
 \begin{align}\label{Jtwisting}
|J| =\bigg|\det\left(\frac{\partial (\lambda, \bar{\lambda})}{\partial \lambda_A}\right)\bigg|= \frac{1}{2}\frac{1}{(\i \bar{\epsilon}\epsilon)^2} \neq 0\,.
\end{align}

After choosing the twisted variable \eqref{twistinglambda}, we further reorganize variables following the algorithm we presented at the beginning of this subsection.  We  obtain the following set of cohomological variables
\be\ba{ll}
\Phi  =  \{A_{a}\,,  \P\,\}\,, \hspace{5mm} \quad\quad&\qeq \Phi =\{ \qeq A_a \, ,  \qeq \P\,\}\,, \label{eq:VQarray} \\
\Psi  =  \{\lambda_4\,, c\,, \bar{c}\}\,,\hspace{5mm} &\qeq \Psi = \{\qeq \lambda_4\,, \qeq c \, ,\qeq \bar{c}\}\,,\\
\Phi_0 = \{{b}_0\,, {\Bar{\Lambda}}_0\} \,,\hspace{5mm} \quad\quad&\qeq \Phi_0 =\{ \,{Q}_{eq}{b}_0 \,,{Q}_{eq}{\Bar{\Lambda}}_0\}\, ,\quad \quad \left\{\Lambda_0 \right\}\,,
\ea\ee
where the explicit expression of the variables in $\qeq \Phi$, $\qeq \Psi$ and $\Lambda_0$ are given as
\be\ba{lll}
\qeq A_{a}&=&% \delta A_{\mu} +  Q_{\text{brst}}A _{\mu}= 
\lambda_{a} + e_{a}{}^{\mu}\partial_{\mu} c \,, \label{eq:QLamb4} \\
\qeq \P  & = &%\delta \P  + Q_{\text{brst}} \P  = \delta \P  = 
\lambda_3\,, \\
\qeq \lambda_4 &=&\i \bar{\epsilon}\epsilon \left( D  -\i H \S  -\i G\P  \right) + \i  \bar\epsilon \gamma_3 \epsilon  \left(-\i F_{12} +  G \S  -  H \P  \right) + \i \bar\epsilon \gamma_3 \gamma^\mu \epsilon \partial_\mu \P  \, , 
\\
\qeq c & =&  -\Lambda+ \Lambda_{\text{0}}\,, \quad \quad \qquad \mbox{where } \quad\Lambda \equiv -\xi^\nu A_\nu -\bar\epsilon \epsilon \S  - \i \bar\epsilon \gamma_3 \epsilon\, \P \,, \\ 
\qeq \bar{c} & =& b\,,
\quad \quad
\qeq b_0 = c_0\,,
\quad \quad
\qeq \Bar{\Lambda}_0 = \Bar{c}_0 \,,
\ea\ee
and the variable $\Lambda_0$ that was introduced in \eqref{L0ads2} and \eqref{a0ghostofghost} for \ads2 and \ss2 respectively. Note that the $\Lambda_0$  is  singlet in $\qeq$ as explained in \eqref{qeqGGmultiplet}, and  therefore it is  outside of the cohomology complex given by the set $\Phi$ and $\Psi$ and their descendant $\qeq$ variation.

The expression \eqref{eq:QLamb4} shows the change of variables essentially from $\{ \l_a\,, \l_3\,, D\,, \S \,,$  $ b \}$ to $\{\qeq A_a\,,$ $\qeq \P\,,$ $\qeq \l_4\,, \qeq c\,,\Lambda_0\,, $ $\qeq \bar{c}\,\}$. Now, we find the total Jacobian associated to the change of variables from the original variables \eqref{offshellVM} to the cohomological variables \eqref{eq:VQarray}. 
The non-trivial part of the Jacobian 
is  given as follows:
 \begin{align}\label{JacobialFull}
|J| =\bigg|\det\left(\frac{\partial (\lambda\,, \bar{\lambda}\,, D\,, \S\,,b)}{\partial ( \lambda_4\,,\qeq A_a\,,\qeq\rho\,, \qeq \lambda_4\,, \qeq \bar{c}\,, \Lambda_0\,, \qeq c )} \right)\bigg|= \frac{1}{2} \neq 0\,.
\end{align}
Thus it guarantees the invertibility of the change of variables. 
We note that the algebra is closed only with the symmetry generated by the Killing vector, as the cohomological variables are neutral under $R$-symmetry. 
~\\
From \eqref{IntVariableVM} and \eqref{IntVariableCM}, we can see the actual integration variables in terms of cohomological variables are  
\beqa\label{IntVectorCoho}
\wt{\Phi}  &\=&  \left\{\wt{A}_{a}\,,  \wt\P\,\right\}\=\left\{\frac{A_a L}{\gym L_0}\,, \frac{\P L}{\gym L_0}\, \right\}\, , \nn \\
\wt{\Phi}_0 & \=& \left\{\, \wt{b}_0\,, \wt{\Bar{\Lambda}}_0\right\} \= \left\{\, \frac{b_0 L}{\gym}\,, \frac{\bar{\Lambda}_0 L^2 L_0}{\gym} \right\}\, ,\nn \\
\wt{Q}_{eq} \wt{\Phi}  &\=&\left\{ \wt{Q}_{eq} \wt{A}_a \, , \qeq \wt{\P}\, \right\} \,\= \left\{  \frac{\qeq {A}_a \sqrt{L}}{\gym \sqrt{L_0}}\, , \frac{\qeq {\P}\sqrt{L}}{\gym \sqrt{L_0}} \,\right\}\,,\nn
\\
\wt{Q}_{eq} \wt{\Phi}_0 &\=& \left\{\,\wt{Q}_{eq}\wt{b}_0 \,,\wt{Q}_{eq}\wt{\Bar{\Lambda}}_0 \right\}\, \=
\left\{   \frac{\qeq b_0 \sqrt{L L_0}}{\gym}\,, \frac{\qeq \Bar{\Lambda}_0 (L L_0)^{3/2}}{\gym}\right\} \,,   \nn \\
\wt\Psi  &=&  \left\{\wt\lambda_4\,, \wt{c}\,, \wt{\bar{c}}\,\right\}\= \left\{\frac{\lambda_4 \sqrt{L}}{\gym \sqrt{L_0}}\,, \frac{c \sqrt{L}}{\gym L_0^{3/2}}\,, \frac{\Bar{c}L^{3/2}}{\gym \sqrt{L_0}}\,\right\}
\,,\hspace{5mm} 
\\
 \wt{Q}_{eq} \wt\Psi &\=& \left\{ \wt{Q}_{eq}  \wt{\lambda}_4\,,
 \wt{Q}_{eq} \wt{c} \, ,\wt{Q}_{eq}
\wt{\bar{c}}\,\right\}  \= \left\{\frac{ \qeq \lambda_4 }{\gym} \,, \frac{\qeq c}{\gym L_0} \, ,\frac{\qeq \bar{c} L}{\gym}\right\}\,,\nn
\\ 
\wt{\Lambda}_0 &=& \frac{\Lambda_0 }{\gym L_0}\,.\nn
\eeqa
where $\wt{Q}_{eq}$ is the rescaled equivariant supercharge that was defined in \eqref{tildeQeq} as
\be\label{Qeqtilde}
\wt{Q}_{eq} \;\equiv \;\sqrt{\frac{L_0}{L}} \qeq \,.
\ee
We note that the cohomological arrangement of fields makes the  mapping between bosonic and fermionic integration variables through the supercharge \eqref{Qeqtilde} manifest.  
With the definition of rescaled equivariant supercharge \eqref{Qeqtilde}, the algebra is closed as
\be
(\wt{Q}_{eq})^2 \= \frac{L_0}{L}\partial_\theta\,.
\ee

%%%%%%%%%%%%%%%%%%%%%%
\subsection*{The chiral multiplet}\label{chiralmultiplet}
Let us consider the following  $\mathcal{N}=(2,2)$ chiral multiplets  of the form
\be
 \Phi^{\text{c.m.}} = \{\phi\,, \bar{\phi}\,, \psi\,, \bar{\psi}\,, F \,,\bar{F}\}\,.
\ee
The candidate set of twisted variables in this case is given by the collection of bispinors
\be
\epsilon \psi\,, \epsilon \bar{\psi}\,, \bar{\epsilon} \psi\,,\bar{\epsilon} \bar{\psi}\,.
\ee
The invertibility of changing to this set of variables can be checked by directly looking at the inverse relation, 
\begin{equation}\label{eq:psiTw}
 \psi \= \frac{1}{\bar{\epsilon} \epsilon} \left( \epsilon (\bar{\epsilon} \psi )-\bar{\epsilon} (\epsilon \psi)\right)\, ,\qquad 
\bar{\psi} \= \frac{1}{\bar{\epsilon} \epsilon} \left( \epsilon (\bar{\epsilon} \bar{\psi} )-\bar{\epsilon} (\epsilon \bar{\psi})\right),   
\end{equation}
with the Jacobian given by
\be
|J| = \bigg|\det\left(\frac{\partial (\psi \, , \bar \psi)}{\partial (\bar \epsilon \psi\, , \epsilon \psi\, , \bar \epsilon \bar \psi \, , \epsilon \bar \psi }\right) \bigg| = \frac{1}{(\bar \epsilon \epsilon)^2} \neq 0 \,.
\ee
 Following the algorithm we presented at the beginning of this subsection, we obtain the following set of cohomological variables
\be\ba{ll}
\Phi  =  \{\phi\, ,\bar\phi\}\, , \hspace{5mm} &\qeq \Phi =\{ \qeq \phi \, ,\qeq \bar{\phi}\}\,  \label{eq:CQarray} ,\\ 
\Psi  =  \{\epsilon \psi \, , \bar{\epsilon} \bar{\psi}\}\,, \hspace{5mm}\quad&\qeq \Psi = \{\qeq \left(\eps \psi\right)\, , \qeq \left( \bar{\epsilon} \bar{\psi}\right)\}\, .
\ea\ee
The explicit expression of the variables in $\qeq \Phi$ and $\qeq \Psi$ are
\be\ba{lll} \label{eq:ChBPS3}
\qeq\phi& =& \bar{\epsilon} \psi \,, 
\\
\qeq \bar{\phi} & =& \epsilon \bar{\psi},
\\
\qeq(\epsilon \psi) & =& \epsilon \bar{\epsilon} F + \i \epsilon \gamma^{\mu} \epsilon D_{\mu} \phi+ \epsilon \gamma_3 \epsilon \left(\P+ \frac{\r}{2}G\right) \phi, \\
 \qeq(\bar{\epsilon} \bar{\psi}) & =& \bar{\epsilon}\epsilon \bar{F} + \i \bar{\epsilon} \gamma^{\mu}  \bar{\epsilon} D_{\mu} \bar{\phi}- \bar{\epsilon} \gamma_3 \bar{\epsilon}\left(\P+ \frac{\r}{2}G\right)\bar{\phi} .\\
\ea\ee
The Jacobian of the transformation is clearly non singular since we have:
\begin{align}
|J| = \bigg{|}\det\left(\frac{\partial  ( \psi, \bar{\psi}, F, \bar{F})}{\partial (  \qeq \bar{\phi} ,\qeq \phi,\epsilon \psi,\bar{\epsilon} \bar{\psi}, \qeq \epsilon \psi, \qeq \bar{\epsilon}\bar\psi ) }\right)\bigg{|} =1\neq 0.
\end{align}
This ensures the invertibility of the change of variables. We note that differently from vector multiplet case, the cohomological variables of chiral multiplets have non-trivial $R$-charge. 
In table \ref{tab:CHM0}, we display  the $R$-charge assignment of elementary cohomological variables in chiral multiplet.
\begin{table}[ht]
\begin{center}
\begin{tabular}{| c ||c | c |  c | c   | }
\hline
-& \quad  \qquad  $\Phi$ \qquad \quad  &  \qquad  $Q_{eq} \Phi$ \quad  \qquad  &   $\Psi$ &  $Q_{eq} \Psi $    \\ \hline
 -& \quad $\phi$ \qquad $\bar{\phi}$ \qquad  & \quad $Q_{eq}\phi$ \quad $Q_{eq}\bar{\phi}$ \quad & $\epsilon \psi$ \qquad\, $\bar{\epsilon}\bar{\psi}$ &  $Q_{eq}(\epsilon \psi) $ \quad $Q_{eq}(\bar{\epsilon}\bar{\psi)}$ \\ \hline
 $G$& ~~ $+1$ \quad $-1$ \qquad & $+1$ \quad $-1$  & $+1$ \,\,\qquad $-1$ & \!\! $+1$ \,\,\,\qquad $-1$\,\, \\ \hline
 $R$& ~~ $+r$ \quad $-r$ & \,$+r$ \quad $-r$ \qquad  & \, $r-2$ \,\, $-(r-2)$ &\, $r-2$ \,\, $-(r-2)$ \\ \hline
\end{tabular}
\end{center}
\caption{Gauge ($G$) and $R$-charge assignment for cohomological fields associated to a chiral multiplet.}
\label{tab:CHM0}
\end{table}

The functional integration measure for the cohomological variables is then given as
\beqa 
\wt{\Phi} & \= & \left\{\wt{\phi}\, ,\wt{\bar\phi}\right\} \= \left\{\frac{\phi L}{L_0}\,, \frac{\Bar{\phi}L }{L_0}\, \right\} \,,\nn
\\
\wt{Q}_{eq} \wt{\Phi} &\=& \left\{ \wt{Q}_{eq} \wt{\phi} \, ,\wt{Q}_{eq} \wt{\bar{\phi}}\,\right\}\= \left\{ \frac{\qeq \phi \sqrt{L}}{\sqrt{L_0}}\,, \frac{\qeq \Bar{\phi}\sqrt{L}}{\sqrt{L_0}}\,\right\}   \label{eq:CQarray1} ,\\ 
\wt{\Psi} & =&  \left\{\wt{\epsilon \psi} \, , \wt{\bar{\epsilon} \bar{\psi}}\,\right\}\= \left\{ \frac{\epsilon \psi \sqrt{L}}{\sqrt{L_0}} \,, \frac{\Bar{\epsilon }\Bar{\psi} \sqrt{L}}{\sqrt{L_0}}\,\right\}\,,\nn
\\
\wt{Q}_{eq} \wt{\Psi} &\=& \left\{\wt{Q}_{eq} (\wt{\eps \psi})\, , \wt{Q}_{eq} (\wt{ \bar{\epsilon} \bar{\psi}})\right\}\= \left\{ \qeq (\eps \psi)\,, \qeq (\bar{\epsilon} \bar{\psi})\, \right\}\, .\nn
\eeqa
Again, the map between boson and fermion by the $\wt{Q}_{eq}$ \eqref{Qeqtilde} is manifest. 
Even though we should bear in mind that the path integral is being carried over these variables, throughout the rest of this paper, unless otherwise stated, we shall omit the tildes in order to avoid clutter during explicit manipulations. This implies that we will effectively omit $\gym L_0$  as well as $\frac{L}{L_0}$ dependence that we will restore in the final results.  
%%%%%%%%%%%%%%%%%%%%%%%%%%%%%%%%%%%%%%%%%%%%%%%%%%%%%%%%%%%%%%%%%%%%%%
%%%%%%%%%%%%%%%%%%%%%%%%%%%%%%%%%%%%%%%%%%%%%%%%%%%%%%%%%%%%%%%%%%%%%%%%%%%%%%%%%%%%

%%%%%%%%%%%%%%%%%%%%%%%%%%%%%%%%%%%%%%%%%

\section{Asymptotic boundary condition on \ads2 }\label{bdycondition}

Since the \ads2  space has the boundary at $\eta \rightarrow \infty$,  to define the theory in this background we need to specify the asymptotic  boundary condition on each field. We devote this section to discuss the boundary conditions that we impose.  To fix boundary conditions we first ensure compatibility with the variational principle given the supersymmetric boundary terms \eqref{eq:bdryLambda}, \eqref{eq:bdryPhi} and \eqref{eq:bdryPsi}. There could be other set of supersymmetric boundary terms, but we do not explore them in this paper. Furthermore, we will need to guarantee a well-defined action of supersymmetry on the field space. What we mean by this is that the supersymmetric transformation of a given field (say a bosonic field) with specified boundary condition should dictate the boundary condition of the superpartner field. That is, boundary conditions on bosons and fermions have to be consistently related by supersymmetry. Supersymmetric boundary condition have been studied in \cite{David:2018pex, David:2019ocd} for \ads2$\times$S$^1$ and~\cite{Sakai:1984vm,Correa:2019rdk} for Lorentzian \ads2.  

Even after following our criteria, there may be more than one consistent choice of boundary conditions. However, we select one of them and leave a more systematic study of more general boundary conditions for the future. In this paper, we will choose normalizable boundary conditions for bosonic fields and let the behavior of fermionic fields be dictated by supersymmetry, and then we will show that it is  compatible with the variational principle.  This may require non-normalizable modes for some fermions due to the fact that the supersymmetry transformation changes the asymptotics of the states through the ${\rm e}^{\frac{\eta}{2}}$ behavior of the Killing spinor. 
 We reserve a more thorough analysis of how to guarantee normalizability of superpartners of normalizable modes for our followup paper~\cite{GonzalezLezcano:2023uar}.

%%%%%%%%%%%%%%%%%%%%%%%%%%%%%%%%%%%%%%%%%%%%%%%%%%%%%%%%%%%%%%%
%%%%%%%%%%%%%%%%%%%%%%%%%%%%%%%%%%%%%%%%
\subsection{Asymptotic boundary condition and supersymmetry} \label{subsec:asympt}
To specify the asymptotic boundary condition, we need to consider an asymptotic expansion called asymptotic expansion of fields at the  boundary $\eta\rightarrow \infty$. The expansion is determined by solving the asymptotic equations of motion. However, since we know the classical solution in~\eqref{solutionads2}, we can assume the asymptotic expansion for the bosonic 
fields in the vector multiplet fields as follows, 
 \footnote{This expansion can also be written in terms of cohomological variables. In particular, we note that $\Lambda$ appearing in \eqref{eq:QLamb4} as a superpartner of ghost field $c$ has the following asymptotic expansion $$\Lambda = \Lambda_{(1)} + \Lambda_{(2)} {\rm e}^{- \eta}\,,$$ because the leading term $\Lambda_{(0)}{\rm e}^\eta $ vanishes due to the asymptotic equation of motion. Thus the $\Lambda_{(1)}$ is identified with $\Lambda_0$ defined in \eqref{L0ads2}.  }
\begin{align} 
\begin{split}
\label{eq:VMexpansion}
 A_\eta & = a_{\eta(0)} + a_{\eta(1)} {\rm e}^{-\eta} + \, \cdots\,,\qquad A_\theta  \={\rm e}^{\eta}( a_{\theta(0)}  + a_{\theta(1)}{\rm e}^{-\eta} + \, \cdots)\,, \\
 \S & = \S_{(0)} + \S_{(1)} {\rm e}^{-\eta} + \, \cdots\,,\qquad \quad \,\, \, \P \= \P_{(0)} + \P_{(1)} {\rm e}^{-\eta} + \, \cdots\,, 
\\
 D & ={\rm e}^{-\eta} \left(d_{(0)} + d_{(1)} {\rm e}^{-\eta} + \, \cdots\right) \,, 
 \end{split}
\end{align}
where each expansion coefficient can in general be function of the angular variable.  
By the asymptotic equation of motion, we set 
\begin{align} \label{eq:aaSP0}a_{\eta(0)} =  a_{\eta,0}\,, \quad a_{\theta(0)}= \i L  \S_{(0)} = \i L \S_0  \,, \quad  \P_{(0)} = \P_0\,, 
\end{align}
where all the leading coefficients are angle independent constants and further  $a_{\theta (0)}$ and $\S_{(0)}$ are identified. This setting can be confirmed by looking at the classical solution \eqref{solutionads2}. Later, in the subsection~\ref{subsec:Varprin}, we will see that we have to further require $\S_0=0$ for consistency with the variational principle. Now, as the normalizable boundary condition for bosons, we demand that the leading modes do not fluctuate, i.e. 
\begin{eqnarray} \label{eq:VMbc}
\delta a_{\eta,0}\=  \delta a_{\theta,0}\=\delta \S_0\=\delta \P_0\=0 \, .
\end{eqnarray}
Any mode that does not appear in \eqref{eq:aaSP0} and \eqref{eq:VMbc} is allowed to fluctuate freely.

The asymptotic expansion of fermion  is also in principle  determined by the asymptotic equation of motion.  However, a more direct way of obtaining the fall off behavior of each component for fermion fields will be in terms of the cohomological variables defined in subsection \ref{subsec:cohomVar}.
Then, it is easy to see that  the expansion for  fermions are dictated by the supersymmetry transformation rules~\eqref{eq:deltaA}.
 In fact, from \eqref{eq:LambdaTW} we can see that, if we express the superpartners of bosonic  fields through the  twisted variables $\lambda_A$, we have that
\begin{align} 
\begin{split}
\label{eq:lambdaAsymp}
    \qeq A_{a} & \= \qeq \left(a_{a, (0)} + a_{a,(1)}{\rm e}^{-\eta}+ \cdots \right) =\lambda_{a,(0)} + \lambda_{a,(1)} {\rm e}^{- \eta} + \cdots\, ,\\
    \qeq  \P & \= \qeq  \left(\P_{(0)} + \P_{(1)} {\rm e}^{-\eta} + \, \cdots\,\right) = \lambda_{3,(0)} + \lambda_{3,(1)} {\rm e}^{- \eta} + \cdots\, ,\\
  \bar{\epsilon}\epsilon  D & \=   d_{(0)} + d_{(1)} {\rm e}^{-\eta} + \, \cdots  = \qeq \left( \lambda_{4,(0)} + \lambda_{4,(1)} {\rm e}^{- \eta }+ \cdots \right)\, .
    \end{split}
\end{align}
Imposing normalizable boundary conditions for the superpartners of the bosonic modes requires us to set to zero the fluctuations of the leading (non-normalizable) modes of $\lambda_{1,2,3}$, that is
\begin{align} \label{eq:lambdaAbc}
    \delta \lambda_{1,(0)}\, = \,\delta \lambda_{2,(0)}\, = \, \delta \lambda_{3,(0)} \, = \, 0\,, \qquad   \lambda_{1,(0)} \, = \, 0 \,,\quad \lambda_{2,(0)} \, = \, \lambda_{2,0} \, , \quad \lambda_{3,(0)} \, = \, 0\,,
\end{align}
where $\lambda_{2,0}$ can be any constant value and $\lambda_{4,(0)}$ is free to fluctuate. This is consistent with the fact that $\lambda_{4,(0)}$ is  the superpartner of $d_{(0)}$ which we have also allowed to fluctuate.
We will see later that consistency with the variational principle can be also attained by setting $\delta \lambda_{4,(0)} = 0$ instead of $\lambda_{1,(0)} = \lambda_{3,(0)} = 0$ in \eqref{eq:lambdaAbc}. However,  if we pick this condition, we have to set $d_{(0)}$ fixed according to supersymmetry in~\eqref{eq:lambdaAsymp}.  Moreover, $\lambda_{1,(0)}$ is the superpartner of $a_{1,(0)}$, which we have set to a fixed constant value that does not fluctuate ( see \eqref{eq:aaSP0} and \eqref{eq:VMbc}), and similarly for $\lambda_{3,(0)}$ is superpartner of a fixed constant value $\P_0$. Therefore,  we  select \eqref{eq:lambdaAbc}.  As we will see in subsection \ref{subsec:Varprin}, this is consistent with the variational principle and it means that a non-normalizable mode for $\lambda_4$ can be allowed. Finally, since the gauge field is normalizable, the ghost field must be also normalizable.  Therefore their expansion reads 
\begin{align}
    c\, = \, c_{(1)} {\rm e}^{- \eta} + \cdots, \, \quad \bar c \,  = \,  \bar c_{(1)}{\rm e}^{-\eta} + \cdots, \, \quad     b\, = \, b_{(1)} {\rm e}^{- \eta} + \cdots\, . 
\end{align}

Let us now focus on the chiral multiplet. For the bosonic fields, let us begin with generic ansatz for the asymptotic expansion as
\begin{align} \label{eq:phiexp0}
    \phi & = {\rm e}^{- \Delta \eta} \left(\phi_{(0)} + \phi_{(1)} {\rm e}^{- \eta} + \cdots\right) \,, \qquad \,
    \bar \phi = {\rm e}^{- \Delta \eta} \left(\bar \phi_{(0)} + \bar \phi_{(1)} {\rm e}^{- \eta} + \cdots\right) \,,\\
        F &= {\rm e}^{- \wt \Delta \eta}\left( f_{(0)} + f_{(1)}{\rm e}^{-\eta}+\, \cdots \right)\,,\quad \, ~~~\, \, \,\Bar{F} \= {\rm e}^{- \wt \Delta \eta}\left( \Bar{f}_{(0)} + \Bar{f}_{(1)}{\rm e}^{-\eta}+\, \cdots\right)\, .
\end{align}
Here, the Weyl weight $\Delta$ for the $\phi $ and $\bar{\phi}$ is determined by the asymptotic equations of motion. The $\wt{\Delta}$ for the auxiliary fields remains undetermined for now. Without loss of generality, we allow $f_{(0)}, \, \bar f_{(0)}$ to   parameterize non-normalizable modes whereas $f_{(1)}, \, \bar f_{(1)}$ parameterize normalizable modes. This can be ensured if we restrict $\wt \Delta$ to be $\frac{1}{2}> \wt \Delta  > 0$. We will see that this window can be stretched after analyzing the variational principle. In practice we demand that both $f_{(0)}$ and $\bar f_{(0)}$ have to vanish in accordance with our normalizable boundary conditions.
Inserting \eqref{eq:phiexp0} in the asymptotic equation, we have
\beqa 
(- \nabla^2 + 2\i A^\mu \partial_\mu + A^\mu A_\mu + M^2_\phi )\phi \label{eq:Aeom}
\sim\left[ - (\partial_\eta^2 + \partial_\eta) + \Bigl(L\P_0 -\frac{r-1}{2}\Bigr)^2 -\frac{1}{4}\right] \phi = 0\, , \quad
\eeqa
where the ``$\sim$'' sign indicates that we are only keeping the leading terms in the asymptotic expansion of the kinetic operator. Holding $\rho_0$ fixed and solving \eqref{eq:Aeom} yields
\be \label{eq:asymp}
\begin{split}
\phi &  = {\rm e}^{-\Delta_-\eta}(\phi^-_{(0)} + \, \cdots ) + {\rm e}^{-\Delta_+ \eta}(\phi_{(0)}^+ + \, \cdots)\,,\\
\bar \phi & = {\rm e}^{-\Delta_-\eta}(\bar \phi^-_{(0)} + \, \cdots ) + {\rm e}^{-\Delta_+ \eta}(\bar \phi_{(0)}^+ + \, \cdots)\, ,
\end{split}
\ee
with
\be \label{eq:range}
\Delta_\pm  = \half \pm \biggl|L\P_0 -\frac{r-1}{2} \biggr|\,.   
\ee
Note that $\Delta_-$ terms in \eqref{eq:asymp} provide the leading asymptotic behavior and are associated to the non-normalizable modes. 
Now, based on the classical solutions~\eqref{classicalsaddles}, 
\begin{align} \label{eq:phiBC0}
\begin{split}
\phi^-_{(0)} &\=  \Bar{\phi}^{-}_{(0)} \= f_{(0)} \= \Bar{f}_{(0)}\=0\, .
\end{split}
\end{align}
 As for the  normalizable boundary conditions for the bosonic fields, we set
\begin{align} \label{eq:phiBC}
\begin{split}
\delta\phi^-_{(0)} &\= \delta \Bar{\phi}^{-}_{(0)} \=\delta f_{(0)} \=\delta \Bar{f}_{(0)}\=0\, ,\\
\phi^+_{(0)} & \;\neq 0\,, \quad \delta \phi^+_{(0)} \neq 0 \, , \quad  \bar \phi_{(0)}^+ \neq 0\, , \quad \delta \bar \phi_{(0)}^+ \neq 0 \,  .
\end{split}
\end{align}
The fall off condition for  fermions will be dictated by the supersymmetry transformation rules~\eqref{deltachiral}.
Similar to the case of the vector multiplet, a more direct way of obtaining the fall off behavior of each components for fermion fields is in terms of the cohomological variables defined in subsection \ref{subsec:cohomVar}. We can see this explicitly if we express the fermionic fields $\psi$ and $\bar \psi$ in terms of the twisted variables $\epsilon \psi, \, \bar \epsilon \psi, \, \epsilon \bar \psi, \, \bar \epsilon \bar \psi$ as given in \eqref{eq:psiTw}. Then, expanding the twisted variables according to \eqref{eq:asymp}, we have the asymptotic expansion
  \begin{align} \label{eq:twsc1}
 \begin{split}
     \bar \epsilon  \psi  & = {\rm e}^{- \Delta_- \eta} \left(\left( \bar \epsilon  \psi\right)^{-}_0 + \, \cdots\right) + {\rm e}^{- \Delta_+ \eta} \left(\left( \bar \epsilon  \psi\right)^+_0+\, \cdots\right) \,,\\ 
     \epsilon \bar \psi & = {\rm e}^{- \Delta_- \eta} \left(\left(\epsilon \bar \psi\right)^{-}_0 + \, \cdots\right) + {\rm e}^{- \Delta_+ \eta} \left(\left(\epsilon \bar \psi\right)^+_0 + \, \cdots\right)\,,
     \end{split}
 \end{align}
 with the same scaling dimensions $\Delta_{\pm}$ as $\phi$ and $\bar \phi$. This is a crucial step in our prescription, since it will be the key to respect supersymmetry while ensuring the validity of the variational principle.

 To set the boundary conditions, we demand that the superpartners of scalar fields $\phi$ and $\bar \phi$ follow the same type of normalizable boundary conditions. Therefore we have
      \begin{align} \label{eq:twbc1}
      \begin{split}
          \left(\bar \epsilon \psi\right)^-_0  & = 0\,, \quad \delta       \left(\bar \epsilon \psi\right)^-_0 = 0\,, \quad  \left( \epsilon \bar \psi\right)^-_0 = 0\,, \quad \delta  \left( \epsilon \bar\psi\right)^-_0 = 0\,,\\
              \left( \bar \epsilon  \psi\right)^+_0  & \neq 0\,, \quad    \delta \left( \bar \epsilon  \psi\right)^+_0 \neq 0\,, \quad  \left(  \epsilon \bar \psi\right)^+_0 \neq 0\,, \quad \delta  \left(   \epsilon \bar\psi\right)^+_0 \neq 0\, .
      \end{split}
  \end{align}

 For the elementary variables $\epsilon \psi$ and $\bar \epsilon \bar \psi$, since they are independent variables,  we use different asymptotic expansion as
  \begin{align} \label{eq:twsc2}
 \begin{split}
    \epsilon \psi  & = {\rm e}^{- \widetilde{\Delta}_- \eta} \left(\left(    \epsilon \psi \right)^{-}_0 + \, \cdots\right) + {\rm e}^{- \widetilde{\Delta}_+ \eta} \left(\left(    \epsilon \psi \right)^+_0+\, \cdots\right) \,,\\ 
     \bar \epsilon \bar \psi  & = {\rm e}^{- \widetilde{\Delta}_- \eta} \left(\left( \bar \epsilon \bar \psi\right)^{-}_0 + \, \cdots\right) + {\rm e}^{- \widetilde{\Delta}+ \eta} \left(\left( \bar \epsilon \bar \psi
     \right)^+_0 + \, \cdots\right)\,,
     \end{split}
 \end{align}
  where since they are scalars we have $\widetilde{\Delta}_{-} + \wt \Delta_+= 1 $. Without loss of generality, we assume  that  $\wt\Delta_-$ is the smaller one, i.e., $\wt\Delta_- < \frac{1}{2}$
  such that the terms associated to the $\wt\Delta_-$ are non-normalizable modes. 
The elementary fermionic variables $\epsilon \psi$ and  $ \bar \epsilon \bar \psi $ admit non-normalizable modes. Therefore, the leading modes are allowed to fluctuate and we set to zero the fluctuations of the sub-leading one, viz., 
   \begin{align} \label{eq:twbc2}
      \begin{split}
                   \left( \epsilon  \psi\right)^+_0  & = 0\,, \quad    \delta \left( \epsilon  \psi\right)^+_0 = 0\,, \quad  \left( \bar \epsilon \bar \psi\right)^+_0 = 0\,, \quad \delta  \left( \bar  \epsilon \bar\psi\right)^+_0 = 0\,,\\
                        \left( \epsilon \psi\right)^-_0  & \neq 0\,, \quad \delta       \left( \epsilon \psi\right)^-_0 \ne 0\,, \quad  \left( \bar \epsilon \bar \psi\right)^-_0 \neq 0\,, \quad \delta  \left( \bar \epsilon \bar\psi\right)^-_0 \neq 0\,.
      \end{split}
  \end{align}
  We choose these boundary conditions since they are compatible with the ones selected for $F$ and $\bar F$ through the supersymmetric transformations
\begin{align}
\begin{split}
  \bar \epsilon \epsilon F + \cdots   & \= {\rm e}^{-( \wt \Delta -1) \eta}\left(  f_{(1)}{\rm e}^{-\eta}+\, \cdots \right)\=  \qeq \left({\rm e}^{- \widetilde{\Delta}_- \eta} \left(\left(    \epsilon \psi \right)^{-}_0 + \, \cdots\right) \right) \,,\\
  \bar \epsilon \epsilon \bar F + \cdots  & \= {\rm e}^{- (\wt \Delta -1)\eta}\left(  \Bar{f}_{(1)}{\rm e}^{-\eta}+\, \cdots\right)\= \qeq  \left({\rm e}^{- \widetilde{\Delta}_- \eta} \left(\left( \bar \epsilon \bar \psi\right)^{-}_0 + \, \cdots\right) \right)\,,
    \end{split}
\end{align}
  upon identifying $\wt \Delta  = \wt \Delta_-$.
 In the next subsection we will verify that this choice is compatible with the variational principle provided  $\wt \Delta_-$ satisfies $\frac{1}{2} > \wt \Delta_- > - \frac{1}{2}$.
 
\subsection{Variational principle} \label{subsec:Varprin}
The variational principle requires that the variation of the total action including boundary action around the on-shell saddle vanishes. More specifically, the variation of the action upon application of equations of motion generates boundary terms, which should vanish when imposing boundary conditions. In this subsection, given the action with boundary terms \eqref{eq:bdryLambda}, \eqref{eq:bdryPhi} and \eqref{eq:bdryPsi}, we check the boundary conditions that we impose in  subsection \ref{subsec:asympt} are consistent with the variational principle.

The explicit form of variation of the total action is
%%%%%%%%%%%%%%%%%%%%%%%%%%%%
%%%%%%%%%%%%%%%%%%%%%%%%%%%%
%%%%%%%%%%%%%%%%%%%%%%%%%%%%
\begin{align} \label{eq:deltaS}
\begin{split}
\delta S&\=  \frac{1}{\gym^2} \int {\rm d}\theta \sqrt{\gamma} \biggl\{  \delta A_2 \Bigl[\bigl(F_{12} - \frac{\i}{L}\S \bigr) + \gym^2 (\i \bar \epsilon \epsilon)\Bar{\phi}\phi \Bigr] + \delta \S   \bigl(  \partial_1 \S - \gym^2 \Bar{\epsilon}\gamma_3\gamma_1 \epsilon\, \Bar{\phi}\phi \bigr)
\\
&\qquad \qquad\qquad\quad   + \delta \P \, \partial_1  \P -\frac{1}{4} \lambda \delta \lambda \bar \epsilon \gamma^2 \bar \epsilon -\frac{1}{4} \bar \lambda \delta \bar \lambda \epsilon \gamma^2 \epsilon + \frac{\i}{4} \delta \bar \lambda \gamma^1  \lambda -
\frac{3}{4}\i \delta  \lambda   \gamma^1 \bar \lambda \biggr\}  
\\
& \quad \qquad +\int {\rm d}\theta \sqrt{\gamma}  \biggl\{ \delta  \Bar{\phi} \bigl[  D_1 \phi + \i (\i \bar \epsilon \epsilon) D_2 \phi - \bar{\epsilon}\gamma_3 \gamma_1 \epsilon \phi \S \bigr]  
\\
&
\qquad \qquad \qquad \qquad  
+\delta \phi \bigl[D_1 \Bar{\phi} -\i (\i \bar \epsilon \epsilon)  D_2 \Bar{\phi}- \bar{\epsilon}\gamma_3 \gamma_1 \epsilon \bar{\phi} \S \bigr] 
\\
& \qquad \qquad \qquad \qquad +\frac{1}{2} \delta \bar{\psi}\bigl[  \gamma_2 \psi  (\i\Bar{\epsilon}\epsilon)+\psi  (\i \Bar{\epsilon}\gamma_2\epsilon)+ \gamma_1\psi (\Bar{\epsilon}\gamma_3 \epsilon)\bigr]
\\
&
\qquad \qquad \qquad \qquad +\frac{1}{2} 
\bigl[ \i \Bar{\epsilon}\epsilon \Bar{\psi}\gamma_2 + (\i \bar \epsilon \gamma_2 \epsilon)\Bar{\psi} + \frac{3}{2}\Bar{\epsilon}\gamma_3 \epsilon \Bar{\psi} \gamma_1
\bigr]
\delta{\psi} \biggr\} + \mbox{e.o.m.}\,.
\end{split}
\end{align}
In what follows we separately study the contributions form different multiplets using the asymptotic expansions and boundary conditions presented in subsection \ref{subsec:asympt}.

\subsection*{The vector multiplet}

We consider the contribution from the vector multiplet.
The bosonic part of the vector multiplet contribution to \eqref{eq:deltaS} is
\begin{align} \label{eq:deltaSvmB}
    \begin{split}
        \delta S^{\text{v.m.}}\Big{|}_{\text{boson}} & =  \frac{1}{\gym^2} \int {\rm d}\theta ~L {\rm e}^{\eta}\left(\delta A_{\theta} \left(F_{12} -  \i \frac{\S}{L}\right){\rm e}^{-\eta}+\delta \S \partial_{\eta} \S  + \delta \P \partial_\eta \P \right) \,.
    \end{split}
\end{align}
Here, we have ignored coupling to the chiral multiplet. 
The effect of this interaction will be considered later when studying the chiral multiplet.
If we impose the asymptotic expansion given in~\eqref{eq:VMexpansion} with the boundary condition given in~\eqref{eq:aaSP0} and \eqref{eq:VMbc}, we find that %\begin{align}
\beqa
\delta S^{\text{v.m.}}\Big{|}_{\text{boson}} \longrightarrow -\i \S_0 \delta a_{\theta,(1)}\,,\qquad\mbox{as }~\eta \rightarrow \infty\,.
\eeqa
Therefore, we set $\S_0=0$ to satisfy the variational principle.

 To analyze the fermionic part of the vector multiplet contribution to \eqref{eq:deltaS} given by
 \begin{align} \label{eq:deltaSvmF}
    \begin{split}
        \delta S^{\text{v.m.}}\Big{|}_{\text{fermion}} & =  \frac{1}{\gym^2} \int {\rm d}\theta ~L {\rm e}^{\eta}\left( -\frac{1}{4} \lambda \delta \lambda \bar \epsilon \gamma^2 \bar \epsilon -\frac{1}{4} \bar \lambda \delta \bar \lambda \epsilon \gamma^2 \epsilon + \frac{\i}{4} \delta \bar \lambda \gamma^1  \lambda -
\frac{3}{4}\i \delta  \lambda   \gamma^1 \bar \lambda\right) \,,
    \end{split}
\end{align}
 we consider the asymptotic expansion of the gaugini according to the definition of twisted fermions~\eqref{eq:LambdaTW} and their asymptotic expansion given in~\eqref{eq:lambdaAsymp}.  Thus, we have
\begin{align}
\begin{split}\label{eq:fermAsymp}
   \lambda & = - \frac{1}{\i \bar \epsilon \epsilon} \left[\gamma^A \epsilon \left(\lambda_{A,(0)} + \lambda_{A,(1)}{\rm e}^{- \eta} + \cdots \right)\right] \, , \\
   \delta \lambda & = - \frac{1}{\i \bar \epsilon \epsilon} \left[\gamma^A \epsilon \left(\delta\lambda_{A,(0)} +  \delta \lambda_{A,(1)}{\rm e}^{- \eta} + \cdots \right)\right]\,, \\
   \bar \lambda & =  \frac{1}{\i \bar \epsilon \epsilon} \left[\gamma^A \bar \epsilon \left(\lambda_{A,(0)} + \lambda_{A,(1)}{\rm e}^{- \eta} + \cdots \right)\right]\,,\\
   \delta \bar \lambda &  = \frac{1}{\i \bar \epsilon \epsilon} \left[\gamma^A \bar \epsilon \left(\delta \lambda_{A,(0)} + \delta \lambda_{A,(1)}{\rm e}^{- \eta} + \cdots \right)\right]  .
   \end{split}
\end{align}
After imposing the boundary condition~\eqref{eq:lambdaAbc}, we insert the expansion~\eqref{eq:fermAsymp} into the boundary contribution in~\eqref{eq:deltaSvmF}, we can check that  $S^{\text{v.m.}}\Big{|}_{\text{fermion}}  \rightarrow 0$ as $\eta \rightarrow \infty$, thus the variational principle is satisfied.

\subsection*{ The chiral multiplet}

We now move on to study the chiral multiplet with  coupling with the vector multiplet. The bosonic part of the chiral multiplet contribution to \eqref{eq:deltaS} is

 \begin{align} \label{eq:deltaScmB}
    \begin{split}
        \delta S^{\text{c.m.}}\Big{|}_{\text{boson}} & =  \int {\rm d}\theta {\rm e}^{\eta}\Biggl( \bar \phi \phi\left(\delta A_{\theta}  -\delta \S   {\rm e}^{\eta}\right)  +\delta \phi  \left(D_{\eta} \bar \phi - \i D_{\theta}\bar \phi- {\rm e}^{\eta} \S \bar \phi \right) \\
        & \qquad \qquad \quad+\delta \bar \phi \left(D_{\eta} \phi + \i  D_{\theta} \phi - {\rm e}^{\eta} \S  \phi\right) \Biggr) \, .
    \end{split}
\end{align}
According to boundary condition give in~\eqref{eq:phiBC0} and \eqref{eq:phiBC}, the fluctuation of scalar fields $\phi$ and $\bar \phi$ take the following form
\begin{align} \label{eq:deltacmcexp}
    \begin{split}
        \delta \phi & =  {\rm e}^{-\Delta_+ \eta}( \delta \phi_{(0)}^+ + \, \cdots)  - \delta \Delta_+ {\rm e}^{-\Delta_+ \eta}(\phi_{(0)}^+ + \, \cdots) \,, \\
        \delta \bar \phi & =  {\rm e}^{-\Delta_+ \eta}( \delta \bar \phi_{(0)}^+ + \, \cdots)  - \delta \Delta_+ {\rm e}^{-\Delta_+ \eta}(\bar \phi_{(0)}^+ + \, \cdots)   \,,
    \end{split}
\end{align}
 where $\delta \Delta_{+}$ takes into account the fact that the $\Delta_+$ depends on the $\P_0$ as shown in~\eqref{eq:range} which can in principle fluctuate. 
 From \eqref{eq:range} we have that $\delta \Delta_{\pm} \sim \delta \P_0$. Therefore, recalling that the boundary condition $\delta \P_0 =0$ given in~\eqref{eq:VMbc}, we have $\delta \Delta_+ = 0$.
We then insert  the expansion~\eqref{eq:deltacmcexp} into  the boundary contribution~\eqref{eq:deltaScmB} and obtain that $\delta S^{\text{c.m.}}\Big{|}_{\text{boson}} \rightarrow 0$ as $\eta \rightarrow \infty$, which is consistent with variational principle.

The chiral multiplet fermions are better analyzed in terms of the cohomological variables defined in~\eqref{eq:psiTw}.
We can write the fermionic contribution of the chiral multiplet to the variation of the action~\eqref{eq:deltaS} as
  \begin{align} \label{eq:deltaScmF}
     \begin{split}
          \delta S^{\text{c.m.}} \Big{|}_{\text{fermion}} \!\!& = \int {\rm d}\theta\frac{\sinh \eta}{2} \biggl(\cosh \eta \left(\delta \bar \psi \gamma_2 \psi - \delta \psi \gamma_2 \bar \psi\right) + \sinh \eta \left(\delta \bar \psi \psi-\delta \psi \bar \psi\right)
      \\
        &    \qquad \qquad \qquad \quad + \i \Bigl(\,\frac{3}{2} \delta \psi \bar \psi - \delta \bar \psi  \psi\Bigr)\biggr)\,.
        \end{split}
  \end{align}
Writing the physical fermions in terms of twisted fermions defined in~\eqref{eq:psiTw} and using the asymptotic expansion~\eqref{eq:twsc1} and \eqref{eq:twsc2}, we insert the result into the boundary contribution~\eqref{eq:deltaScmF}. Then, we find that imposing the boundary conditions given in \eqref{eq:twbc1} and \eqref{eq:twbc2}
    implies that $\delta S^{\text{c.m.}}\Big{|}_{\text{fermion}} \rightarrow 0$ as $\eta \rightarrow \infty$, thus  variational principle is satisfied.

%%%%%%%%%%%%%%%%%%%%%%%%%%%%%%%%%%%%%%%%%%%%%%%%%%%%%%%%%%%%%%%%%%%
%%%%%%%%%%%%%%%%%%%%%%%%%%%%%%%%%%%%%%%%%%%%%%%%%%%%%%%%
%%%%%%%%%%%%%%%%%%%%%%%%%%%%%%%%%%%%%%%%%%%%%%%%%%%%%%%%

\section{Supersymmetric localization} \label{sec:susyloc}

In this section, we use the method of supersymmetric localization to compute the partition function of $\cN=(2,2)$ theories with an Abelian vector multiplet and a chiral multiplet placed on S$^2$ and~AdS$_2$ .

The general principle of the localization method is as follows:
We first deform the action by adding a $\qeq$-exact term with real parameter $t$:
\be\label{QVdeformtoS}
S[\varphi] \rightarrow S[\varphi] + t \,\qeq \cV\,.
\ee
Here $\qeq$ is a supercharge defined in \eqref{qeqDef} whose algebra is equivariantly closed to a compact  bosonic symmetry $\cH$  in the following way:  $\qeq^2 \equiv \cH$, where $\cH$ has the form given in~\eqref{EqAlgebra}. The fermionic quantity $\cV$ is chosen such that $\qeq \cV |_{\text{boson}} \geq 0$ and $\cH \cV =0$.

Since the action $S[\varphi]$ and the functional integration measure are $\qeq$-invariant as a supersymmetric theory, the partition function turns out to be  independent of $t$, thus we can take $t \rightarrow \infty$. In this limit, new saddle points are obtained from locus of the $\qeq \cV$,
\be
\cM_{\text{loc}}= \left\{\varphi_{\text{loc}} \Big{|} \hspace{2mm} \qeq \cV(\varphi_{\text{loc}})\Big{|}_{\text{boson}}= 0, \hspace{2mm} \chi =0\right\}\,. \label{eq:Mloc}
\ee
This solution of the locus~\eqref{eq:Mloc} is called localization saddle since evaluation of the partition function is localized to the integration along this locus,  yielding an exact result, i.e., 
\beqa\label{LocalizedZ}
Z & =& \lim_{t \rightarrow \infty}Z _{t} 
 = \int_{\mathcal{M}_{\text{loc}}} \cD \varphi_{\text{loc}}   {\rm e}^{- S(\varphi_{\text{loc}})} Z^{\prime\qeq \cV}_{\text{1-loop}}\,,\label{eq:ZintLoc}
\eeqa
where $Z^{\prime \qeq \cV}_{\text{1-loop}}$ is 1-loop determinant of the quadratic kinetic operator in $\qeq \cV$ action. Note here that the ``prime'' denotes exclusion of the zero modes in evaluation of the 1-loop.

Throughout this section,  we explicitly evaluate \eqref{LocalizedZ} for our theory.
In subsection~\ref{subsec:locsad}, we choose the supersymmetric deformation action $\qeq\CV$ and present the solution of the localization equation~\eqref{eq:Mloc} on \ss2 and \ads2, which is classified in terms of cohomological variables.  This solution   yields the induced measure~$ \cD \varphi_{\text{loc}}$ and the action on the saddle $S(\varphi_{\text{loc}})$ of the partition function~\eqref{LocalizedZ}.
In subsection~\ref{subsec:IT}, we discuss  how the computation of the 1-loop determinant, $Z^{\prime \qeq \cV}_{\text{1-loop}}$, turns into the computation of an index. Here, we pay special attention to the effect of excluding zero modes in the 1-loop. For \ss2 case, we keep track of the effect of zero modes in ghost and anti-ghost field. For \ads2, we treat the boundary 1-form zero modes, where we show that the fermionic superpartner of the boundary zero modes exist in normalizable fermionic space, and further they are also zero modes. The effect of both zero modes are analyzed. 
After showing that the operator for the index is transversally elliptic, we evaluate, in subsection~\ref{subsec:1loop},
the 1-loop partition function through  the explicit computation of the index.
During this process we classify possible ways of expansion of the index, and single out one according to our boundary condition. An appropriate regularization of infinite products will give the exact 1-loop determinant, which contains the overall dependence on the size of manifold $L$ as the local anomaly contribution as well as global zero mode contribution. 
Towards the end of subsection \ref{subsec:1loop} we show the results of the full partition function for both S$^2$ and AdS$_2$. 

\subsection{Localization saddle} \label{subsec:locsad}
%%%%%%%%%%%%%%%%%%%%%%%%%%%%%%%%%%%%%%%%%%%%%%%%%%%%%%%%%%%%%%%%%  

%%%%%%%%%%%%%%%%%%%%%%%%%%%%%%%%%%%%%%%%%%%%%%%%%%%%%%%%%%%%%%%%%%
We first choose a $\qeq$-exact deformation of the action that satisfies the conditions of being~$\qeq$ invariant and has a positive definite bosonic part, as we have discussed in~\eqref{QVdeformtoS}. Since the vector multiplet action in~\eqref{Actionvector} and the chiral multiplet action in~\eqref{Actionchiral} are $\qeq$-exact, they themselves are good candidates to be used as the $\qeq$-exact deformation for the localization. Indeed, they were used for the case of S$^2$ in \cite{Benini:2012ui, Doroud:2012xw}. However, when dealing with \ads2, we need to be cautious about the positive definiteness of the  action. 
In fact, for the background value of $H$ and $G$ satisfying \eqref{auxiliaryads2}, the bulk Lagrangian given by \eqref{eq:LYM} is no longer positive definite. 
Hence we rule out the action $S_{\text{v.m.}}$ as a permitted deformation to apply localization on AdS$_2$. 
As far as the $\qeq$-exact action is positive definite, we have freedom to choose. In this paper, we use the `canonical' choice of the $\qeq$-exact deformation action whose bosonic part is manifestly positive definite both for vector and chiral multiplet on \ss2 and \ads2, that is 
\beqa
\qeq \cV &=&\qeq \cV^{\text{v.m.}} + \qeq \cV^{\text{c.m.}} + \qeq\cV^{\text{ghost}} 
\label{eq:lag12}
\\ \nonumber
& = & \int \textrm{d}^2 x \sqrt{g}\biggl[ \frac{1}{2 }\qeq \left((\qeq \lambda)^{\dagger} \lambda + \bar{\lambda} (\qeq \bar{\lambda})^{\dagger}\right)  + \dfrac{\i}{g^2_\textrm{YM}}\qeq \left(\bar{c}\nabla_{\mu}A^{\mu}+ \Bar{c}b_0 + c \Bar{\Lambda}_0 \right)\\  \nonumber 
 & &\quad \quad \quad \quad \quad  +\frac{1}{2 }\qeq \left((\qeq \psi)^{\dagger} \psi+ \bar{\psi}(\qeq \bar{\psi})^{\dagger}\right)\biggr] \,.
\eeqa
Notice that, for the case of \ss2, this canonical prescription produces $S_{\text{v.m.}}$ itself as the deformation action. 
In \eqref{eq:lag12} we have included the $\qeq$-exact form of the ghost action presented in~\eqref{actionghost} for the localization, where we again remind that $b_0$ and $\bar \Lambda_0$ are zero on \ads2.

Demanding that the bosonic part of \eqref{eq:lag12} vanishes yields the localization equations. To facilitate the process of finding the solution of the localization equations, we can write $\qeq \cV$  as a sum of squares. 
For the vector multiplet, we have\footnote{ \label{foot:factor}We could divide by $\epsilon^{\dagger} \epsilon$  in  \eqref{eq:lag12} as a choice to obtain  standard kinetic term in~\eqref{eq:Total01}. However, dividing by this factor would cause trouble for  transversal ellipticity of $D_{10}$ in  \ref{subsubsec:TE} (specifically in equation \eqref{eq:Symbolads2}) as $\eta \rightarrow \infty$.   }
\beqa \label{eq:Total01}
\qeq \cV^{\text{v.m.}}\bigg|_{\text{bos.}}
& =& 2\int \textrm{d}^2 x \sqrt{g}\epsilon^{\dagger} \epsilon \Biggl[  \bigg( F_{12}- \i H \P \bigg)^2 + \left(D+ G\S\frac{\epsilon^{\dagger} \gamma_3 \epsilon}{\epsilon^{\dagger} \epsilon}\right)^2 \nn 
\\ 
& &\quad\quad \quad\quad \quad\quad\quad+\left( D_{\mu} \S +  \i G\S  \frac{\epsilon^{\dagger} \gamma_3 \gamma_{\mu}\epsilon}{\epsilon^{\dagger} \epsilon}\right)^2  +\left(D_{\mu} \P \right)^2\Biggr] \,,
\eeqa
and for the chiral multiplet we have
\begin{align}
\qeq \cV^{\text{c.m.}} \bigg{|}_{\text{bos.}} & =  \int \textrm{d}^2 x \sqrt{g} 
\left((\qeq \psi)^{\dagger} \qeq \psi + (\qeq \bar{\psi})^{\dagger} \qeq \bar{\psi}\right)\label{eq:QVcm01} \\ \nonumber
& = \int \textrm{d}^2 x \sqrt{g} 
\biggl\{ \left| \left(
\i \gamma^\mu  D_\mu \phi  
-\i  (\S +\frac{\r}{2} H)\phi   +\gamma_3  (\P +\frac{\r}{2}G)\phi \right)\epsilon \right|^{2}   \\
& \left. \left. \quad \quad \quad \quad \qquad  + \left| \left(\i \gamma^\mu  D_\mu \bar\phi  
-\i( \S +\frac{\r}{2} H)\bar\phi   -\gamma_3 ( \P +\frac{\r}{2}G)\bar\phi\right)\Bar{\epsilon}\right|^{2} + \bar{F}F\biggr\} \right. \right.\,.\nn 
\end{align}
Here, when we take conjugation, $\dagger$,  we have made use of background value of $H$ and $G$ given in \eqref{auxiliaryS2} for \ss2 or in~\eqref{auxiliaryads2} for \ads2. Moreover, we also have used the reality conditions on the fluctuations of fields given in ~\eqref{eq:ReVec} and~\eqref{eq:ReChiral}.  We point out that, although it is not manifestly expressed, the \eqref{eq:Total01} and \eqref{eq:QVcm01} are in terms of the fluctuations of the fields in the theory.

From the sum of squares in \eqref{eq:Total01}, we find  the localization saddle for the vector multiplet: On \ss2,  the solution is parameterized by a real constant $\S_0$ and an integer $\mathbf{m}$ as 
\beqa \label{localS2}
&&\S= \S_0\,,\quad \quad \quad D = 0\,,
\\
&& \P  =  \frac{ \mathbf{m}}{2 L}\,,\quad \quad  F_{12} = \frac{\mathbf{m}}{2 L^2}\, \quad \Rightarrow \quad A =-\frac{\mathbf{m}}{2 } \left(\cos \psi \mp 1\right)d \theta\,.
\nn \eeqa
Note that this saddle coincide with the classical solution  in~\eqref{GNOs2} of the case of $\xi=0$. 

On \ads2, we have off-shell solution on top of the classical configuration \eqref{solutionads2}. The off-shell solution is parameterized by a set of real constants,  $\alpha^{(\ell)}_{\text{bdry}}$ with $ \ell = \pm 1\,, \pm 2\,,\cdots$, and a real constant $\S_1$ as \footnote{  If we impose $\hat{D}\equiv D- \frac{\i}{L}  \P$ to be real instead of $D$ real as in~\eqref{eq:ReVec}, then we find additional solution $$\P = \frac{\P_1}{\cosh \eta}\,, \quad F_{12} = \frac{\P_1}{L \cosh^2 \eta}\,.$$ However, this reality condition and the new solution may spoil  the positivity of the action as explained in~\eqref{Dredef}. Thus we exclude this new solution by setting $D$ real instead of $\hat{D}$ real. In the context of \ads2$\times$ \ss2, there are analogous solutions. The solution found in~\cite{Dabholkar:2010uh,Gupta:2012cy} is analogous to the $\S_1$ and  the additional one found in~\cite{Gupta:2015gga} is analogous to the $\P_1$. This existence of the additional solution has been puzzling, but we expect that this can also be excluded demanding positivity of the action by imposing appropriate reality condition of some auxiliary field in $\cN=2$ supergravity.  
}
\beqa \label{localads2}
&&\S  \=   \S_0 + \frac{\S_1}{\cosh\eta} \,,\quad \hspace{5mm} D\=  \frac{\S_1}{L \cosh^2\eta } \,, \hspace{5mm}\quad
\P\=  \P_0  \,,
 \\
 && F_{12}\= \i \frac{\S_0}{L}  \quad \Rightarrow \quad A \= \i\, \S_0 L \,(\cosh\eta -1)d\theta + \sum_{\ell \neq 0}\alpha^{(\ell)}_{\text{bdry}}  A^{(\ell)}_{\text{bdry}}  \,,\nn
\eeqa 
where $A_{\text{bdry}}^{(\ell)} $  are the boundary zero modes as already defined in \eqref{nonnormalizablescalarsads2}.
We note that the off-shell configuration parameterized by $\S_1$ and $\alpha^{(\ell)}_{\text{bdry}}$ are  regular everywhere and square integrable. From our choice of normalizable boundary conditions for bosons on \ads2, it follows that the constants $\S_0$ and $\P_0$ are fixed values. 
We also note that the gauge field $A_\theta$ vanishes at the origin of \ads2, i.e. at~$\eta=0\,$, and thus regular everywhere.  

For the chiral multiplet, the sum of squares form in \eqref{eq:QVcm01} with generic vector multiplet saddle given in \eqref{localS2} and \eqref{localads2} implies that there is no non-trivial saddle solution. This result can be sketched as follows.
For \ss2, the  non-trivial locus of \eqref{eq:QVcm01} is given by 
\beqa
	\phi = C \left(\tan\frac{\psi }{2}\right)^{\frac{\mathbf{m}}{2} } \ \left(\sin \psi \right)^{-\frac{r}{2}-\i L \sigma_0 } {\rm e}^{L \theta  (\sigma_0 \mp \i \frac{\mathbf{m}}{2 L}  )-\frac{\i r \theta }{2}+  \i \mathbf{m} \cos \psi} \,.  \label{eq:PhiS2}
\eeqa
The $\mp$ in the exponent of \eqref{eq:PhiS2} corresponds to solutions for the patch containing the north and south pole of \ss2 respectively.
The~$\phi$ solution blows up at the north and south poles unless $\sigma_0=0$ and $ \mathbf{m} \geq  r$, which is violated for some values of $\mathbf{m}$. Therefore we conclude that $C=0$.
For \ads2, the  non-trivial locus of \eqref{eq:QVcm01} is given by
\beqa 
	\phi &=& C  \left(\tanh \frac{\eta}{2} \right)^{- \i \S_0 L }\left( \sinh 2 \eta\right) ^{\frac{\P_0 L}{2}-\frac{r}{4}} \left(\tanh\eta\right)^{\frac{\P_0 L}{2}-\frac{r}{4}-\i  \S_1 L} {\rm e}^{\i \theta \left((\P_0 L-\i  \S_1 L- \i \S_0 L)-\frac{r}{2}\right)}\, .  \label{eq:newPhiads}
\eeqa
Note that for all $\P_0 L \neq \frac{r}{2}$ the $\phi$ solution in \eqref{eq:newPhiads} blows up either in the limit $\eta \rightarrow \infty$ or  $\eta \rightarrow 0$. If $\P_0 = \frac{r}{2}$, then the function \eqref{eq:newPhiads} becomes ill-defined at the origin because of the term $\tanh(\eta)^{- \i L \S_1}$. Therefore, we conclude that the only valid solution for $\phi$ corresponds to $C=0$.

\subsection*{In terms of cohomological variables}\label{BCsaddle}
It will be instructive to see how the localization saddles can be also reproduced if we write the  $\qeq \cV$-action in terms of the twisted variables. In what follows we present these results for the vector multiplet and the chiral multiplet.

For vector multiplet, we can write the $\qeq \cV^{\text{v.m.}}+\qeq \cV^{\text{ghost}}$ in terms of the twisted variables by simply replacing \eqref{eq:LambdaTW} in the vector multiplet contribution of \eqref{eq:lag12}  to obtain  
\be
\qeq \cV^{\text{v.m.}} +\qeq \cV^{\text{ghost}} = \int \textrm{d}^2 x \sqrt{g} \left[\frac{1}{2 (\i \bar \epsilon \epsilon) } \qeq \left(\left(\qeq \lambda_A\right)^{\dagger} \lambda_A  \right) + \dfrac{\i}{g^2_{\textrm{YM}}}\qeq \left(\bar{c}\nabla_{\mu}A^{\mu}+\Bar{c}b_0 + c \Bar{\Lambda}_0\right)\right]. \label{eq:QVeqvm}
\ee
where on \ads2, $b_0$ and $\Bar \Lambda_0$ are zero. Setting to zero the bosonic part of \eqref{eq:QVeqvm} leads to the localization saddle. In particular, the vanishing of the ghost multiplet contribution automatically fixes the gauge whereas the vector multiplet contribution yields  
\begin{align}
\qeq \lambda_A =0\, ,  \qquad A =1\, ,    \cdots, \, 4\, .
\end{align}
Let us now analyze them explicitly
\begin{align}
\begin{split}
\qeq \lambda_{a} &  =  {\rm e}^{\mu}_a \cL_{\xi} A_{\mu} - \partial_a (\qeq c) = \xi^{b}F_{b a} -  \partial_{a}(\bar{\epsilon} \epsilon \S )-  \i \left[\partial_{a}(\bar{\epsilon} \gamma_3\epsilon \P)\right]  =0\, ,\label{eq:BPStw}\\
\qeq \lambda_3 & =  \cL_{\xi} \rho =0\, ,\\
\qeq \lambda_4 & = \i \bar \epsilon\epsilon \, D   + \i  \bar \epsilon \gamma_3 \epsilon  \left(-\i F_{12} +  G \S  -  H \P  \right) + \i \bar\epsilon \gamma_3 \gamma^b \epsilon \,\partial_b \P  =0\,. 
\end{split}
\end{align}
To solve these equations, we separate them into their real and imaginary parts where we use the reality condition on fields \eqref{eq:ReVec} and the values of the bispinors \eqref{bispinorsS2} and \eqref{bispinorsAdS2}. Then it is straightforward to see that the localization saddles coincide with those found directly in terms of the physical fields \eqref{localS2} for \ss2 and \eqref{localads2} for \ads2.

For chiral multiplet, we now  rewrite $\qeq \cV^{\text{c.m.}}$ in terms of the twisted variables. Using~\eqref{eq:psiTw} we can write:
\beqa
\qeq \cV^{\text{c.m.}} & = & \int \textrm{d}^2 x \sqrt{g} ~\frac{1}{2 } \qeq \left[\left(\qeq (\epsilon \psi)\right)^{\dagger} \epsilon \psi + (\qeq (\epsilon \bar{\psi}))^{\dagger} \epsilon\bar{\psi} + (\qeq \bar{\epsilon} \bar{\psi})^{\dagger} \bar{\epsilon} \bar{\psi}
\right.    \nonumber\\ 
&  &\quad\quad\quad\quad\quad\quad\quad \quad\left.  +(\qeq \bar{\epsilon} \psi)^{\dagger} \bar{\epsilon} \psi \right]\,.\label{eq:QVcmTW}
\eeqa
Setting to zero the bosonic part of \eqref{eq:QVcmTW} implies the vanishing of \eqref{eq:ChBPS3}.
With the reality $\bar{F} = F^{\dagger}$ and $\bar{\phi} = \phi^{\dagger}$ , we  deduce trivial solution, $F = \bar{F} = 0$ and $\phi =\bar{\phi} = 0$, which agrees with the result obtained in \eqref{eq:PhiS2} and \eqref{eq:newPhiads}.
%%%%%%%%%%%%%%%%%%%%%%%%%%%%%%%%%%%%%%%%%%%%%%%%%%%%%%%%%%%%%%%%%%%%%%
%%%%%%%%%%%%%%%%%%%%%%%%%%%%%%%%%%%%%%%%%%%%%%%%%%%%%%%%%%%%%%%%%%%%%%%%%%%%%%%%%%%%

\subsubsection*{Cohomological classification}
The localization solution \eqref{localS2} and \eqref{localads2} can now be   organized according to  the cohomological classification \eqref{eq:VQarray} and \eqref{eq:CQarray}. 

As is shown in \eqref{localS2},  the localization solution on \ss2  is parameterized by  magnetic fluxes $\mathbf{m}$ and one real constant parameter~$\S_0$. In terms of the cohomological classification, the mode $\S_0$ is encoded in the $\qeq$-singlet $\Lambda_0$. i.e. 
\beqa\label{saddleCohoS2}
 \S_0 \;\in\; \Lambda_0\,,
\eeqa
  as its explicit form is given by
\beqa
\Lambda_0
\=\i \S_0 L \mp \frac{\mathbf{m}}{2}  \label{eq:L0S2} \,   \qquad \qquad \textrm{for S$^2$}\,,
\eeqa
where we remind that the $\mp$ in \eqref{eq:L0S2} are for the north and the south poles on \ss2 respectively.

Also, as was shown  in \eqref{localads2}, the localization solutions on \ads2 are parameterized by  infinite constants $\alpha^{(\ell)}_\text{bdry}$ and one constant, $\S_1$. They are respectively classified in  the elementary boson $\Phi$ and the $\qeq$-singlet $\Lambda_0$, i.e. 
\beqa\label{saddleCohoAdS2}
\alpha^{(\ell)}_\text{bdry} \;\in \; \Phi\,, \qquad\qquad\S_1 \;\in\; \Lambda_0\,.
\eeqa
This is because the  $\alpha^{(\ell)}_\text{bdry}$ are the modes in the 1-form gauge field which is in elementary boson $\Phi$. As for the $\S_1$ mode, although it appears in two fields $\S$ and $D$ as shown in \eqref{localads2}, in the expression of the $\qeq \lambda_4$ in \eqref{eq:BPStw}, the mode $\S_1$ is canceled within it. The only place where  the mode  $\S_1$ appears in the cohomological classification is the $\qeq$-singlet $\Lambda_0$ as its explicit form is given by 
\beqa
%\textrm{AdS$_2$:}&&\quad 
\Lambda|_{\text{loc.}}\=\Lambda_0|_{\text{loc.}} \= - A_{\theta}^{\text{bdry}}+ \i  \left( \S_0 +\S_1\right)L  - \P_0 L \label{eq:L0Ads2}\,    \qquad \qquad \textrm{for AdS$_2$}\,,
\eeqa
where we remind that we set $\S_0 =0$  to ensure  well defined variation principle in section~\ref{bdycondition}.

It is important to note that while the variable $\Lambda_0$ does not have its superpartner since it is singlet of $\qeq$ as was mentioned in~\eqref{qeqGGmultiplet} and after~\eqref{eq:QLamb4}, the variable $\Phi$ has its superpartner $\qeq \Phi$ by  construction of  the cohomological variables. Therefore, while the mode 
$\S_0$ or $\S_1$  does not have its superpartner, the modes $\alpha^{(\ell)}_\text{bdry}$ have their superpartners in the $\qeq \Phi$ of the cohomological classification.
These two classes of the localization solution, one in $\Phi$ and the other in $\Lambda_0$,  will differently play their role later in section~\ref{subsubsec:ZM}, when we analyze the effect of zero modes in the method of index for 1-loop computation.

\subsection*{Integration measure and action } 
By the argument of supersymmetric localization, the functional integration measures defined in terms of the variables \eqref{IntVariableVM} and \eqref{IntVariableCM} are now reduced along the localization saddle \eqref{localS2}
 for \ss2 and \eqref{localads2} for \ads2. Thus we have the following finite dimensional measure
 \be \label{eq:IntMeasure}
\sum_{\mathbf{m}\in \mathbb{Z}}\int \frac{{\rm d}(\S_0 L)}{\left(\gym L_0\right)^2}\qquad\mbox{for \ss2}\,,\qquad\qquad\int \frac{{\rm d}(\S_1 L)}{\gym L_0}\qquad\mbox{for \ads2}\,.
\ee

In the case of \ads2, the contribution to the measure from the boundary zero modes is trivial as can be seen as follows 
\be\label{MeasureonAdS2zm}
\int {\cD \varphi_{\text{zm}}^{\text{AdS}_2}} \= \int {\cD A^{\text{bdry}}} \= \int \prod_{\ell}{d\left( e_a^{\mu}A_\mu^{(\ell)}L\right)}\sim 1\,,
\ee
where  the $L$ factor in the measure is canceled by the $1/L$ dependence of $e_a^\mu\,$.

At the localization saddle  given in  \eqref{localS2} or~\eqref{localads2}, the value of the  vector multiplet action $S_{\vm}$ in \eqref{eq:LYM} and chiral multiplet action $S_{\cm}$ \eqref{mattermultiplet} all vanish and only non-trivial contribution in the total action comes from the FI-term and the topological term. For \ss2, 
\begin{align} \label{eq:SclaS2}
   S^{\text{S}^2}_{\text{tot}}\Bigr|_{\text{loc.}}\=  S^{\text{S}^2}_{\text{FI}} +  S^{\text{S}^2}_{\textrm{top.}}\Bigr|_{\text{loc.}}&= - \i  4 \pi \xi (\S_0 L) +  \i \vartheta \mathbf{m} \,.
\end{align}
For \ads2, the total action on localization saddle vanishes since we set $\xi=\vartheta = 0$ due to the variational principle.

%%%%%%%%%%%%%%%%%%%%%%%%%%%%%%%%%%%%%%%%%%%%%%%%%%%%%%%%%%%%%%%%%%%%%%%%%%%%%%
%%%%%%%%%%%%%%%%%%%%%%%%%%%%%%%%%%%%%%%%%%%%%%%%%%%%%%%%%%%%%%%%%%%%%%%%%%%%%%%%%%%%%%%%%%%

\subsection{Index method 
}\label{subsec:IT}
In order to evaluate the 1-loop determinant  $Z^{\prime \qeq \cV}_{\text{1-loop}}$, we use the index method\footnote{There is a central issue to the applicability of the index theorem  to non-compact spaces like the \ads2. The theorem in its original formulation was shown to hold in compact spaces. In \cite{Atiyah:1974obx}, the applicability of the theorem was further enhanced by considering non-compact spaces where a part was excised to make it \textit{effectively} compact. However, our case is dissimilar in the sense that the fields actually do continue up to the conformal boundary at $\eta \rightarrow \infty$. Therefore, as discussed, there are boundary conditions on the field. However, we use the compact version of the theorem assuming that the boundary conditions do not alter the final results. This is a priori not obvious at all, but the agreement with heat kernel results gives us a consistency check that this might be a plausible assumption. We leave further work in this regard for the future . }. According to this method, evaluation of the 1-loop determinant is reduced to calculating the index of an operator denoted by $D_{10}$, where the $D_{10}$ is a map from the elementary boson $\Phi$ into the elementary fermion $\Psi$ of the cohomological variables as can be seen in the schematic form of the $\qeq \cV$ action in~\eqref{Vcoho}.  However, since the computation of the 1-loop determinant  
excludes the zero modes of the quadratic part of $\qeq \cV$ action\footnote{In our case,  the $\qeq \cV$ action  itself is quadratic: it is clear for vector multiplet, and the $\qeq \cV_{\text{c.m.}}$ for chiral multiplet is also quadratic in the Coulomb branch localization.} ,  we have to take this effect into account in this reduction process. This effect of zero modes  provides extra contribution in addition to the index contribution.

In the following subsection \ref{subsubsec:ZM}, 
we analyze the index method and consider  the effect of the exclusion of zero modes in the 1-loop determinant. 
For both the \ss2 and \ads2 case, the zero mode in the $\qeq$-singlet $\Lambda_0$, which is the mode of scalar along localization saddle,  does not affect the 1-loop as it is factored out from the index computation. 
 For \ss2, we keep track of how the absence of zero modes and the anti-ghost affects the index analysis.
 For \ads2, we treat the boundary zero modes of gauge field and in particular their fermionic superpartner following the prescription devised in \cite{Jeon:2018kec}. 
 We fill a gap in the literature by proving that those superpartners exist in normalizable fermionic field space, which was instead assumed in~\cite{Jeon:2018kec}. It turns out that, in \ads2, they are also zero modes of kinetic term of physical action. However,  the prescription which we will apply provides an appropriate way of regularization.  
  For the actual evaluation of  the index of $D_{10}$ that will be done in section~\ref{subsec:1loop},  we use Atiyah-Bott fixed point formula. Since this formula requires the $D_{10}$ operator to be  transversally elliptic, we devote subsection \ref{subsubsec:TE}  to prove that this is indeed the case here. 

\subsubsection{The effect of zero modes 
} \label{subsubsec:ZM}
To analyze the index method with effect of zero modes of the $\qeq\cV$ action, we begin by noting that the zero modes are also accommodated either into the $\qeq$-complex $(\Phi\,,$ $ \qeq \Phi\,,$ $ \Psi\,,$ $\qeq\Psi)$ or the $\qeq$-singlet $\Lambda_0$. Let us briefly review the classification of zero modes. 

For \ss2 case,  we have a constant zero mode $\s_0$ from the scalar $\S$ which is along localization saddle obtained in~\eqref{localS2} and it is classified in the $\qeq$-singlet $\Lambda_0$ as explained in~\eqref{saddleCohoS2}. Note that there are additional zero modes which are constant modes from ghost and anti-ghost field and they are classified in elementary fermion~$\Psi$. 

For the case of~\ads2, we  have a zero mode parameterized by~$\S_1$ from the scalar $\S$ and the boundary 1-form zero modes $A^{\text{bdry}}$ which are localization solution obtained in~\eqref{localads2}  and they are respectively encoded in the $\qeq$-singlet $\Lambda_0$ and
the elementary boson $\Phi$ as explained in \eqref{saddleCohoAdS2}. We also note that the superpartner of the boundary zero mode $\qeq A^{\text{bdry}}$, whose existence will be proven shortly, are also zero modes. 
Among them, the zero mode $\Lambda_0$ is not involved in the index analysis as its integration is completely factored out from the 1-loop computation. The effect of the other zero modes in $\Phi$ or $\Psi$ with their superpartners  in the index analysis will be discussed in what follows. 

%%%%%%%%%%%%%%%%%%%%%%%%%%%%%%%%%%%%%%%%%%%%%%%%%%%%%%%%%%%%%%%%%%%%%%%%
%%%%%%%%%%%%%%%%%%%%%%%%%%%%%%%%%%%%%%%%%%%%%%%%%%%%%%%%%%%%%%%%%%%%%%%%
%%%%%%%%%%%%%%%%%%%%%%%%%%%%%%%%%%%%%%%%%%%%%%%%%%%%%%%%%%%%%%%%%%%%%%%

\subsection*{Analysis of the $Z^{\prime \qeq \cV}_{1\text{-loop}} $}
In order to convert the 1-loop computation into the computation of the index together with effect of zero modes, we look at the formal expression of the total $\qeq \cV$ action. They are the   $\qeq$-exact terms in~\eqref{eq:lag12} plus the extra $\qeq$-exact term \eqref{eq:ActionBM}, given in terms of the cohomological variables  $(\Phi\oplus\Phi_0 \,, \, \qeq (\Phi\oplus \Phi_0)\,,\,  \Psi\,,\, \qeq \Psi) %\oplus (\Phi_0\,, \, \qeq \Phi_0 )  
$. 
Here, $\Phi_0$ and $\qeq\Phi_0$ are constant multiplet in~\eqref{eq:VQarray} and they are absent for \ads2. Let us for now denote $\cB \equiv \Phi \oplus \Phi_0$.
Then the formal expression for the fermionic functional~$\cV$ can be written as
 \beqa \label{Vcoho}
 \cV & = & \int  d^2x  \sqrt{g}  \Biggl[\bigl(\ba{cc} \qeq \cB & \quad  \Psi  \ea  \bigr)  \Biggl( \ba{cc} D_{00} & D_{01}  \\ D_{10} & D_{11} \ea  \Biggr) \Biggl( \ba{cc}  \cB \\ \qeq \Psi \ea    \Biggr) \Biggr],
 \eeqa
which implies
  \be
 \qeq \cV  = \int  \textrm{d}^2 x \sqrt{g} \Biggl[ \bigl(  \cB \quad \qeq \Psi  \bigr) \mathcal{K}_b \Biggl( \ba{cc}  \cB \\ \qeq \Psi \ea \Biggr)  +\bigl(   \qeq \cB \quad \Psi  \bigr)\mathcal{K}_f \Biggl( \ba{cc} \qeq  \cB \\  \Psi \ea \Biggr) \Biggr], \label{eq:QVcohom}
 \ee
where, by using $\cH^T = -\cH$ the bosonic and fermionic kinetic operators can be written as
\beqa \label{eq:Kb}
\mathcal{K}_b &  = & \frac{1}{2}\Biggl( \ba{cc} - \cH & 0 \\
0 & 1 \ea \Biggr) \Biggl( \ba{cc} D_{00} & D_{01} \\
D_{10} & D_{11} \ea \Biggr) +\frac{1}{2} \Biggl( \ba{cc} D^{\text{T}}_{00} & D^{\text{T}}_{10} \\
D^{\text{T}}_{01} & D^{\text{T}}_{11} \ea \Biggr) \Biggl( \ba{cc}  \cH & 0 \\
0 & 1 \ea \Biggr)\, ,  \\ \nn
\mathcal{K}_f & = & \half \Biggl( \ba{cc} 1& 0 \\
0 & - \cH \ea \Biggr) \Biggl( \ba{cc}  D^{\text{T}}_{00} & D^{\text{T}}_{10} \\
D^{\text{T}}_{01} & D^{\text{T}}_{11} \ea \Biggr) -\half \Biggl( \ba{cc} D_{00} & D_{01} \\
D_{10} & D_{11} \ea \Biggr) \Biggl( \ba{cc} 1& 0 \\
0 &  \cH \ea \Biggr)\,.
\eeqa
From this, we want to compute the ratio of determinant of  bosonic and fermionic kinetic operators. Note that for this computation, we should work in terms of the rescaled variables defined in \eqref{IntVectorCoho} and \eqref{eq:CQarray1}. Thus the result will be in terms of  rescaled operator, in particular rescaled $\cH$, i.e. $\tilde{\cH}$, according to the \eqref{tildeQAlgebra}.   As was  explained after~\eqref{tildeQAlgebra}, this is how the scale dependence appears in the 1-loop result.  From now on, we will omit the tildes for ease of notation. One should note however that whenever determinant or index is written, it will mean that the operators and their eigenvalues are all in terms of the rescaled one.

First, we consider the $\cH=0 $ sectors. Then we can see that the determinants are, up to overall sign and some numerical pre-factor, 
\beqa\label{D10nondegenerate}
\det \CK_b \bigr|_{\cH=0}\= (\det D_{10})^2\,, \qquad \det \CK_f \bigr|_{\cH=0} = (\det D_{10})^2\,.\,
\eeqa
Note here that $D_{10}$ is non-degenerate as we are separating out all the zero modes and considering 1-loop determinant for non-zero modes of $\qeq\cV$ action. Any mode in  $\text{Ker}\left(D_{10}\right)$ will make $\qeq \cV$-action vanish and it should have been treated as the mode along localization saddle. Once we separate out all the zero modes correctly, then the $D_{10}$ should be non-degenerate on orthogonal space to the zero modes.\footnote{Conversely, we can utilize the $D_{10}$ operator to find all the localization saddle solution by solving the  kernel of $D_{10}$ \cite{Sen:2023dps}. }
Therefore,
the bosonic and fermionic contribution to the 1-loop determinant of $\cH=0$ modes cancel each other \cite{Gupta:2015gga}.  

Now, we look at $\cH\neq0$ sectors. 
From the expression in~\eqref{eq:Kb}, we can see that the kinetic operators satisfy
\be
\biggl( \ba{cc} 1 & ~0 \\
0 & - \cH \ea \biggr) \mathcal{K}_b = \mathcal{K}_f \biggl(\ba{cc} \cH & 0\\ 
0 & ~1  \ea \biggr)  \label{eq:KbKf}\,.
\ee
Taking the determinant of both sides of \eqref{eq:KbKf}, the ratio of determinants of $\mathcal{K}_f$ and $\mathcal{K}_b$  would reduce to a ratio of determinant over $\cH\,$. In what follows let us consider \ss2 and \ads2 case separately.  

\paragraph{ The case of \ss2:}
Here, we first note that, in the $\cH=0$ sector, the effect of the zero modes of ghost $c$ and anti-ghost~$\bar{c}$ in $\Psi$ and their superpartners in $\qeq\Psi$ are canceled by the constant multiplet $\Phi_0$ and $\qeq \Phi_0$ through the argument in \eqref{D10nondegenerate}. This is in fact expected as  the role of ghost of ghost multiplet is to  eliminate the zero modes of ghosts. Let us then denote by  $\Psi'$ and $\qeq\Psi'$ the space without the zero mode of ghosts and their superpartners respectively. Since the cohomology complex $(\Phi\,, \qeq \Phi\,, \Psi'\,, \qeq \Psi')$, by its structure, has one-to-one correspondence between bosonic and fermionic variables, we note that the size of functional space with $\cH\neq 0$ onto which the left and right hand side of~\eqref{eq:KbKf} act is the same. Therefore,  we can utilize the~\eqref{eq:KbKf} and reduce the 1-loop determinant as  
\beqa
 \label{eq:prelogNozm}
Z'^{ \qeq \cV}_{1\text{-loop}} &=& \sqrt{\frac{\det_{\qeq \Phi\,, \Psi'} \mathcal{K}_f}{\det_{\Phi , \qeq \Psi'} \mathcal{K}_b}}  =\sqrt{\frac{\text{det}_{\qeq \Psi^{\prime}}\cH}{\text{det}_{\qeq \Phi}\cH}} \= \sqrt{\frac{\text{det}_{\Psi^{\prime}}\cH}{\text{det}_{\Phi }\cH}} \,.
\eeqa
To arrive at the last equality, we have used the fact that $\cH$ and $\qeq$ commute. The reduced determinant~\eqref{eq:prelogNozm} can be obtained by calculating  the following quantity, 
\beqa \label{eq:TrNozm}
  \text{Tr}_{\Psi^{\prime}}{\rm e}^{  t \cH} -  \text{Tr}_{\Phi} {\rm e}^{ t \cH}&=& -\text{Tr}_{\Psi_{\text{zm}}}\mathbb{I}+ \text{Tr}_{\Psi}{\rm e}^{  t \cH} -  \text{Tr}_{\Phi} {\rm e}^{ t \cH}
  \\ \nn
 & = & -n^\Psi_{\text{zm}}  -\text{ind}(D_{10}) (t)\,,
\eeqa
where we have added and subtracted the trace over zero modes in $\Psi$, which are in our case constant modes of ghost and anti-ghost and thus the number of zero modes is  $n^\Psi_{\text{zm}}=-n^{c,\bar{c}}_{\text{zm}}=2\,$. Now, we can regard the traces over $\Psi$ and $\Phi$ in~\eqref{eq:TrNozm} are over complete basis. This is because we can freely add the contribution of trace over all the $\cH=0$ modes without changing the value  since those modes in $\Psi$ and $\Phi$ are uniquely paired by the map $D_{10}$ which is  non-degenerate as was argued in~\eqref{D10nondegenerate}.
In the second line of~\eqref{eq:TrNozm}, the difference between two traces is identified with the equivariant index of $D_{10}$ with respect to the $U(1)$ generated by $\cH$, which is defined as
\be
\text{ind}(D_{10}) (t):= \text{Tr}_{\text{Ker}(D_{10})}{\rm e}^{t \cH} - \text{Tr}_{\text{Coker}(D_{10})}{\rm e}^{t\cH} \,. \label{eq:indDeg1}
\ee 
This can be seen from the fact that only the kernel of $D_{10}$ in $\Phi$ and cokernel of $D_{10}$ in $\Psi$  are not paired and therefore their contribution remains. 

The index given in~\eqref{eq:indDeg1} admits an expansion in 
%Once we obtain the index, it can be expressed  as a series in 
terms of the eigenvalues of $\cH$, $\lambda_n$, and their degeneracies, $a(n)$, in the following way
\be 
{\rm ind}( D_{10})(t) = \sum_n a(n){\rm e}^{ \lambda_n t}. \label{eq:indDeg}
\ee
From this, we can read off the contribution from index to the 1-loop determinant. We  note that coefficient of the zero eigenvalue in this expansion accounts for the number of zero modes because this zero mode is added in \eqref{eq:TrNozm} to define index.  ( In general, the coefficient is the difference between number of bosonic and fermionic zero modes ). So, from the $t$-independent constant in~\eqref{eq:indDeg} we should obtain $-n^{\Psi}_{\text{zm}}= -2\,$.   To read off the contribution to the 1-loop from the zero mode effect, i.e. $-n^{\Psi}_{\text{zm}}= -2\,$, from the first term in~\eqref{eq:TrNozm}, 
we can use the regularization  in the following way
\begin{align} \label{tr2}
- \frac{1}{2}\int_{\bar\varepsilon }^{\infty} \frac{dt}{t}  (-  n^{\Psi}_{\text{zm}} )\Big{|}_{\text{reg}} \quad \text{contributes as} \quad \left(\frac{L}{L_0}\right)^{ \frac{1}{2}n^{\Psi}_{\text{zm}}}\,, 
\end{align}
where $\bar{\varepsilon} \equiv \varepsilon \left(\frac{L_0}{L}\right)$ is the UV cutoff.
Therefore, collecting \eqref{tr2} and by looking at the eigenvalue and corresponding degeneracy from the form in~\eqref{eq:indDeg} of the index,  we can calculate the 1-loop determinant which is given by
\be \label{eq:degeneracies}
 Z^{\prime \qeq \cV}_{1\text{-loop}}   \=\left(\frac{L}{L_0}\right)^{ \frac{1}{2}n^{\Psi}_{\text{zm}}}\prod_n \lambda_n^{-\frac{1}{2}a(n)}\,.
\ee

\paragraph{The case of \ads2:} Here, we begin by listing properties concerning the boundary zero modes. 
First,  $\qeq$ is nilpotent over the   1-form boundary zero modes.
 By applying the equivariant algebra~\eqref{EqAlgebra}, we find that 
\be\label{Nilpotent}
\qeq^2 A_{\mu}^{\text{bdry}} =\cL_{\xi} A^{\text{bdry}}_{\mu} + \partial_{\mu}\Lambda_0^{\text{bdry}}=\cL_{\xi} A^{\text{bdry}}_{\mu} + \partial_{\mu}\left(-\xi^{\nu}A_{\nu}^{\text{bdry}} \right) =\xi^{\nu} F^{\text{bdry}}_{\nu \mu} =0\,.
\ee
This nilpotency was also  reported in~\cite{Jeon:2018kec}, however the interpretation is now clarified. 
The reason for the nilpotency is that the boundary zero modes obey the equivariant algebra given in~\eqref{EqAlgebra} with new interpretation of the parameter $\Lambda_0$ explained in~\eqref{L0ads2}, whereas ~\cite{Jeon:2018kec} interpreted  the nilpotency as the boundary modes not obeying the algebra.

Second, for each 1-form boundary zero mode, there exists a fermionic superpartner in the normalizable fermionic basis. 
Although it is already suggested by its cohomological structure, one still needs to confirm its existence and, in particular, whether it belongs to a normalizable basis. Before proving it, let us first provide a plausibility argument for the existence of a $\qeq A^{\text{bdry}}$ with non zero norm.  To this end we first note that the operator $\qeq$ is not hermitian since the supersymmetry variation does not preserve the reality condition as was explained after~\eqref{eq:ReVec} (See also \cite{Benini:2012ui}). Therefore, even though $A^{\text{bdry}}$ satisfies $\qeq^2 A^{\text{bdry}}=0$, this  does not imply that $\qeq A^{\text{bdry}}=0\,$.  To explicitly prove it, it is enough to consider inner product between the basis of the boundary zero modes \eqref{nonnormalizablescalarsads2} and  twisted variables~\eqref{twistinglambda} constructed out of the delta function normalizable fermionic basis \eqref{EigenAdS2}  which are the corresponding fermionic bi-linear from the supersymmetry variation of 1-form field as in ~\eqref{eq:deltaA}.  One can show that the inner product gives
\begin{align} \label{eq:existence}
      \int \textrm{d}^2 x \sqrt{g} \, \bigl(\partial_a \Lambda_{\text{bdry}}^{(\ell)}\bigr)^\ast \left(\bar{\epsilon} \gamma^a \psi_{k,\lambda} + \epsilon \gamma^a \bar{\psi}_{k,\lambda}\right) \sim \delta_{k,|\ell|-s}\delta(\lambda) \, , \quad s= 0\, \, \text{or} \, \, 1\,,
\end{align}
 where the $\psi_{k,\lambda} $ is collective notation of the fermionic eigenfunctions of the Dirac operator on \ads2 \eqref{EigenAdS2}. In the result, the Dirac delta function selects the fermionic eigenfunction with $ \lambda=0$ eigenvalue. Since the spectral density for this fermionic basis with $\lambda=0$ is non-zero as can be seen appendix \eqref{Fermionatorigin}, the superpartners of 1-form boundary zero modes are in the physical normalizable spectrum. 

Lastly, the superpartner $\qeq A^{\text{bdry}}$ is also a zero mode of $\qeq\cV$ action as well as the kinetic term of the physical action. Since a zero mode $A^{\text{bdry}}$ makes the fermionic quantity $\cV$ vanishing,  not only the bosonic part of $\qeq \cV$ action is zero on the  $A^{\text{bdry}}$, but also the fermionic part of $\qeq \cV$ action is zero on the superpartner $\qeq A^{\text{bdry}}$. Since those modes correspond to $\lambda=0$, their eigenvalue for the Dirac operator is zero. This is a specific feature that we encounter for the case of pure \ads2. When we study, for example, \ads2$\times$\ss2, the $\qeq A^{\text{bdry}}$ are not zero modes because the Dirac operator along \ss2 does not have zero eigenvalue. If they are zero modes of physical action, the path integral over these zero modes will yield a vanishing result. However, 
this will yield regularized result in the spirit that the gravitino zero modes would do in the quantum entropy function \cite{Banerjee:2011jp}.

As worked out in \cite{Jeon:2018kec}, we treat those fermion zero modes by  promoting the zero modes to a non-zero modes by adding an extra $\qeq$-exact term to the deformation action $\qeq \cV$.   
We choose this term to be
 \beqa
 \qeq \cV^{\qeq A^{\text{bdry}}}&=&  \int \textrm{d}^2 x \sqrt{g} \,\qeq\left( A^{\text{bdry}} ~  \cL_{\xi}(\qeq A^{\text{bdry}})\right)  \label{eq:ActionBM}
 \\
 &=&  \int \textrm{d}^2 x \sqrt{g} \,(\qeq A^{\text{bdry}} )~  \cL_{\xi} (\qeq A^{\text{bdry}})\nn
 \, .
 \eeqa
   Since  $\qeq^2 A^{\text{bdry}}  =0$, there is no bosonic contribution in \eqref{eq:ActionBM}, which means that only the fermionic mode $\qeq A^{\text{bdry}}$ is promoted to be non-zero mode of the total $\qeq \cV$ action. 
 This  term localizes the superpartner of the boundary zero modes. In our pure \ads2 case, this is regarded as a supersymmetric regulator to obtain the finite result out of the zero modes of fermions. 
  With all this data in hand, we go back and analyze the determinant in \eqref{eq:ZintLoc}.

%%%%%%%%%%%%%%%%%%%%%%%%%%
In the 1-loop determinant, the integration over localization saddles, $\Lambda_0$ and $A^{\text{bdry}}$, are already factored out as they are zero modes.  Now, since the superpartners of the boundary zero modes $\qeq A^{\text{bdry}}$ are no longer zero modes due to the lifted action \eqref{eq:ActionBM}, we can separately compute their 1-loop effect from this action. Let us then denote $\Phi'$ and $\qeq\Phi'$ as the space without the boundary zero mode  and their superpartners respectively. Among the cohomological complex $(\Phi'\,, \qeq \Phi'\,, \Psi\,, \qeq \Psi)$, the 1-loop effect of the $\cH=0$ sectors is canceled  as is argued in \eqref{D10nondegenerate}, and the effect of $\cH\neq 0$ sectors can be reduced by utilizing the~\eqref{eq:KbKf} as was done in \eqref{eq:prelogNozm}. In this way, we implement the reduction process of  the 1-loop determinant  as follows:
\beqa\label{eq:prelogBdry}
 Z^{\prime \qeq \cV}_{1\text{-loop}} &=& 
 \sqrt{\text{det}_{ \qeq A^{\text{bdry}}  }\cL_\xi}\sqrt{\frac{\det_{\qeq \Phi'\,, \Psi} \mathcal{K}_f}{\det_{\Phi' , \qeq \Psi} \mathcal{K}_b}}   \=\sqrt{\text{det}_{ %\qeq
\qeq A^{\text{bdry}}  }\cL_\xi}\sqrt{\frac{\text{det}_{\Psi}\cH}{\text{det}_{\Phi' }\cH}}\, . 
\eeqa
Now, the reduced determinant in~\eqref{eq:prelogBdry} can be obtained by calculating  the following quantity 
\beqa \label{eq:Trnzmbdry}
\text{Tr}_{ \qeq A^{\text{bdry}}}{\rm e}^{ t\, \cL_\xi} + \text{Tr}_{\Psi}{\rm e}^{  t \cH} -  \text{Tr}_{\Phi'} {\rm e}^{ t \cH}& = & \text{Tr}_{  \qeq A^{\text{bdry}}}{\rm e}^{ t\, \cL_\xi }+ \text{Tr}_{  A^{\text{bdry}}} \mathbb{I} +\text{Tr}_{\Psi}{\rm e}^{  t \cH} -  \text{Tr}_{\Phi} {\rm e}^{ t \cH}\nn\\ 
& = & \text{Tr}_{\qeq  A^{\text{bdry}}}{\rm e}^{ t\, \cL_\xi } +n^{\Phi}_{\text{zm}} -\text{ind}(D_{10}) (t)\,.
\eeqa
Here, we have added and subtracted the trace over zero modes in $\Phi$, which are the 1-form  boundary zero modes whose number is given as $n^{\Phi}_{\text{zm}} =n^{\text{bdry}}_{\text{zm}} = -1$ ( see \eqref{vectorbdymode}). Thus the traces over $\Psi$ and $\Phi$ can now be regarded as over complete basis and the difference of two is identified with the equivariant index. Expansion of the index  in the form of \eqref{eq:indDeg} reproduces the local contribution of 1-loop determinant, we note again that the $t$-independent constant term of the expansion should obtain $n^\Phi_{\text{zm}}= -1$.  The first two terms in \eqref{eq:Trnzmbdry} provide the global zero mode contribution to the 1-loop determinant. From the constant term $n^{\Phi}_{\text{zm}}$, we read off the contribution $(L/L_0)^{-\frac{1}{2}n^{\Phi}_{\text{zm}}}$ in the same way as in ~\eqref{tr2}. In the first term, $\,\text{Tr}_{ \qeq A^{\text{bdry}}}{\rm e}^{ t\, \cL_\xi }$, which comes from the superpartner of the boundary zero modes,  we know the  eigenvalue of the operator $\cL_\xi$ (actually rescaled operator $\wt{\cL}_\xi \equiv (L_0/L) \cL_\xi$  through the rescaling of variables \eqref{IntVectorCoho}) for each mode $\qeq A^{(\ell)}_{\text{bdry}}$ that is $\ell L_0/L$. Thus, we obtain their contribution as
\begin{align} \label{eq:zetafunczm}
    \prod_{\ell \neq 0\,,\ell \in \mathbb{Z}} \left(\frac{L_0}{L}\ell \right)^{\frac{1}{2}}\Bigg{|}_{\text{reg}} = \left(\frac{L}{L_0}\right)^{-\frac{1}{2}n_{\text{zm}}^{\Phi}}\, ,
\end{align}
where we have used Zeta function regularization \eqref{eq:regen} to define the regularized number of the boundary zero modes $n_{\text{zm}}^{\Phi}=-1$. This definition of number of the zero modes is also justified using heat kernel method in~\cite{Jeon:2018kec}.
Therefore, we can put all the results together, including the index contribution and the global contribution, to finally yield the following $1$-loop determinant 
\begin{align}
\begin{split} \label{eq:1loopgen}
  Z^{\prime \qeq \cV}_{1\text{-loop}}  & =%\prod_{ A^{\text{bdry}}} 
  \left(\frac{L}{L_0}\right)^{- n^{\Phi}_{\text{zm}}}\prod_n \lambda_n^{-\frac{1}{2}a(n)} \, .
   \end{split}
\end{align}

In summary, the 1-loop determinant is reduced to the global zero mode contribution and local index contribution as in~\eqref{eq:degeneracies} for \ss2 and in~\eqref{eq:1loopgen} for \ads2. The zero mode is identified as constant modes of ghost and anti-ghost for \ss2 whose number is $n^\Psi_{\text{zm}}= 2$, and  as the boundary 1-form zero modes for \ads2 whose number is $n^\Phi_{\text{zm}}= -1$. Now, we are left with evaluating  the index of the operator $D_{10}$ and finding its expansion in the form of~\eqref{eq:indDeg}, which we will do in section~\ref{subsec:1loop}. As a  preparation,  we verify that the operator $D_{10}$ is transversally elliptic in what follows.

\subsubsection{Transversally elliptic $D_{10}$ }\label{subsubsec:TE}
In order to evaluate the index using  fixed point formula, we need to show that $D_{10}$ is transversally elliptic. For this purpose, we need to define the Symbol of $D_{10}$, which is obtained by replacing $\partial_{a} \rightarrow \i p_a$  in the highest derivative terms in $\Psi D_{10} \Phi$.  From \eqref{eq:QVeqvm} and \eqref{eq:QVcmTW} we see that we need to rewrite the conjugated variations of elementary fields in terms of elementary fields and then identify higher derivatives terms. 

\subsubsection*{The vector multiplet}

Let us consider the conjugation of the variations of elementary twisted fields. 

\beqa \label{QPsiDagger}
(\qeq c)^\dagger &=& - \qeq c + 2 \xi^a A_a + 2\i \bar{\epsilon}\gamma_3\epsilon \rho ~+~ 2~\text{Re}( \Lambda_0), \,\,\,\,\,\quad\quad (\qeq \bar c)^{\dagger} =b, \,\nonumber
\\
(\qeq \lambda_a )^\dagger 
&=& \cL_\xi A_a   + \partial_a ( \qeq c - 2 \xi^b A_b - 2\i \bar{\epsilon}\gamma_3 \epsilon \rho +2~ \i~\text{Im}(\Lambda_0)) ,
\\ \nonumber
( \qeq \lambda_3)^\dagger &=& \cL_\xi \rho,
\\ \nonumber
( \qeq \lambda_4)^\dagger &=&     \qeq \lambda_4 + 2 \Big[ \i \bar\epsilon \epsilon ( \i G\rho)  + \i \bar\epsilon \gamma_3 \epsilon ( \i F_{12} + H\rho) - \i \bar\epsilon \gamma_3 \gamma^a \epsilon D_a \rho \Big],
\eeqa
where we used \eqref{eq:QLamb4}.
Replacing $\partial_a \rightarrow \i p_a$, the highest derivative terms in $\Psi D_{10} \Phi$ are of the form:
\be
\frac{1}{2 (\i \bar\epsilon \epsilon)}  \begin{pmatrix} \lambda_4 \\ c \\ \bar{c}  \end{pmatrix}^T\!\! 
\begin{pmatrix} 
2(\i \bar\epsilon \gamma_3 \epsilon) p_2 & - 2(\i \bar\epsilon \gamma_3 \epsilon) p_1 & 2 ( \bar\epsilon \gamma_3 \gamma^a \epsilon) p_a \\
2 p^a p_a \xi^1 \!- \!p_1 \xi^a p_a  ~&~ 2 p^a p_a \xi^2 \!-\! p_2 \xi^a p_a ~&~ 2(\i\bar\epsilon \gamma_3 \epsilon) p^a p_a \\
  2(\i \bar\epsilon \epsilon) \i  p_1 &   2 (\i \bar\epsilon \epsilon) \i  p_2 & 0 
  \end{pmatrix}
  \begin{pmatrix}  A_1  \\ A_2 \\ \rho \end{pmatrix} \label{eq:matSymb}\,.
\ee
The matrix appearing in \eqref{eq:matSymb}, including the $\frac{1}{2\i \bar \epsilon \epsilon}$ factor, is what we call the symbol of $D_{10}$ which we denote as $\boldsymbol{\sigma} (D_{10})$  and has the following determinant on \ss2
 \beqa 
\text{det}\left[\boldsymbol{\sigma}(D_{10}) \right]\bigg|_{\text{S}^2} = -\, \dfrac{\i}{2} \left(p^2_{1} + ~p_{2}^2  \right) \left(p^2_{1} + \cos^2 \psi ~p_{2}^2  \right) ,
 \eeqa
 whereas on \ads2, it is
 \beqa \label{eq:Symbolads2}
\text{det}\left[\boldsymbol{\sigma}(D_{10}) \right]\bigg|_{\text{AdS}_2}=-\, \dfrac{\i}{2}   \left(p^2_{1} + p_{2}^2  \right) \left(p^2_{1} + ~ \frac{1}{\cosh^2 \eta}~p_{2}^2  \right) .
 \eeqa
 We note that at $\psi =\frac{\pi}{2}$ on \ss2 or $\eta = \infty$ on AdS$_2$, $\text{det}\left[\boldsymbol{\sigma}(D_{10}) \right] =0$ for the non-zero  value of~$p_2$ as long as $p_1=0$, i.e.  the symbol $\boldsymbol{\sigma}(D_{10})$ is not invertible for non-zero $p_a$. Therefore, the operator $D_{10}$ is not elliptic. However, if we restrict the momentum to be orthogonal to   the Killing vector $\xi^{\mu}$, viz., $p_2 \equiv 0$,  then the non-invertibility of the symbol implies $p_1=0$ and vice versa. Therefore, $D_{10}$ for the case of the vector multiplet is transversally elliptic with respect to the symmetry~$\cL_\xi$. 
\subsubsection*{The chiral multiplet}
Let us now analyze the case of the chiral multiplet. Then, from \eqref{eq:CQarray}, we write the complex conjugation of the variation of elementary fields as
\beqa
\left( \qeq \left(\epsilon \psi \right)\right) ^\dagger  = \qeq\left( \bar{\epsilon }\bar{\psi} \right) + 2 \Bigg[ \left(\P+\dfrac{r}{2} G\right)\bar{\epsilon}\gamma_3 \bar{\epsilon}  \bar{\phi}  - \i \bar{\epsilon}\gamma^a \bar{\epsilon} \left( -i p_a \bar{\phi }\right)  \Bigg] ,\\
\left( \qeq \left(\bar{\epsilon} \bar{\psi} \right)\right) ^\dagger = \qeq\left( \eps  \psi\right)  - 2 \Bigg[  \left(\P+\dfrac{r}{2} G\right)\epsilon\gamma_3 \epsilon  \phi  - \i \epsilon\gamma^a \epsilon \left(i p_a \phi  \right)  \Bigg] ,
\eeqa
where, we have used 
\beqa
\partial_a \phi = \i p_a \phi\, , \qquad  \partial_a \bar{\phi} = - \i p_a \bar{\phi}\, ~.
\eeqa 
The highest derivative terms in $\Psi D_{10} \Phi$ are
\be
\frac{1}{\bar{\i\epsilon} \epsilon} \bigl( \ba{cc} \eps \psi \,\quad & \bar{\eps}\bar{\psi} \ea \bigr) 
\biggl( \ba{cc}
0 & - 2 \i\bar{\eps} \gamma^a \bar{\eps}p_a \\ 
2 \i\eps \gamma^a \eps p_a & 0 
\ea \biggr) \biggl( \ba{cc} \phi \\ \bar{\phi} \ea \biggr) ,
\ee
hence:
\beqa
\det \left[\boldsymbol{\sigma}\left(D_{10}\right)\right] && = \frac{4}{(\i\bar{\epsilon} \epsilon)^2}\left(\i \bar{\eps} \gamma^a \bar{\eps}\right) \bigl(\i \eps \gamma^b \eps \bigr)  p_a p_b .
\eeqa
We then have:
\be
\det \left[\boldsymbol{\sigma}\left(D_{10}\right)\right]\Big{|}_{\text{S}^2}  =\,-4 \left(  p^2_1 + \cos ^2 \psi~ p^2_2\right)\label{eq:CsymS},
\ee
on S$^2$, whereas on AdS$_2$ we have:
\be
\det \left[\boldsymbol{\sigma}\left(D_{10}\right)\right]\Big{|}_{\text{AdS}_2} =\, - 4  \left( ~ p^2_1 +\frac{1}{\cosh ^2 \eta} ~p^2_2\right). \label{eq:CsymAdS}
\ee
From \eqref{eq:CsymS} and \eqref{eq:CsymAdS} we see that, following a reasoning completely parallel to the one presented for the vector multiplet, the operator $D_{10}$ associated to the chiral multiplet is transversally elliptic.
%%%%%%%%%%%%%%%%%%%%%%%%%%%%%%%%%%%%%%%%%%%%%%%%%%%%%%%%%%%%%%%%%%%%%%%%%%%%%%%%%%%%%%%%%%%%%%%%%%%%%%%%%%%%%%%%%%%%%%%%%%%%%%%%%%%%
\subsection{Evaluation of $1$-loop determinant } \label{subsec:1loop}

\begin{table}
\begin{center}
\begin{tabular}{|c| c | c|| c | c|}
\hline 
-& $\Phi$ & $\mathcal{H}$-charge  &  $\Psi$ & $\cH$-charge \\
\hline 
 & $A_z$ & $\i$ &      $\lambda_4$ & 0\\
Vector &  $A_{\bar{z}}$ & $-\i$  &    $c$ & 0   \\ 
 & $\P  $ & 0  &   $\bar{c}$ & 0  \\
\hline 
\hline 
 \vspace{-2mm} & $\phi$ & $ \i  \Lambda_0 +  \frac{\i}{2} \r  $ &   $\epsilon \psi$ & $\i  \Lambda_0+\frac{\i}{2} (\r-2) $ \\
  \vspace{-2mm} Chiral &  &  & & \\
 & $\bar{\phi}$ & $-\i  \Lambda_0 - \frac{\i}{2}\r $ &   $\bar{\epsilon} \bar{\psi}$ & $ -\i  \Lambda_0 - \frac{\i}{2}(\r-2)$ \\
\hline 
\end{tabular}
\end{center}
\caption{The table shows the fixed point values of the $\cH$-charge of the elementary fields associated with the vector and chiral multiplet. Here $\Lambda_0$ is respectively given by \eqref{eq:L0S2} on \ss2 and \eqref{eq:L0Ads2} evaluated at the center of \ads2. }
\label{tab2}
\end{table}
Recall that the 1-loop determinant is given  in the form \eqref{eq:degeneracies} for \ss2 and \eqref{eq:1loopgen} for \ads2, where the global zero mode contributions were identified. 
We are now in conditions to complete the 1-loop by evaluating the local contribution through the index~\eqref{eq:indDeg1}. After obtaining the 1-loop  for each multiplet for \ss2 and \ads2,
we will present the full result of the partition function by combining the global and local contributions.

To evaluate the index, we apply  the Atiyah-Bott fixed point formula \cite{atiyah1966lefschetz,10.2307/1970694,10.2307/1970721}  given by
\be
\text{ind}\left(D_{10}\right)(t) = \sum_{\left\{x | f(x) =x \right\}} \frac{\text{Tr}_{\Phi}{\rm e}^{ t \cH} - \text{Tr}_{\Psi} {\rm e}^{ t \cH}}{\text{det} \left(1 - \frac{\partial f(x)}{\partial x}\right)} \, . \label{eq:fpform}
\ee
 This formula reduces the trace of the operator ${\rm e}^{ t \cH}$ to a sum over the trace evaluated only at the fixed points of the operator $\cH: x  \mapsto f(x) = x$. On \ss2 there are two fixed points, viz., the north and south pole, whereas on \ads2 there is only one fixed point viz., the center. Although we do not provide a rigorous proof of  the validity of \eqref{eq:fpform} on non-compact spaces, we have tested it in appendix \ref{app:Atiyah} using a simple example of de-Rham cohomology. With this in mind, we apply \eqref{eq:fpform} on both \ads2 and \ss2.  After expanding the result of the index we read off the 1-loop contribution, where we use Zeta function regularization that allows us to keep the dependence on the size of the manifold.

For the case of chiral multiplet, the fact that $D_{10}$ is transversally elliptic implies an ambiguity in expanding  the index with respect to the equivariant parameter.  We will devise a systematic way of eliminating such an ambiguity.

\subsubsection{ S$^2$}  \label{subsubsec:S2}
\subsubsection*{Contribution from the vector multiplet}
In order to evaluate the index using the fixed point formula \eqref{eq:fpform}, we first define the equivariant parameter $q\equiv {\rm e}^{ \tfrac{L_0}{L} t }$. Then the action of $\cH$ on the complex coordinate is given by $z \mapsto f(z)=q^{\i} z , \quad  \bar z \mapsto f(\bar z) = q^{- \i} \bar z$, from which we can write down the denominator of~\eqref{eq:fpform} as $\left(1- q^\i\right)\left(1- q^{-\i}\right)$. To obtain the numerator we read off the $\cH$-charges of $\Phi$ and $\Psi$ at the fixed points, the north pole and the south pole of \ss2, which we summarize in table \ref{tab2}. Adding the two fixed point contributions we have
\begin{align}\label{eq:indD10s2}
\begin{split}
\text{ind} \left(D_{10}\right)(t) 
& =2 \frac{1 +q^{\i}+ q^{-\i} -(1+1+1)}{(1-q^{\i})(1- q^{-\i})}  \=-2\,.
\end{split}
\end{align} 
Note that the $t$-independent term in the expansion of the index \eqref{eq:indDeg} is $a(0) = -2 $ which is indeed expected as explained after \eqref{eq:indDeg}.  It is because, to define the index  in \eqref{eq:TrNozm},
 we have added the contribution of the zero modes, $- n^{\Psi}_{\text{zm}}$.
 The contribution from $a(0)$ to the 1-loop determinant can be obtained using the same regularization as in~\eqref{tr2}. Now, combining with the contribution from zero modes \eqref{eq:degeneracies}, we obtain
\begin{align} \label{zvms2}
Z^{\prime\,\text{v.m.}}_{1\text{-loop}} & \= \left(\frac{L}{L_0}\right)^{\frac{1}{2}n_{\text{zm}}^\Psi  + \frac{1}{2}a(0) } = \left(\frac{L}{L_0}\right)^{1-1} = \left(\frac{L}{L_0}\right)^{0}\,. 
\end{align}

%%%%%%%%%%%%%%%%%%%%%%%%%%%%%%%%%%%%%%%%%%%%%%%%%%%%%%%%%%

\subsubsection*{Contribution from the chiral multiplet}

For the chiral multiplet on S$^2$ there are no zero modes to deal with. Therefore we only have to  evaluate the index  applying the fixed point formula \eqref{eq:fpform}. Using the charge assignment given in table \ref{tab2}
we have 
\begin{align}
\begin{split}\label{indChiralS2}
\text{ind} \left(D_{10}\right)(t) 
& = \biggl( \frac{q^{\i \left(\frac{\r}{2}+\Lambda_0\right)}}{1-q^{\i}} + \frac{q^{-\i \left(\frac{\r}{2}+\Lambda_0\right)}}{1-q^{-\i }} \biggr) \Biggr{|}_{\text{NP}}+  \biggl( \frac{q^{\i \left(\frac{\r}{2}+\Lambda_0\right)}}{1-q^{\i}} + \frac{q^{-\i \left(\frac{\r}{2}+\Lambda_0\right)}}{1-q^{-\i }} \biggr)\Biggr{|}_{\text{SP}}.
\end{split}
\end{align}
To extract the eigenvalues of $\cH$ and their degeneracies,  we have to expand the above in terms of the equivariant parameter~$q$ and express it in the form of \eqref{eq:indDeg}.  However, for each fixed point, there is the ambiguity of whether to expand $\text{ind}\left(D_{10}\right)(t)$ in terms $q^{\i}$ or $q^{-\i}$ for each term of \eqref{indChiralS2}.  The ambiguity has been resolved in  \cite{Hama:2011ea, Alday:2013lba, Closset:2013sxa} and the prescription is that we should expand in $q^{\i}$ on the north pole and in $q^{-\i}$ in the south pole (equivalently, $q^{-\i}$ in the north and $q^{\i}$ in the south pole). 
We then expand as follows
\begin{align} \label{NP}
\begin{split}
 \biggl( \frac{q^{\i \left(\frac{\r}{2}+\Lambda_0\right)}}{1-q^{\i}} + \frac{q^{-\i \left(\frac{\r}{2}+\Lambda_0\right)}}{1-q^{-\i }} \biggr)\Biggr{|}_{\text{NP}} & =  q^{\i \left(\frac{\r}{2}+\Lambda^{\text{NP}}_{0}\right)}\sum_{n=0}^{\infty} q^{ \i n }-q^{-\ \i\left(\frac{\r}{2}+\Lambda^{\text{NP}}_{0}\right)} \sum_{n=1}^{\infty} q^{ \i n }\, .
\end{split}
\end{align}
Analogously for the south pole where we replace $\Lambda^{\text{NP}}_{0} \rightarrow \Lambda^{\text{SP}}_{0}$ and expand in terms of $q^{-1}$, i.e., 
\begin{align}
\begin{split} \label{SP}
 \biggl( \frac{q^{\i \left(\frac{\r}{2}+\Lambda_0\right)}}{1-q^{\i}} + \frac{q^{-\i \left(\frac{\r}{2}+\Lambda_0\right)}}{1-q^{-\i }} \biggr)\Biggr{|}_{\text{SP}} & =  -q^{\i \left(\frac{\r}{2}+\Lambda^{\text{SP}}_{0}\right)}\sum_{n=1}^{\infty} q^{-\i n }+q^{-\i \left(\frac{\r}{2}+\Lambda^{\text{SP}}_{0}\right)} \sum_{n=0}^{\infty} q^{-\i n }\, .
\end{split}
\end{align}
We note that in the vanishing $R$-charge and gauge coupling limit, the constant parts of \eqref{NP} and \eqref{SP} give the contribution of zero modes $a(0) = 1+1$ that accounts for the constant mode of the scalar field appearing in this limit. This is true only for the correct expansion of the index. Hence we will use this as a criteria to resolve the ambiguity in the context of~\ads2. 

From \eqref{NP} and \eqref{SP} we can read off the 1-loop determinant contribution of the chiral multiplet
\be \label{eq:infprod}
Z^{\text{c.m.}}_{1\text{-loop}}\=  \frac{  \prod_{n=1}^\infty\sqrt{L_0( n-\left(\frac{\r}{2}+\Lambda^{\text{NP}}_{0}\right))/(- \i L)  }}{\prod_{n=0}^\infty\sqrt{L_0( n+\left(\frac{\r}{2}+\Lambda^{\text{NP}}_{0}\right))/(- \i L)}}  \frac{  \prod_{n=1}^\infty\sqrt{L_0( n-\left(\frac{\r}{2}+\Lambda^{\text{SP}}_{0}\right))/(\i L)  }}{\prod_{n=0}^\infty\sqrt{L_0( n+\left(\frac{\r}{2}+\Lambda^{\text{SP}}_{0}\right))/( \i L)}}.
\ee
At this point we only need to specify the value of the gauge parameter $\Lambda_0$ at the north and south pole respectively. This is given in \eqref{eq:L0S2}. 
The infinite products in \eqref{eq:infprod} require to be regularized. To keep track of the scale dependence, we use a regularization scheme that has been previously prescribed in \cite{quine1993zeta} and whose main idea we sketch in appendix~\ref{subsubsec:Reg}. 
Using \eqref{eq:regen}, we have
\begin{align}
    Z^{\text{c.m.}}_{1-\text{loop}} = (-1)^{\tfrac{\mathfrak{m}}{2}}\left(\frac{L}{L_0}\right)^{1-r-(\Lambda_0^{\textrm{NP}} + \Lambda_0^{\textrm{SP}})}\Bigg(\dfrac{\Gamma (\tfrac{r}{2} + \Lambda^{\textrm{NP}}_0)\Gamma (\tfrac{r}{2} + \Lambda^{\textrm{SP}}_0)}{\Gamma (1-\tfrac{r}{2} - \Lambda^{\textrm{NP}}_0)\Gamma (1-\tfrac{r}{2} - \Lambda^{\textrm{SP}}_0)}\Bigg)^{\tfrac{1}{2}}\, .
\end{align} 
As expected, the above expression shows a manifest symmetry between the interchange of $\Lambda^{\text{NP}}_0 \Longleftrightarrow \Lambda^{\text{SP}}_0 $. We can simplify the above further using Euler's reflection formula\footnote{For $z \in \mathbb{C}$ the Euler's reflection formula is  $
       \Gamma(z)\Gamma(1-z) = \frac{\pi}{\sin ( \pi z)}.$} and obtain
\begin{align} \label{eq:1loopcmS2}
    Z^{\text{c.m.}}_{1-\text{loop}} = \left(\frac{L}{L_0}\right)^{1-r-2 \i \sigma_0}\frac{\Gamma\left( \frac{r}{2} + \i \S_0 L - \frac{\mathbf{m}}{2 }\right)}{\Gamma\left(1- \frac{r}{2} - \i \S_0 L - \frac{\mathbf{m}}{2 }\right)}\, , 
\end{align} 
The ratio of the Gamma functions of the above result matches the result stated in \cite{Benini:2012ui}, as expected.
\subsection*{Result of full partition function on S$^2$}
We are now in conditions to put together all the ingredients we have collected so far and state that
\begin{align} \label{eq:ZS2reg0}
\begin{split}
    Z_{\text{\ss2}} & = \sum_{\mathbf{m}=- \infty}^{\infty}\int \frac{{\rm d}(\S_0 L)}{\left(\gym L_0 \right)^2} \exp\left(- S^{\text{S}^2}_{\text{tot}}\Bigr|_{\text{loc.}}\right)Z^{\prime \,\text{v.m.}}_{1-\text{loop}} Z^{\text{c.m.}}_{1-\text{loop}} ,
    \end{split}
\end{align}
where  we have used the measure for \ss2 given in \eqref{eq:IntMeasure}. The one loop contributions $Z^{\prime \,\text{v.m.}}_{1-\text{loop}}$ and $ Z^{\text{c.m.}}_{1-\text{loop}}$  are given in \eqref{zvms2} and \eqref{eq:1loopcmS2} respectively. Using the saddle value of the action given in \eqref{eq:SclaS2}, then  explicit evaluation of \eqref{eq:ZS2reg0} yields
\begin{empheq}[box=\fbox]{equation}
\begin{split} \label{s2result}
\vspace{1mm} & \\ 
   Z_{\text{\ss2}} & =\! \left(\frac{L}{L_0}\right)^{1-r}\!\!\!\!\!\!\sum_{\mathbf{m}=-\infty}^{\infty} \!{\rm e}^{-\i \vartheta \mathbf{m}}\!\int \frac{{\rm d}(\S_0 L)}{\left(\gym L_0\right)^2}\!\left(\frac{L}{L_0}\right)^{-2\i \S_0L}\!\!\!\!{\rm e}^{ 4 \pi \i \xi \S_0 L  }\frac{\Gamma\left( \frac{r}{2} + \i \S_0 L - \frac{\mathbf{m}}{2 }\right)}{\Gamma\left(1- \frac{r}{2} - \i \S_0 L - \frac{\mathbf{m}}{2 }\right)}\,,\\
   \vspace{1mm} &
  \end{split}
\end{empheq}
 where we can absorb the $\S_0 L$ into the renormalized FI parameter as $\xi_{\text{ren}}= \xi - \frac{1}{2 \pi} \log\left(\frac{L}{L_0}\right)$. 
 The super-renormalizability of $\gym$ implies that it is not a running coupling. Hence the $\left(\gym L_0\right)^{-2}$ contribution in \eqref{s2result} is a purely numerical factor that will be irrelevant and we only preserve it to be able to emphasize our use of Fujikawa's prescription to deal with the measure. We see that \eqref{s2result} reproduces the expected result obtained in \cite{Benini:2012ui, Doroud:2012xw} and includes the scale dependent pre-factor that consists of $-1 +(1-r)$ from the scaling anomaly \cite{Zamolodchikov:1986gt, Gerchkovitz:2014gta} and $1$ from the zero modes of ghost fields. 

%%%%%%%%%%%%%%%%%%%%%%%%%%%%%%%%%%%%%%%%%%%%
%%%%%%%%%%%%%%%%%%%%%%%%%%%%%%%%%%%%%%%%%%

\subsubsection{AdS$_2$} \label{sec:AdS2calc}

\subsubsection*{Contribution from the vector multiplet}

To obtain the numerator of \eqref{eq:fpform}, we read off the $\cH$-charge of $\Phi$ and $\Psi$ in the chiral multiplet given in table~\ref{tab2} and apply the fixed point formula. Thus we obtain
\begin{align}\label{eq:indAdS}
\begin{split}
\text{ind} \left(D_{10}\right)(t)
& = \frac{1 +q^{\i}+ q^{-\i } -(1+1+1)}{(1-q^{\i})(1- q^{-\i})} \=- 1.
\end{split}
\end{align}
This constant is expected as explained after \eqref{eq:Trnzmbdry}, where to define the index the number of boundary 1-form zero modes $n^{\Phi}_{\text{zm}}= -1$ was added . 
Now, combining with the constant contribution $a(0)$ from zero modes in \eqref{eq:1loopgen}, we obtain 
\begin{align}
Z^{\prime \,\text{v.m.}}_{1\text{-loop}}= \left(\frac{L}{L_0}\right)^{-n^\Phi_{\text{zm}} +\frac{1}{2}a(0) } \=\left(\frac{L}{L_0}\right)^{1 -\frac{1}{2} }= \left(\frac{L}{L_0}\right)^{\frac{1}{2}}. \label{eq:Lhalf}
\end{align}

%%%%%%%%%%%%%%%%%%%%%%%%%%%%%%%%%%%%%%%%%%%%%%%%%%%%%%%%%%%%%%%%%%%%%%%%%%%%%%%%%%%%%%%%%%%%%%%%%%%%%%%%%%%%%
\subsubsection*{Contribution from the chiral multiplet}

For the chiral multiplet on AdS$_2$ there are no zero modes to deal with. Therefore we only have to evaluate the index using the fixed point formula \eqref{eq:fpform}.
Recalling that there is only one fixed point at the center of \ads2, and reading off the $\cH$-charges of $\Phi$ and $\Psi$ of the chiral multiplet from table \ref{tab2}, we have
\begin{align}
\begin{split}\label{eq:indAdSch}
\text{ind} \left(D_{10}\right)(t)
& =   \frac{q^{-\i \left(\frac{\r}{2}+\Lambda_0\right)}+q^{\i \left(\frac{\r}{2}+\Lambda_0\right)}-\left(q^{-\i \left( \left(\frac{\r}{2}+\Lambda_0-1\right)\right)} +q^{\i \left( \left(\frac{\r}{2}+\Lambda_0\right)-1\right)}\right)}{(1-q^{\i})(1- q^{- \i})} \\
& = 
 \left(\frac{q^{\i \left(\frac{\r}{2}+\Lambda_0\right)}}{1-q^{\i }} + \frac{q^{-\i \left(\frac{\r}{2}+\Lambda_0\right)}}{1-q^{-\i }}\right)\, . 
\end{split}
\end{align}
 Similar to the \ss2 case, to extract the eigenvalues of $\cH$ and their degeneracies,  we have to expand the above in terms of the equivariant parameter~$q$ and express it in the form of~\eqref{eq:indDeg}.  
 Since $D_{10}$ is transversally elliptic, we are faced with an ambiguity in how to express $\text{ind}\left(D_{10}\right)(t)$ as a power series. In practice, we have two terms in~\eqref{eq:indAdSch} each of which can be expanded in powers of $q^{\i}$ or $q^{- \i}$ and we need a valid criteria to pick the appropriate combination.\footnote{In the context of localization on AdS$_3$ in \cite{Assel:2016pgi} the resolution of the ambiguity  was tackled with a different approach related to another method used to evaluate the index itself, which is called the ``unpaired eigenmodes'' method.} 
 In what follows, let us explore all possible ways of expanding the index:
\begin{itemize}
    \item[\textbf{a)}]Expanding the first term in powers of $q^{-\i}$ and the second in powers of $q^{\i}$. 
    
    \item[\textbf{b)}] Expanding the first term in powers of $q^{\i}$ and the second in powers of $q^{-\i}$.
    
    \item[\textbf{c)}] Expanding both terms in powers of $q^{\i}$. 
    
    \item[\textbf{d)}]  Expanding both terms in powers of $q^{-\i}$.
\end{itemize}
\paragraph{a)} This option gives us
 \begin{align} \label{eq:optionA}
  \text{ind} \left(D_{10}\right)(t)    & =  -q^{\i \left(\frac{\r}{2}+\Lambda_0\right)}\sum_{n=1}^{\infty} q^{-\i n }-q^{-\i \left(\frac{\r}{2}+\Lambda_0\right)} \sum_{n=1}^{\infty} q^{ \i n }\, ,
 \end{align}
where the first  and second terms can be associated to the contribution of the chiral and anti-chiral multiplets respectively.
We note that, in the limit of zero $R$-charge and gauge coupling, expansion \eqref{eq:optionA} does not have constant term. This
implies the absence of zero modes for both the chiral and the anti-chiral multiplets. This is compatible with the normalizable boundary condition of $\phi$ and $\bar{\phi}$ because they do not admit the constant zero modes  in their spectrum in the free massless limit. 
 From \eqref{eq:optionA} we can read off the eigenvalues of $\cH$ and write
 \begin{align}
 Z^{\text{c.m.}}_{1\text{-loop}}& =\sqrt{  \prod_{n=1}^\infty\frac{L_0\left( n-\left(\frac{\r}{2}+\Lambda_0\right)\right)}{ \i L}  \prod_{n=1}^\infty\frac{L_0\left( n-\left(\frac{\r}{2}+\Lambda_0\right)\right)}{-\i L}}\,. \label{eq:logZch2}    
 \end{align}
  We can now regularize the infinite products and use the value of $\Lambda_0$ given in~\eqref{eq:L0Ads2} at the fixed point to obtain
\begin{align} \label{eq:Z1loopCMAdS2}
Z^{\text{c.m.}}_{1\text{-loop}} & = \left(\frac{L}{L_0}\right)^{\frac{1}{2}\left(1- r - 2\i \S_1L +2 \P_0 L\right)}\frac{1}{\Gamma\left(1- \frac{r}{2}- \i \S_1L   + \P_0 L\right)}, \,
\end{align}
where we set $\S_0 =0$ in \eqref{eq:L0Ads2} according to our choice of boundary conditions. This result resembles the one obtained in \cite{Honda:2013uca} for the partition function on the hemisphere with Dirichlet boundary conditions.
\paragraph{b)} This option yields
 \begin{align} \label{eq:optionB}
  \text{ind} \left(D_{10}\right)(t)   & =  q^{\i \left(\frac{\r}{2}+\Lambda_0\right)}\sum_{n=0}^{\infty} q^{\i n }+q^{-\i \left(\frac{\r}{2}+\Lambda_0\right)} \sum_{n=0}^{\infty} q^{ -\i n }\, ,
 \end{align}
 which implies the presence of zero modes for both $\phi$ and $\bar \phi$ in the vanishing $R$-charge and gauge coupling limit. This is compatible with non-normalizable boundary conditions as both scalars admit constant zero modes in their spectrum in the massless free limit. Although we do not choose these boundary conditions, this may also be a possible choice upon an analysis along the lines of section \ref{bdycondition}.
We then have
 \begin{align}
   Z^{\text{c.m.}}_{1\text{-loop}} & =\sqrt{  \prod_{n=0}^\infty\frac{-\i L}{\left( n+\left(\frac{\r}{2}+\Lambda_0\right)\right)}  \prod_{n=0}^\infty\frac{\i L}{\left( n+\left(\frac{\r}{2}+\Lambda_0\right)\right)}}\,. \label{eq:logZch}    
 \end{align}
Regularizing \eqref{eq:logZch} and using \eqref{eq:L0Ads2} we can write:
\begin{align} \label{eq:Z1loopCMAdS2b}
    Z^{\text{c.m.}}_{1\text{-loop}} & = \left(\frac{L}{L_0}\right)^{\frac{1}{2}\left(1- r - 2\i \S_1L  +2\P_0 L \right)}\Gamma\left( \frac{r}{2} - \i \S_1L +   \P_0 L  \right). \,
\end{align}
 This result is similar to the result of the theory on hemisphere with Neumann boundary conditions on the chiral multiplet studied in  \cite{Honda:2013uca}. 
 
  \paragraph{c)}  This option leads to the following expansion
\begin{align} \label{choiceC}
      \text{ind} \left(D_{10}\right)(t)   & =  q^{\i \left(\frac{\r}{2}+\Lambda_0\right)}\sum_{n=0}^{\infty} q^{\i n }+q^{-\i \left(\frac{\r}{2}+\Lambda_0\right)} \sum_{n=1}^{\infty} q^{ \i n } \, ,
\end{align}
which implies the presence of zero mode for $\phi$ and absence of zero mode for $\bar \phi$ in the vanishing $R$-charge and gauge coupling limit. This means that $\phi$ follows non-normalizable boundary conditions and $\bar \phi$ follows normalizable ones which is in conflict with these two field configurations being conjugate to each other. Therefore, this choice is not allowed. In fact, if we perform the evaluation, we will be lead to inconsistent results as follows: from~\eqref{choiceC} we obtain
\begin{align} 
\begin{split}
Z^{\text{c.m.}}_{1\text{-loop}} & =\sqrt{\frac{  \prod_{n=1}^\infty( n-\left(\frac{\r}{2}+\Lambda_0\right))/(- \i L)  }{\prod_{n=0}^\infty( n+\left(\frac{\r}{2}+\Lambda_0\right))/(-\i L)}} \label{eq:logZch1} \, .\\
\end{split}
\end{align}
Using regularization and the value of $\Lambda_0$ given in \eqref{eq:L0Ads2} evaluated at the fixed point, we obtain
\begin{align} \label{eq:chiralA}
     Z^{\text{c.m.}}_{1\text{-loop}} & = \left(- \i \frac{L}{L_0}\right)^{\frac{1}{2}\left(1- r - 2\i \S_1 L +2 \P_0 L\right)}\sqrt{\frac{\Gamma\left(\frac{r}{2}+\i \S_1 L\right)}{\Gamma\left(1- \frac{r}{2}- \i \S_1 L+ \P_0 L\right)}}\,.
\end{align}
This result is not well-defined because of an ambiguous choice of phase for $\i= {\rm e}^{ \i m \frac{\pi}{2}}\,, m \in 2\mathbb{Z}+1 $. i.e. the overall factor of the 1-loop includes 
\be
Z^{\text{c.m.}}_{1\text{-loop}} \sim {\rm e}^{\i m \frac{\pi}{2}( \i \S_1 + \P_0)}\, .
\ee
 Even if a non-zero value of $m$ could be fixed, the integration over $\S_1$ would make the partition function  divergent. 
\paragraph{d)} This option gives the same result  as option c) with $-\i L \rightarrow \i L$ in the pre-factor in~\eqref{eq:chiralA}, and hence it is not allowed. 

As a final observation,  we point out that  all four options yield the same scale dependence.

 \subsection*{Result of full partition function on AdS$_2$}

Let us now put together all the results for \ads2. The contribution form the vector multiplet~\eqref{eq:Lhalf} combined with the regularized contribution from the chiral multiplet~\eqref{eq:Z1loopCMAdS2}, yields
\begin{align} \label{anomalycoefficientads}
Z_{\text{AdS}_2} & = \int \frac{{\rm d}(\S_1 L)}{\gym L_0} \exp\left(-S^{\text{AdS}_2}_{\text{tot}}\Bigr|_{\text{loc.}}\right)Z^{\prime \,\text{v.m.}}_{1\text{-loop}} Z^{\text{c.m.}}_{1\text{-loop}}  \, ,
\end{align}
where the 1-loop contributions $Z^{\prime \,\text{v.m.}}_{\text{1-loop}}$ and $ Z^{\text{c.m.}}_{\text{1-loop}}$  are given \eqref{eq:Lhalf} and \eqref{eq:Z1loopCMAdS2} respectively. Explicit evaluation in \eqref{anomalycoefficientads} yields
\begin{empheq}[box=\fbox]{equation}
\begin{split} \label{Ads2result} 
\vspace{0.25mm} & \\
Z_{\text{AdS}_2} & = \left(\frac{L}{L_0}\right)^{1-\frac{r}{2}+ \P_0 L}\int \frac{{\rm d}(\S_1 L)}{\gym L_0} \left(\frac{L}{L_0}\right)^{-  \i \S_1 L}  \frac{1}{\Gamma\left(1-\frac{r}{2}- \i \S_1 L+ \P_0 L\right)} \,. \\
\vspace{0.25mm} & \,
\end{split}
\end{empheq}
The scale dependent pre-factor in \eqref{Ads2result} contains a contribution $-\frac{1}{2}+ \frac{1}{2}(1-r) $ from the conformal anomaly as well as a $1$ coming from boundary zero modes. Note that the dependence on the $\P_0L$  can be absorbed into the definition of $R$-charge. 
 We also note that \eqref{Ads2result} is equivalent to the hemisphere partition function \cite{Hori:2013ika, Honda:2013uca, Sugishita:2013jca} up to the overall scaling factor and Chan-Paton factors.

%%%%%%%%%%%%%%%%%%%

%%%%%%%%%%%%%%%%%%%%%%%%%%%%%%%%%%%%%%%%%%%%%%%%%%%%%%%%%%%%%%%%%

%%%%%%%%%%%%%%%%%%%%%%%%%%%%%%%%%%
%%%%%%%%%%%%%%%%%%%%%%%%%%%%%%%%%%%%%%%%%%%%%%%%%%%%%%%%%%%%%%%

\section{Heat kernel} \label{sec:HK}

  In this section, we use the heat kernel method \cite{Vassilevich:2003xt, Fursaev:2011zz} for perturbative  evaluation of partition function around the classical saddle. 
 We compare  this  result with the scaling dependence of the partition function obtained using supersymmetric localization in the previous section.
For this purpose,  we consider the classical Coulomb branch saddle obtained in \eqref{GNOs2} and \eqref{solutionads2}, where  it is enough to focus on flat connection. This means we set $\mathfrak{m}=0$ on \ss2 
and  $\S_0 =0$ on \ads2 in the classical saddle. 
 For  this computation,
we use standard normalizable boundary conditions both for bosons and fermions.
Although this boundary condition is different from the supersymmetric boundary condition selected in the localization computation of section \ref{bdycondition}, we expect this computation to still agree with the localization one. 
This expectation is justified on grounds of similar such agreement obtained, for example, in the calculations of logarithmic corrections to black hole entropy~\cite{ Sen:2012cj,Bhattacharyya:2012ye,Banerjee:2011jp,Keeler:2014bra,Sen:2012kpz}
%This  can be justified from the examples of logarithmic corrections to black hole entropy~\cite{Sen:2008vm, Sen:2012kpz},  
because the results obtained using heat kernel method with normalizable boundary conditions successfully match with the ones from the corresponding supersymmetric theory in UV. 
Rigorous justification will be discussed in our upcoming paper~\cite{GonzalezLezcano:2023uar}.

In the following subsection \ref{HKresult}, we  present the analysis of the method of heat kernel and directly state the results in \eqref{eq:s2anomalyZ} and \eqref{ads2HKresult} . The next subsection \ref{a0computation} is dedicated to detailed computations to arrive at these results, first by placing the theory on \ss2 and then on \ads2.

\subsection{Heat kernel method: Analysis and Result} \label{HKresult}

The standard saddle point approximation of the partition function instructs us to expand the action around the classical saddle, and from the quadratic action we compute the 1-loop determinant by integration over the fluctuation of fields.  Since the quadratic action can have zero modes, we divide the fluctuations into two orthogonal directions, viz., the zero modes $\varphi_{\text{zm}}$ and non-zero modes $\varphi_{\!\perp}$ . Then the partition function is approximated by  
\beqa\label{Zapprox}
Z \;\sim\; {\rm e}^{- S_{\text{cl}}}  \int {\cD \varphi_{\text{zm}}}\cD \varphi_{\!\perp} \,{\rm e}^{- S''\bigr|_{\text{cl}}  \varphi_{\!\perp}^2} \=  {\rm e}^{- S_{\text{cl}}} \int {\cD \varphi_{\text{zm}}} \,Z^\prime_{\textrm{1-loop}}\,.
\eeqa
In what follows, we shall analyze first $Z^\prime_{\textrm{1-loop}}$  which is the result of integration over the non-zero modes. Then we will discuss the integration measure over the zero modes and finally directly present the result.

%%%%%%%%%%%%%%%%%%%%%%%%%%%%%%%%%%%%%%%%%

%%%%%%%%%%%%%%%%%%%%%%%%%%%%%%%%%%%%%
%%%%%%%%%%%%%%%%%%%%%%%%%%%%%%%%%%%%%
  Let us now turn to  $Z_{\textrm{1-loop}}^\prime$. 
We begin by  treating the bosonic and fermionic cases separately.  
  Let us call the kinetic operator of a boson $\mathcal{O}_b$. Then the logarithm of the 1-loop determinant for the boson  gives
\beqa 
\log Z_{\textrm{1-loop}}^{\prime (\text{boson})} &=& 
-\dfrac{1}{2} \log {\det}^\prime \mathcal{O}_b \= \dfrac{1}{2} \int^\infty_{\varepsilon} \frac{{\rm d}s}{s}\, \Tr^ \prime {\rm e}^{-s \mathcal{O}_b} \qquad \quad \nonumber \\ 
&=& \dfrac{1}{2}  \int^\infty_{\varepsilon/L^2} \frac{{\rm d}\bar{s}}{\bar{s}} ~\left( \Tr ~{\rm e}^{-\bar{s} \bar{\mathcal{O}}_b} - n^b_{\textrm{zm}}\right)\nn \\
&\equiv& \dfrac{1}{2}  \int^\infty_{\varepsilon/L^2} \frac{{\rm d}\bar{s}}{\bar{s}} \left(K^b\left(\bar{s} \right) - n^b_{\textrm{zm}}  \right)\,.  ~~~~ \label{zprimeoneloopboson}
\eeqa 
Here, in the first line we use the integral representation of the $\log$ function by introducing the UV cutoff $\varepsilon$ which we will take to be zero, and we change the order of trace and the integration. In the second line we introduce the dimensionless parameter $\bar{s}$ by absorbing $1/L^2$. This rescaling makes the eigenvalue of the operator $\bar{\mathcal{O}}_b$ a dimensionless number. Finally we add the zero mode contribution $n^b_{\textrm{zm}}$ to make the trace to be over complete functional basis and then separately subtract it. 
In the last step we have defined $K^b(\bar{s})$ as the trace of the heat kernel.

Likewise, we call the  kinetic operator of Dirac fermions in the Gaussian path integration~$\mathcal{O}_f$ .   Then logarithm of the 1-loop  for the fermions gives 
\beqa 
\log Z_{\textrm{1-loop}}^{\prime(\text{fermion})} & =& \log \det \mathcal{O}_f \=  \dfrac{1}{2} \log \det \mathcal{O}^2_{f}  \=-\dfrac{1}{2} \int^\infty_{\varepsilon} \frac{{\rm d}s}{s}\, \Tr^ \prime {\rm e}^{-s\, \mathcal{O}^2_f}  \nonumber \\ 
&=& \dfrac{1}{2}  \int^\infty_{\varepsilon/L^2} \frac{{\rm d}\bar{s}}{\bar{s}} ~\left( - \Tr ~{\rm e}^{-\bar{s} \bar{\mathcal{O}}^2_f} + n^f_{\textrm{zm}}\right)\nonumber \\
&\equiv& \dfrac{1}{2}  \int^\infty_{\varepsilon/L^2} \frac{{\rm d}\bar{s}}{\bar{s}} \left(K^f\left(\bar{s}\right) + n^f_{\textrm{zm}}  \right)\,.
\label{zprimeoneloopboson11}
\eeqa 
Here, we use square of the fermionic kinetic operator $\cO^2_f$ to define the heat kernel and
  in the last step we defined the~$K^f(\bar{s})$ as  trace over the heat kernel including a minus sign. Now the analysis runs parallel to the bosonic case giving the same normalization factor~$\half$ in front of the $K^f(\bar{s})$.   If we use  Majorana fermions instead of Dirac fermions, we would have an additional  $\tfrac{1}{2}$ in the normalization.

For both the boson and fermion, the trace of the heat kernel can be expanded as 
\beqa \label{Kexpansion}
K \left( \bar{s}\right)  = \sum^\infty_{n=- \tfrac{d}{2}} \dfrac{a_n }{\bar{s}^n}\, ,
\eeqa 
where $d$ is the spacetime dimension which is $d=2$ for our case. 
Next, we need to find out the contribution to the log of the partition function coming from the path integral measure. The  expansion coefficient $ a_n $ is  the spacetime integration of the Seeley-DeWitt coefficients, which is well-known for the case of massless fields~\cite{10.2307/2373078,10.2307/2373309}. 
 With this expansion, we collect the $\varepsilon$-independent terms in the integrations in  \eqref{zprimeoneloopboson} and   \eqref{zprimeoneloopboson11} and we obtain the overall scaling behavior as 
\be\label{1-loop'}
\log Z'_{\text{1-loop}} \= \left(-  n^b_{\textrm{zm}}+  n^f_{\textrm{zm}} + a_0 \right) \log L + \mathcal{O}\left(1/L\right) \, . 
\ee 
The full partition function in \eqref{Zapprox} has  integrations over the zero modes.  Therefore,  generically, the $L$ dependence of the partition function is given by   
\beqa \label{partitionfunctionLdependence}
Z\, \sim\, \int \cD\varphi_{\textrm{zm}}   ~ L^{- n^b_{{\textrm{zm}}} +n^f_{{\textrm{zm}}} + a_0}~ \cO(1/L) \,.
\eeqa The measure contributes additional $L$ factors to the final result   which are not captured in \eqref{1-loop'}. In the following, we obtain the overall $L$ dependence by identifying the contribution of the zero modes as well as calculating the Seeley-DeWitt coefficients on \ss2 and then on~\ads2. 

\paragraph{\ss2: }
We have the constant mode of scalar $\S$ as a bosonic zero mode and the constant modes of the ghost anti-ghost pair as two fermionic zero modes. Therefore, in~\eqref{partitionfunctionLdependence}, we have $n^b_\textrm{zm} = n^{\S}_\textrm{zm} = 1$ and $n^f_\textrm{zm} = n^{c,\bar c}_\textrm{zm} = 2$. Their measure contribution is the following: For zero modes of the ghost fields, there are no integration measures  as they are not in the physical spectrum. The measure for the zero mode of the scalar is given in~\eqref{eq:IntMeasure}. Therefore, the total measure is given by
\beqa\label{zmmeasureS2}
\int \cD \varphi_\textrm{zm} \sim \int d \left(\sigma_0 L \right) \, .  
\eeqa
The total contribution of the Seeley-DeWitt coefficient $a_0$ is a sum of contribution from the vector and chiral multiplet, i.e. 
$
a_0 \= a_0^{\text{v.m.}} + a^{\text{c.m.}}\,
$. They are obtained in \eqref{resultHKs2a} and \eqref{resultHKs2acm} respectively in the next section and are %. Here, we summarize the results as 
\beqa \label{a0s2}
 &&a^{\textrm{v.m.}}_0 = -1 \, , \quad\qquad a^{\textrm{c.m.}}_0 = 1-r -2\i \sigma_0 L  \, .
\eeqa
With the measure \eqref{zmmeasureS2} and these values, we obtain 
\beqa \label{s2measure}
 Z_{\textrm{S}^2} & \;\sim\; & \int d \left(\sigma_0 L \right)~  L^{- n^{\S}_{\textrm{zm}} +n^{c,\bar c}_{\textrm{zm}} + a_0^{\text{v.m.}}+a_0^{\text{c.m.}}} ~\exp \bigl({\cO(1/L)}\bigr)  \, , \nonumber \\
 & \;=\; & L^{1-r} \int d \left(\sigma_0 L \right)~ L^{-2\i \sigma_0 L }  \exp \bigl({\cO(1/L)}\bigr) \label{eq:s2anomalyZ} \, . 
\eeqa 
  We see that the scaling behavior of the partition function is $\left(1-r\right)$ in powers of $L$.  Here, the global contribution is $- n^{\S}_{\textrm{zm}} +n^{c,\bar c}_{\textrm{zm}} = 1$ and the local contribution is given by the constant part of the Seeley-DeWitt coefficients in \eqref{a0s2}, which is $-1 +(1-r)$. These values match with the corresponding ones found using localization, as discussed in the paragraph following \eqref{s2result}.
 \paragraph{\ads2 : } We now do a similar analysis for the theory on \ads2. Here we have 1-form boundary zero modes and their superpartners as the bosonic and fermionic zero modes, where the existence of the latter was explicitly shown in \eqref{eq:existence}. The number of boundary zero modes is given in \eqref{vectorbdymode}.  Since the 1-form boundary zero modes and their superpartners are 
 in one-to-one correspondence, the number of these fermionic zero modes are the same. Therefore, we have $n^b_\textrm{zm} =  n^{A^{\text{bdry}}}_{\textrm{zm}} = -1\,$ and $ n^f_\textrm{zm} = n^{Q_{eq}A^{\text{bdry}}}_{\textrm{zm}} =-1$ in \eqref{1-loop'}. As for the measure contribution of the 1-form boundary zero modes to the $L$ dependence, we see in~\eqref{MeasureonAdS2zm} that it  is trivial, i.e., it does not contribute any $L$ factor to the overall scaling of the full partition function. The evaluation of the measure contribution for the fermionic zero modes requires more care.
 We note in \eqref{IntVariableVM} that a $L^{1}$ factor is present in the integration measure of gaugino. Due to its Grassmann odd nature, for each fermionic zero mode the measure has $L^{-1}$ contribution. Therefore, scaling of the total measure is given by
 \beqa\label{ads2zmmeasure}
\int \cD \varphi_\textrm{zm} \,\sim\, L^{-n^{Q_{eq}A^{\text{bdry}}}_{\textrm{zm}}} \, . 
\eeqa
The total contribution of the Seeley-DeWitt coefficient $a_0$ is a sum of 
 contribution from the vector and the chiral multiplet. They are  obtained  in \eqref{a0vmads} and \eqref{a0cmads}  respectively in the next section and are
\beqa 
\label{a0ads2} &&a^{\textrm{v.m.}}_0 =  - \dfrac{1}{2}\, , \quad\quad a^{\textrm{c.m.}}_0 = \dfrac{1}{2} \left(1-r\right) +\rho_0 L \, .
\eeqa
With the measure \eqref{ads2zmmeasure} and these values, we obtain 
\beqa
 Z_{\textrm{AdS}_2} &\; \sim \;& L^{- n^{A^{\textrm{bdry}}}_{\textrm{zm}} + a_0^{\text{v.m.}}+a_0^{\text{c.m.}}} \exp\bigl({\cO(1/L)}\bigr)\,  = L^{1-\tfrac{r}{2} + \rho_0 L}\exp\bigl({\cO(1/L)}\bigr) \, .\label{ads2HKresult} \qquad 
\eeqa  We see that the scaling behavior of the partition function is $\left( 1-\frac{r}{2} + \rho_0 L \right)$  in powers of $L$.  Here, the global contribution is $- n^{A^{\textrm{bdry}}}_{\textrm{zm}} = 1$ and the local contribution is given by the Seeley-DeWitt coefficients in \eqref{a0ads2}, which is $-\frac{1 }{2} + \frac{1}{2}(1-r)+ \rho_0 L$. These values match with the corresponding ones found using localization, as discussed in the paragraph following~\eqref{Ads2result}.

In conclusion, the results for \ss2 in \eqref{s2measure} and \ads2 in \eqref{ads2HKresult} agree with the corresponding ones obtained using localization computation 
 by reproducing both the global and local contribution to the scaling dependence. 

\subsection{Calculation of Seeley-DeWitt coefficients} \label{a0computation}
In this section, we compute the Seeley-DeWitt coefficients for the theories on \ss2 and \ads2 which were presented as results in equations \eqref{a0s2} and \eqref{a0ads2} in the previous section. 

\subsubsection{Theory on S$^2$} \label{s2theory}

\subsubsection*{The vector multiplet}\label{s2vm}
We obtain the Lagrangian for the vector multiplet on \ss2 by setting the value of the gravity background as shown in \eqref{auxiliaryS2}. To fix the gauge, we work in a setting where the gauge field is in the Lorentz gauge. Therefore, to the  Yang-Mills Lagrangian in \eqref{eq:LYM}, we add the gauge fixing term $ \lag_{\textrm{GF}} =\tfrac{1}{2} \left(\nabla_\mu A^\mu \right)^2 $. This introduces ghost and anti-ghost fields in our theory and we add anti-commuting ghost and anti-ghost field $(\bar{c}\,,c)$ kinetic term given by, $\cL_{\text{ghost}}= \i \bar{c}\,\nabla^2 c$. Thus we have the total Lagrangian as 
\beqa \label{totalvecL}
\cL_{\text{vec}}^{\text{tot.}}&=&
	\lag_{\text{YM}}  + \lag_{\textrm{GF}}+\lag_\textrm{ghost}\nn
	\\
	&=& \half \left( F_{12} + \frac{1}{L}\rho \right)^2   + \half \partial_\mu \rho\, \partial^\mu \rho +  \dfrac{1}{2} \left(\nabla_\mu A^\mu \right)^2  \nn \\
	&& + \half \partial_\mu \sigma\, \partial^\mu \sigma +  \half D^2  + \frac{\i}{2} \overline{\l}\gamma^\mu D_\mu \l +\i \bar{c}\,\nabla^2 c \,.
\eeqa
The contribution to heat kernel from the last line  is easily obtained.
We first note that the auxiliary scalar field $D$ has no kinetic term and can be immediately integrated performing the Gaussian integral. The resulting constant from the integral gives an irrelevant contribution to the partition function as the functional integration measure is chosen  as such in \eqref{Ultralocal}. Next, the heat kernel of the free massless scalar $\S$, massless Dirac fermion $\l, \bar \l$,  and ghost fields $c, \bar c$ fields are given by the well-known result of  heat kernel of scalar and fermion which are summarized in appendix \ref{appendixheatkernel}. Taking massless limits of  \eqref{massivescalars} and \eqref{massivefermions2} , we obtain 
\beqa
K(\bar{s})_{\text{S}^2}^\sigma &=& K_{\text{S}^2}^{sc} (\bar{s})\, = \, \dfrac{1}{\bar{s}} + \dfrac{1}{3}  + \dfrac{\bar{s}}{15} + \cdots \, , 
\nonumber \\
K(\bar{s})_{\text{S}^2}^{\l,\bar{\l}} &=& K_{\text{S}^2}^{f}\left( \bar{s}\right)\, = \, - \dfrac{2}{\bar{s}} + \dfrac{1}{3} + \dfrac{\bar{s}}{30} +   \cdots  \, , \label{Ks2}
\\
K(\bar{s})_{\text{S}^2}^{c,\bar{c}} &=& -2  K_{\text{S}^2}^{sc} (\bar{s}) \, = - \dfrac{2}{\bar{s}} - \dfrac{2}{3} - \dfrac{2\bar{s}}{15} + \cdots \, . \nonumber
\eeqa 
Here, the ghost contribution given in the last equation is understood as they contribute as complex valued Grassmann scalars. Note that these trace evaluations are done over the complete functional space as the zero modes of the scalar $\S_0$ and  the ghost fields $c$ and $\bar{c}$ were added to the heat kernel  and subtracted separately as discussed in~\eqref{zprimeoneloopboson} and~\eqref{zprimeoneloopboson11}.

To obtain the trace of the heat kernel of the remaining fields we need more analysis.  
 To proceed, let us first club the gauge fixing term with the standard kinetic term for the vector field to obtain 
\gathr{
	\dfrac{1}{4} F_{\mu \nu} F^{\mu \nu} + \dfrac{1}{2} \left( \nabla_\mu A^\mu \right)^2 =   \half A_\mu \triangle A^\mu~, \qquad \triangle A^\mu \equiv - \nabla^2 A^\mu + R^{\mu \nu} A_\nu~\, , } where $R^{\mu \nu}$ is the Ricci tensor of \ss2. Then the first line of the Lagrangian in \eqref{totalvecL} is written as 
\gathr{	\mathcal{L}^{\left( A_\mu, \rho\right)} \equiv \half \pmat{\rho &&  A_ \mu} \pmat{ - \nabla^2+ \dfrac{1}{L^2} && 	\dfrac{1}{L}\epsilon^{\mu \nu} \partial_{\mu} \\ 
		\dfrac{1}{L}\epsilon^{\mu \nu} \partial_{\nu}&& g^{\mu \nu} \triangle + \nabla^\mu \nabla^\nu}  \pmat{\rho \\  A_ \nu}  
	- \half A_\mu \nabla^\mu \nabla^\nu A_\nu \, .  
	 \label{matrixdecoms2}
}
Using the standard decomposition of a vector in two dimensions as 
\begin{gather}
	A_\mu = \partial_\mu a_1 + \epsilon_{\mu\nu} \partial^{\nu}a_2 \,,  \label{vectordecomposition1}
\end{gather} where $a_1$ and $a_2$ are normalizable scalars, we obtain
\beqa
	\mathcal{L}^{\left( A_\mu, \rho\right)} &=& \half \pmat{\rho &&  \epsilon_{\mu\sigma} \partial^{\sigma}a_2} \pmat{ - \nabla^2+ \dfrac{1}{L^2} && 	\dfrac{1}{L}\epsilon^{\mu \nu} \partial_{\mu} \\ 
		\dfrac{1}{L}\epsilon^{\mu \nu} \partial_{\nu}&& g^{\mu \nu} \triangle + \nabla^\mu \nabla^\nu}  \pmat{\rho \\  \epsilon_{\nu\lambda} \partial^{\lambda}a_2}  \\  
&&  - \half \partial_\mu a_1  \nabla^\mu \nabla^\nu \partial_\nu a_1  \, .\nn
	 \label{matrixdecomI}
\eeqa
Now, 
we expand all the scalars in the basis of the eigenfunctions of the scalar Laplacian as\footnote{The basis is real, or can be chosen to be real. }
\begin{gather} \label{modedecomps2s2}
	\rho=\sum^\infty _{l=0} \sum^{l}_{m=-l} \rho_{lm} Y_{lm}\, , 	\quad a_1= \sum^\infty _{l=1} \sum^{l}_{m=-l} a_{1,lm} \dfrac{Y_{lm}}{L\sqrt{\kappa_l}}\, , \quad	a_2= \sum^\infty _{l=1} \sum^{l}_{m=-l}  a_{2,lm}\dfrac{ Y_{lm}}{L\sqrt{\kappa_l}}\, ~~ ,
\end{gather} where $Y_{lm}$ are the spherical harmonics and $\kappa_l$ is the eigenvalue of Laplacian operator on \ss2 given by $\kappa_l = \frac{l(l+1)}{L^2}$. We note that while in the  scalar $\P_{lm}$, the modes are labelled by $l\in \left\lbrace 0,1,2, \cdots \right\rbrace $, the modes of the scalars $a_{1,2}$  in the decomposition of the vector in  \eqref{vectordecomposition1} are labelled by $l \in \left\lbrace 1,2,3, \cdots  \right\rbrace $, i.e., the $l=0$ mode is excluded. This is because $Y_{l=0,m=0}\left( x\right) $ is a constant and therefore \eqref{vectordecomposition1} shows that this mode will not give us any gauge field component.
Then, using the orthonormality property of the spherical harmonics, we obtain
\beqa
		\mathcal{L}^{\left( A_\mu, \rho\right)}  &=& \half \Bigg[  \sum^\infty_{l=1}\sum^{l}_{m=-l}\pmat{\rho_{lm} && \dfrac{a_{2,lm} }{L} }   \pmat{ \kappa_l + \dfrac{1}{L^2} && 	\dfrac{1}{L}\sqrt{\kappa_l}  \\ 
		\dfrac{1}{L}\sqrt{\kappa_l} &&\kappa_l}  \pmat{\rho_{lm} \\  \dfrac{a_{2,lm}}{L} }    \nonumber \\ 
		&&
	~\quad +\left( \kappa_0 + \dfrac{1}{L^2}\right) \rho_{00}^2 +	\sum^\infty_{l=1}\sum^{l}_{m=-l} \kappa_l \left(\dfrac{a_{1,lm}}{L}\right)^2
		\Bigg]	\, .  \label{VMmatrix}
\eeqa 
We note that the modes of scalar $\P_{lm}$ with $l \neq 0$ couple to the half of the modes in vector,~$a_{2, lm}$, which is expressed by  a matrix. Another half of vector modes $a_{1,lm}$ and constant mode of scalar $\P_{00}$ are isolated.

The contribution of the modes $a_{1,lm}$ to the heat kernel  is simply  given by the heat kernel of a scalar obtained as the massless limit of \eqref{massivescalars}. Here, since the index $l$ of the modes starts from 1 instead of 0, as discussed,  
the heat kernel is 
\begin{eqnarray} 
	K^{a_{1}   }_{\text{S}^2}  \left(\bar{s}\right) 
	&=&  K_{\text{S}^2}^{sc} (\bar{s}) -1  \= \frac{1}{\bar{s}} - \frac{2}{3} + \frac{\bar{s}}{15} + \dfrac{4 \bar{s}^2}{315} +\cdots\,.
 \label{heatkernels2sigmas}
\end{eqnarray}

\begin{figure}[h!]
\centering
\begin{tikzpicture}
\draw[help lines] (-10,0) -- (-6.9,0) ;
\draw[dotted] (-6.9,0) -- (-6.4,0) ;
\draw[help lines,->] (-6.4,0) -- (-4,0) coordinate (xaxisA);
\draw[help lines,->] (-9,-2.0) -- (-9,2.0) coordinate (yaxisA);

            \node [below] at (xaxisA) {$\text{Re}(\lambda)$};
			\node [above] at (yaxisA) {$ \hspace{1mm}\text{Im}(\lambda)$};
			\node at (-8.1,0) {{\color{red}$\times$}};
				\node at (-7.1,0) {{\color{red}$\times$}};
				\node at (-6.1,0) {{\color{red}$\times$}};
				\node at (-5.1,0) {{\color{red}$\times$}};
		
\draw[thick] (-9,0) -- (-6.9, 0.34125);
\draw[dotted,thick] (-6.9,0.34125) -- (-6.4, 0.4225);
\draw[thick, ->] (-6.4,0.4225) -- (-5.9, 0.51);
\node [above] at (-6, 0.5) {$\mathcal{C}_{+}$};
\draw[thick] (-6,0.5) -- (-5, .65);

\draw[thick] (-9,0) -- (-6.9, - 0.34125);
\draw[dotted,thick] (-6.9,- 0.34125) -- (-6.4, -0.4225);
\draw[thick] (-6,- 0.5)--(-6.4,-0.4225) ;
\node [below] at (-6, -0.5) {$\mathcal{C}_{-}$};
\draw[thick,->](-5, -.65)  -- (-6,-0.5);

\draw[thick, ->] (-6.89,0) arc (-17.1:2.5:0.91cm);
\node [above] at (-6.65, -0.056) {$\theta$};
%\draw[thick] (-8,-0.5) -- (-9, 0);
%\draw[thick,->] (-7, -1) -- (-8,-0.5);

\draw[thick, ->] (-5,0.65) arc (25.5651:0:1.536cm);
\node [right] at (-4.9, 0.35) {$\mathcal{C}_{\infty}$};
\draw[thick] (-4.86393,0) arc (0:-25.5652:1.536cm);

\end{tikzpicture}
\caption{The figure sketches the contour in the complex $\l$ plane with the poles indicated by red crosses distributed along the horizontal axis.\label{fig:C}}
\end{figure}
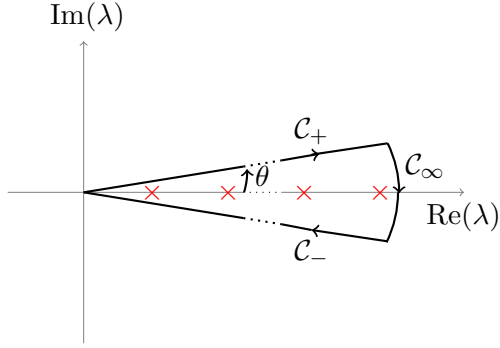

To obtain the heat kernel contribution from $\P_{lm}$ and $a_{2,lm}$ modes, we diagonalize the square matrix in \eqref{VMmatrix} and define the heat kernel with the eigenvalues of the diagonal matrix. The eigenvalues $\lambda_{1\,,2}$ of the matrix are   $\lambda_1 = \dfrac{l^2}{L^2}$ and $\lambda_2 = \dfrac{\left( l+1\right)^2 }{L^2} \,.$ Together with the eigenvalue for the $\P_{00}$ mode, which is $1/L^2$,  the contribution to the trace of the heat kernel  from the $\rho_{lm}$ and $a_{2lm}$ modes is given by
\beqa  K^{\rho+a_2} \left( \bar{s}\right)
	&=& {\rm e}^{ -\bar{s} } +   \sum_{l=1}^{\infty} \Bigg[\left( 2l+1\right) {\rm e}^{ -  \bar{s} l^2}  + \left( 2l+1\right) {\rm e}^{ -  \bar{s}\left(l+1 \right) ^2}\Bigg]\, \nonumber \\
	&=&   \sum_{l=0}^{\infty}(2l+1){\rm e}^{- \bar{s} l^2} +\sum_{l=0}^{\infty
	}(2 l+1) {\rm e}^{-\bar{s}(l+1)^2} -1  \, ,  \label{krhoa2}
\eeqa where the first term in the right hand side of the first line is obtained from $\rho_{00}$ mode and the second term is from the modes in the matrix. In the second line of \eqref{krhoa2}, we have rearranged such that summation start from $l=0$. To evaluate the above series, we use the fact that the function $\lambda \tan \left(\pi \lambda \right) $ has poles at $\lambda = \dfrac{(2l+1)}{2}$ with residue equal to $-\dfrac{(2l+1)}{2\pi}$ for $l = 0,1,2,3, \, \cdots$ . Therefore, by residue theorem, we rewrite the series as 
\begin{align}
	\sum_{l=0}^{\infty}\left(2 l+1\right){\rm e}^{- \bar{s} l^2}  & ={\rm e}^{- \frac{\bar{s}}{4}}\left[\frac{1}{ \i}\left( \oint_{C} d \lambda  \lambda \tan \pi \lambda {\rm e}^{- \bar{s} \lambda(\lambda-1)}\right)\right]\, ,  \label{eq:Im} \\
	\sum_{l=0}^{\infty}\left(2 l+1\right){\rm e}^{- \bar{s} (l+1)^2}  & ={\rm e}^{- \frac{\bar{s}}{4}}\left[\frac{1}{ \i}\left( \oint_{C} d \lambda  \lambda \tan \pi \lambda {\rm e}^{- \bar{s} \lambda(\lambda+1)}\right)\right]  \, ,  \label{eq:Ip}
\end{align}
where the contour $C$ is the sum of contours $C_+, C_-$ and $C_{\infty}$ as depicted in figure \ref{fig:C}.  
Selecting a small  angle $\theta$, we have 
\begin{align}\label{randomeqncont}
	\oint_{C} d \lambda  \lambda \tan \pi \lambda {\rm e}^{- \bar{s} \l\left( \l \pm 1\right) }  & = 2 \i  \hspace{1.5mm}\text{Im} \left( \int_{C_-} d \lambda  \lambda \tan \pi \lambda {\rm e}^{- \bar{s} \lambda(\lambda \pm 1)}\right)\, . 
\end{align}
Let us now expand the trigonometric function $\tan\pi \l$ with  ${\rm e}^{ 2 \pi \i \lambda }$ as an expansion parameter,  
 \begin{gather}
	\tan \pi \lambda =  \dfrac{1}{\i}\Bigg[1 - \dfrac{2 \exp \left( - 2 \pi \i \lambda \right) }{1+ \exp\left(- 2 \pi \i \lambda  \right) } \Bigg]= -\i \Bigg[1+ 2 \sum_{ k=1}^{\infty} (-1)^k \exp \left( -2\pi \i k \lambda\right)\Bigg] \, . 
\end{gather} 
Here, the choice of  expansion parameter is dictated by the contour $C_-$. Namely, on $C_-$  this expansion parameter satisfies $\Re \left[\exp\left(- 2 \pi \i \lambda  \right)\right]<~1$ and  the above series converges.  
Let us take the limit $\theta \rightarrow 0$, then
\beqa
& \text{Im}& \left( \int_{C_-} d \lambda ~ \lambda \tan \pi \lambda ~{\rm e}^{- \bar{s} \lambda(\lambda \pm 1)}\right)  \nonumber \\ 
 &=& -  \lim_{\theta\rightarrow 0} \int^0_{\infty {\rm e}^{-\i \theta}} d \lambda ~\l ~\Bigg[1+ 2 \sum_{ k=1}^{\infty} (-1)^k \exp \left( -2\pi \i k \lambda\right)\Bigg] {\rm e}^{- \bar{s} \lambda(\lambda \pm 1)}\ \,  \nonumber \\ 
 &=&    \int_0^{\infty } d \lambda ~\l ~\Bigg[1+ 2 \sum_{ k=1}^{\infty} (-1)^k \exp \left( -2\pi \i k \lambda\right)\Bigg] {\rm e}^{- \bar{s} \lambda(\lambda \pm 1)}\, .
\eeqa Using the above and  \eqref{randomeqncont}, we have 
\beqa
\sum_{l=0}^{\infty}\left(\dfrac{2 l+1}{2\pi}\right){\rm e}^{- \bar{s} l^2}   =2 {\rm e}^{- \frac{\bar{s}}{4}}  \int_0^{\infty } d \lambda ~\l ~\Bigg[1+ 2 \sum_{ k=1}^{\infty} (-1)^k \exp \left( -2\pi \i k \lambda\right)\Bigg] {\rm e}^{- \bar{s} \lambda(\lambda - 1)} \,, \qquad  \\
\sum_{l=0}^{\infty}\left(\dfrac{2 l+1}{2\pi}\right){\rm e}^{- \bar{s} \left(l+1\right) ^2}   =2 {\rm e}^{- \frac{\bar{s}}{4}}  \int_0^{\infty } d \lambda ~\l ~\Bigg[1+ 2 \sum_{ k=1}^{\infty} (-1)^k \exp \left( -2\pi \i k \lambda\right)\Bigg] {\rm e}^{- \bar{s} \lambda(\lambda + 1)} \, . \qquad
\eeqa The above integrations can directly be performed and gives the following result as a series expansion in $\bar{s}$. Then \eqref{krhoa2} gives us
\begin{align}
	\begin{split}
		K^{\rho+a_2}_{\textrm{S$^2$}}  \left(\bar{s}\right) =  \left( \frac{2}{\bar{s}}-\frac{1}{3}-\frac{\bar{s}}{30} - \frac{\bar{s}^2}{126} + \mathcal{O}\left(\bar{s}^3\right)\right). \label{eq:Kvps}
	\end{split}
\end{align} 

Therefore, collecting  \eqref{Ks2}, \eqref{heatkernels2sigmas} and \eqref{eq:Kvps}, we get the trace of heat kernel for vector multiplet on \ss2 as 
\beqa\label{resultHKs2}
K^{\text{v.m.}}_{\textrm{S}^2}&=&  K_{\textrm{S}^2}^{\rho+a_2}\left( \bar{s}\right) + K_{\textrm{S}^2}^{a_1 } \left( \bar{s}\right)+ K_{\textrm{S}^2}^{\S  }\left( \bar{s}\right)+ K_{\textrm{S}^2}^{c,\bar{c}} \left( \bar{s}\right) + K_{\textrm{S}^2}^{\lambda,\bar{\lambda}}\left( \bar{s}\right)  = -1 + \mathcal{O}\left( \bar{s}^3\right)  \, .
\eeqa
Note here that the  ${\cal O}(1/\bar{s})$ divergences coming from trace of heat kernel of each field  cancel. 
In fact there is no $\bar{s}$ dependent corrections, which we will show below using an alternative derivation.

\subsection*{Alternative derivation } We will now show that in fact the right hand side of \eqref{resultHKs2} receives no $\bar{s}$ dependent  corrections.  Let us recall that the contribution to the trace of the heat kernel due to the ghosts cancels the $\bar{s}$ dependent contributions coming from the scalars $a_1$ and $\sigma$. Explicitly, 
\beqa
K_{\textrm{S}^2}^{a_1 } \left( \bar{s}\right)+ K_{\textrm{S}^2}^{\S  }\left( \bar{s}\right)+ K_{\textrm{S}^2}^{c,\bar{c}} \left( \bar{s}\right) &= &\left( K_{\text{S}^2}^{sc} (\bar{s}) -1 \right) + K_{\text{S}^2}^{sc}\left( \bar{s}\right) - 2 K_{\text{S}^2}^{sc} \left( \bar{s}\right) =-1\, . \quad
\eeqa We are then left with $K_{\textrm{S}^2}^{\rho+a_2}\left( \bar{s}\right) $ and $K_{\textrm{S}^2}^{\lambda,\bar{\lambda}}\left( \bar{s}\right)$. From \eqref{krhoa2} we have, 
\beqa
	K^{\rho+a_2}_{\textrm{S}^2}  \left(\bar{s}\right) + K_{\textrm{S}^2}^{\lambda,\bar{\lambda}}\left( \bar{s}\right)  &=& \Bigg(\sum_{l=0}^{\infty}(2l+1){\rm e}^{- \bar{s} l^2} +\sum_{l=0}^{\infty
	}(2 l+1) {\rm e}^{-\bar{s}(l+1)^2} -1\Bigg)   -4  \sum^\infty_{l=1} l {\rm e}^{-\bar{s}l^2}\nn \\ 
 &=& \Bigg(\sum_{l=0}^{\infty}(2l+1){\rm e}^{- \bar{s} l^2} +\sum_{l=1}^{\infty
}(2 l-1) {\rm e}^{-\bar{s}l^2} -1\Bigg)   -4  \sum^\infty_{l=1} l {\rm e}^{-\bar{s}l^2} \nn \\ 
&=& \Bigg(\sum_{l=1}^{\infty}(2l+1){\rm e}^{- \bar{s} l^2} +\sum_{l=1}^{\infty
}(2 l-1) {\rm e}^{-\bar{s}l^2} \Bigg)   -4  \sum^\infty_{l=1} l {\rm e}^{-\bar{s}l^2} \nn \\
&=& 0 \, . 
\eeqa Altogether, we get
 \beqa
K^{\text{v.m.}}_{\textrm{S}^2}(\bar{s})&=&  K_{\textrm{S}^2}^{\rho+a_2}\left( \bar{s}\right) + K_{\textrm{S}^2}^{a_1 } \left( \bar{s}\right)+ K_{\textrm{S}^2}^{\S  }\left( \bar{s}\right)+ K_{\textrm{S}^2}^{c,\bar{c}} \left( \bar{s}\right) + K_{\textrm{S}^2}^{\lambda,\bar{\lambda}}\left( \bar{s}\right)  = -1  \, .
\eeqa  This is an exact result because the right hand side receives no corrections in higher order of $\bar{s}$.
Therefore, we have obtained the $\bar{s}$ independent Seeley-DeWitt coefficient, $a^{\text{v.m.}}_{0,S^2}$, of the $K^{\text{v.m.}}_{\textrm{S$^2$}}(\bar{s})$ as
\beqa \label{resultHKs2a}
a^{\text{v.m.}}_{0,S^2} = -1\,.
\eeqa

\subsubsection*{The chiral multiplet}\label{s2cm}

As can be seen from the  Lagrangian for the chiral multiplet in \eqref{mattermultiplet}, it has a massive complex scalar and a massive fermion. Therefore, the chiral multiplet has the following contribution to the trace of the heat kernel, given by
\beqa
K^{\text{c.m.}}_{\textrm{S}^2}(\bar{s})&=&  2  K^{sc}_{\textrm{S}^2}\left( \bar{s},\bar{m}_b \right)  +K^{f}_{\textrm{S$^2$}} \left(\bar{s},\bar{m}_D\right)  \, ,
\eeqa
where the auxiliary fields $F $ and $\bar F$ have been integrated out and they do not contribute to the heat kernel.
The masses for the scalars and the fermions are summarised in equations~\eqref{MassScalar} and \eqref{MassFermion} respectively. Following  \eqref{massivescalars} and \eqref{massivefermions2}, we obtain the $\bar{s}$ independent Seeley-DeWitt coefficient $ a^{\text{c.m.}}_{0,\textrm{S$^2$}}$ of $K^{\text{c.m.}}_{\textrm{S$^2$}}(\bar{s})$ as
\begin{eqnarray} \label{resultHKs2acm}
 a^{\text{c.m.}}_{0,\textrm{S}^2}= 1-r -2\i \sigma_0 L  \, .
\end{eqnarray}
Equations \eqref{resultHKs2a} and \eqref{resultHKs2acm} conclude our evaluation of Seeley-DeWitt coefficients of the vector and chiral multiplet for the theory on \ss2. They were collectively presented in~\eqref{a0s2}. 

\subsubsection{Theory on AdS$_2$} \label{ads2theory}

 \subsubsection*{The vector multiplet}\label{ads2vm}
 
From the review of the theory given in section \ref{sec2}, we begin by  setting the value of the gravity background as in \eqref{auxiliaryads2}. As in the  \ss2 case, we work in the Lorentz gauge. So we add the gauge fixing term $ \lag_{\textrm{GF}} =\tfrac{1}{2} \left(\nabla_\mu A^\mu \right)^2 $ to the Lagrangian. This introduces ghosts in our theory and we add the term for the anti-commuting ghost and anti-ghost field $(\bar{c}\,,c)$, $\cL_{\text{ghost}}= \i \bar{c}\,\nabla^2 c$. Thus we have the total Lagrangian as 
 \beqa \label{totalvecLads}
 \cL_{\text{vec}}^{\text{tot.}}&=&
 \lag_{\text{YM}}  + \lag_{\textrm{GF}}+\lag_\textrm{ghost}\nn
 \\
 &=& \half \left( F_{12} - \dfrac{\i \sigma }{L}\right)^2   + \half \partial_\mu \sigma \partial^\mu \sigma   +  \dfrac{1}{2} \left(\nabla_\mu A^\mu \right)^2    \qquad \qquad \nn \\  &&   + \half \partial_\mu \rho \partial^\mu \rho + \half D^2  + \frac{\i}{2} \overline{\l}\gamma^\mu D_\mu \l +\i \bar{c}\,\nabla^2 c \,.
 \eeqa
The subsequent analysis mirrors the \ss2 case. The contribution to the trace of the heat kernel from the second line is easily obtained.
 Again,  the auxiliary scalar field $D$ can be immediately integrated out and gives an irrelevant contribution to the partition function. Next, the heat kernel of the free massless scalar $\rho$, massless Dirac fermion $\l, \bar \l$,  and ghost fields $c, \bar c$ fields are given by the well-known result of  heat kernel of scalar and fermion which are summarized in appendix \ref{appendixheatkernel}. Taking massless limits of  \eqref{massivescalarads} and \eqref{massivefermionads}, we obtain  
 \beqa
 K(\bar{s})_{\text{AdS}}^\rho &=& K_{\text{AdS}}^{sc} (\bar{s})\, = \, -  \dfrac{1}{2\bar{s}} + \dfrac{1}{6} - \dfrac{\bar{s}}{30}  + \cdots \, ,  
 \nonumber \\
 K(\bar{s})_{\text{AdS}}^{\l,\bar{\l}} &=& K_{\text{AdS}}^{f}\left( \bar{s}\right)\, = \, \dfrac{1}{\bar{s}} + \dfrac{1}{6} - \dfrac{\bar{s}}{60}  + \cdots \, , \label{Kads} \\
 K(\bar{s})_{\text{AdS}}^{c,\bar{c}} &=& -2  K_{\text{AdS}}^{sc} (\bar{s})  \, =  \dfrac{1}{\bar{s}} - \dfrac{1}{3} + \dfrac{ \bar{s}}{15}  + \cdots \, . \nonumber
 \eeqa Note that these trace evaluations too are done over the complete functional space as the zero modes of $\l$ and $\bar \l$ which are the superpartners of the 1-form boundary zero modes, were added and subtracted separately. 
 
 As in the case of \ss2, obtaining the heat kernel contribution of the rest of the  fields in~\eqref{totalvecLads} requires more analysis. We first club the gauge fixing term with the standard kinetic term for the vector field to obtain 
\gathr{
	\dfrac{1}{4} F_{\mu \nu} F^{\mu \nu} + \dfrac{1}{2} \left( \nabla_\mu A^\mu \right)^2 =   \half A_\mu \triangle A^\mu~, \qquad \triangle A^\mu \equiv - \nabla^2 A^\mu + R^{\mu \nu} A_\nu~\,,} where $R^{\mu \nu}$ is the Ricci tensor of \ads2. Then the first line of the  total Lagrangian is written as 
\gathr{	\mathcal{L}^{\left( A_\mu, \sigma\right)} \equiv \half \pmat{\sigma &&  A_ \mu} \pmat{ - \nabla^2- \dfrac{1}{L^2} && 	-\dfrac{\i}{L}\epsilon^{\mu \nu} \partial_{\mu} \\ 
		-\dfrac{\i}{L}\epsilon^{\mu \nu} \partial_{\nu}&& g^{\mu \nu} \triangle + \nabla^\mu \nabla^\nu}  \pmat{\sigma \\  A_ \nu}
	- \half A_\mu \nabla^\mu \nabla^\nu A_\nu \, .   	\label{matrixdecomads}
} We use the standard decomposition of a vector on \ads2 as 
\begin{gather}
A_\mu = \partial_\mu a_1 + \epsilon_{\mu\nu} \partial^{\nu}a_2 + \partial_{\mu} \Lambda \,,  \label{vectordecompositionads1}
\end{gather} where $a_1$ and $a_2$   are normalizable scalars and $\Lambda$ is non-normalizable scalar which give rise to boundary zero modes. (See \eqref{1-formExpansion} for more explicit form.) %Plugging this in  \eqref{matrixdecoms2}, 
The zero mode part from the $\Lambda$ disappears from the Lagrangian. 
From the non-zero modes, we obtain
\beqa \label{matrixdecomIads}
\mathcal{L}^{\left( A_\mu, \sigma\right)} &=& \half \pmat{\sigma &&  \epsilon_{\mu\sigma^\prime} \partial^{\sigma^\prime}a_2} \pmat{ - \nabla^2- \dfrac{1}{L^2} && 	-\dfrac{\i}{L}\epsilon^{\mu \nu} \partial_{\mu} \\ 
	-\dfrac{\i}{L}\epsilon^{\mu \nu} \partial_{\nu}&& g^{\mu \nu} \triangle + \nabla^\mu \nabla^\nu}  \pmat{\sigma \\  \epsilon_{\nu\lambda} \partial^{\lambda}a_2}   \\  
&& - \half \partial_\mu a_1  \nabla^\mu \nabla^\nu \partial_\nu a_1  \, .\nn
\eeqa
Let us expand the scalars in the basis of the eigenfunctions of the Laplacian \eqref{AdS2eigenfunction} as
\begin{gather}
\sigma=\sum^\infty_{k=0} \int_{0}^{\infty} d\l~ \sigma_{\l,k} f_{\l,k} (x) \, , \qquad \nn \\ 	 a_1=\sum^\infty_{k=0} \int_{0}^{\infty} d\l~   \dfrac{1}{L\sqrt{\kappa_\l}}a_{1{\l,k}} f_{\l,k} (x) \, , \quad a_2=\sum^\infty_{k=0} \int_{0}^{\infty} d\l~   \dfrac{1}{L \sqrt{\kappa_\l}}a_{2{\l,k}} f_{\l,k} (x) \,  ,\end{gather} where the expansion of $a_1$ and $a_2$ stem from the basis of the vector ~\eqref{vectorlaplacianads} and we have denoted $\k_\l$ as the eigenvalue of the scalar Laplacian, i.e., $\k_\l =\tfrac{1}{L^2 }\left(\l^2+\frac{1}{4} \right) $. Using the orthonormality of the eigenfunctions,  \eqref{matrixdecomIads} reduces to 
\beqa  \label{matrixdecomIads2}
\mathcal{L}^{\left( A_\mu, \sigma\right)} &=& \half \pmat{\sigma_{\l,k} && \dfrac{a_{2{\l,k}}}{L} } \pmat{ \k_\l - \dfrac{1}{L^2} && 	- \dfrac{\i}{L}\sqrt{\k_\l}\\ 
	- \dfrac{\i}{L}\sqrt{\k_\l}&& \k_\l }  \pmat{\sigma_{\l,k} \\  \dfrac{a_{2{\l,k}}}{L} } + \half \k_\l \left( \dfrac{a_{1{\l,k}}}{L} \right) ^2   \, .
\eeqa 

Let us deal with the $a_1$ mode first. The contribution of the mode $a_1$ to the trace of the heat kernel is given by the massless limit of \eqref{massivescalarads}. 
 So,
\begin{eqnarray} \label{heatkernela1}
K^{a_{1}   }_{\text{AdS}}  \left(\bar{s}\right) 
&=&  K_{\text{AdS}}^{sc} (\bar{s}) = -  \dfrac{1}{2\bar{s}} + \dfrac{1}{6} - \dfrac{\bar{s}}{30} + \cdots \, . 
\label{heatkernels2sigmaads}
\end{eqnarray} Next, we treat the $\S$ and the $a_2$ modes which are coupled. To proceed, we diagonalize the square matrix given in  \eqref{matrixdecomIads2}. The eigenvalues of the square matrix are given by $
L^{-2}\left(\l+ \tfrac{\i}{2} \right) ^2 \,$ and $ L^{-2}\left(\l-\tfrac{\i}{2} \right) ^2$. So, the trace of the heat kernel for these modes is given~by
\beqa \label{eq:traceheatkernelads}
K_{\text{AdS}}^{\S+a_2} \left( \bar{s}\right) &=& \dfrac{1}{L^2} \int_{\text{AdS}} \textrm{d}^2 x \sqrt{g}~\sum^\infty_{k=0} \int d\l ~f^*_{\lambda,k}\left(\eta,\theta \right) f_{\lambda,k}\left(\eta,\theta \right) \exp\Big[-\bar{s}\left( \lambda +\tfrac{\i}{2} \right)^2 \Big] \nonumber \\
&+& \dfrac{1}{L^2} \int_{\text{AdS}} \textrm{d}^2 x \sqrt{g}~\sum^\infty_{k=0} \int d\l ~f^*_{\lambda,k}\left(\eta,\theta \right) f_{\lambda,k}\left(\eta,\theta \right) \exp\Big[-\bar{s}\left( \lambda -\tfrac{\i}{2} \right)^2 \Big] \, . \quad 
\eeqa 
Using the homogeneity of \ads2, we evaluate the above at the origin~$\eta=0$. 
Since the eigenfunction at the origin is given by~\eqref{fAtOrigin}, the heat kernel  simplifies to
\beqa \label{integralKAdS}
K_{\text{AdS}}^{\S+a_2} (\bar{s})
&=&  - \int^\infty_0 d\lambda~  \lambda \tanh \pi \lambda   \Bigg[\exp\Big[-\bar{s}\left( \lambda +\tfrac{\i}{2} \right)^2 \Big]+ \exp\Big[-\bar{s}\left( \lambda -\tfrac{\i}{2} \right)^2 \Big]\Bigg] \, ,  \quad
\eeqa where we have used the fact that the regularized volume integral over \ads2 in \eqref{eq:traceheatkernelads} gives a factor of $-2\pi L^2\,$.

To evaluate this integral, we expand hyperbolic function in the integrand as follows:
\begin{gather}
	\tanh \pi \lambda =  1 - \dfrac{2 \exp \left( - 2 \pi \lambda \right) }{1+ \exp\left(- 2 \pi \lambda  \right) } = 1+ 2 \sum_{ k=1}^{\infty} (-1)^k \exp \left( -2k\pi\lambda\right) \, . \label{randomeqn}
\end{gather} 
We compute the constant part and the $k$ dependent part of~\eqref{randomeqn} separately in the integral~\eqref{integralKAdS}.  Plugging the constant part the integration gives 
%We plug in the $k$ independent part of the above series \eqref{randomeqn} into the integrand of~\eqref{integralKAdS}. It can immediately be integrated to give
\footnote{The full integral can actually be done in closed form and is expressed in terms of (imaginary) Error Function Erfi$\left( \bar{s}\right)$, but for our purposes, the given series expansion is sufficient. }
\be
	I_1 =  -\int_{0}^{\infty} d\lambda  ~ \lambda \Bigg[\exp\Big[-\bar{s}\left( \lambda +\tfrac{\i}{2} \right)^2 \Big]+ \exp\Big[-\bar{s}\left( \lambda -\tfrac{\i}{2} \right)^2 \Big]\Bigg] = - \dfrac{1}{\bar{s}} + \dfrac{1}{4} + \dfrac{\bar{s}}{96} + \cdots \,. 
\ee 
We plug  the $k$ dependent terms in~\eqref{randomeqn} into the integrand of~\eqref{integralKAdS} and we have the following:
\beqa
I_2 &=& -2 \sum_{ k=1}^{\infty} (-1)^k \int_{0}^{\infty} d\lambda  ~ \lambda \exp\left( -2k\pi\l\right)  \Bigg[\exp\Big[-\bar{s}\left( \lambda +\tfrac{\i}{2} \right)^2 \Big]+ \exp\Big[-\bar{s}\left( \lambda -\tfrac{\i}{2} \right)^2 \Big]\Bigg] \nonumber \\ 
&=& \dfrac{1}{12} + \dfrac{\bar{s}}{160} + \cdots \, . 
\eeqa Note that to evaluate the above, we first expand the exponential functions in the crotches in a series in $\bar{s}$, perform the integration for each term in the series and then take the sum order by order~\footnote{Tacitly, we have interchanged the order of integration and the summation and it is not a priori clear that the result should be the same. This is understood as follows. We note that in $\l \rightarrow 0$ limit, the integrand diverges as $\exp\left( \tfrac{\bar{s}}{4}\right) $. Therefore, the path integral will not be well defined. However, we can truncate the contribution from the large $\bar{s}$ regime by expanding the exponentials in a series in $\bar{ s}$ and keep terms up to some finite order in $\bar{ s}$. In fact, for our purposes, we would be interested in only the $\bar{ s}$ independent term in the final result. This process is also standard in string theory where the integration over the modular parameter is done last. At the level of the path integral, this seems to correspond to contour deformations where the path integral is well defined. For example, see  \cite{Sen:2012kpz, Banerjee:2010qc}. }.
Then we obtain that  
\beqa
K_{\text{AdS}}^{\S+a_2} (\bar{s}) &=&  I_1 + I_2= - \dfrac{1}{\bar{s}}  + \dfrac{1}{3} +\dfrac{\bar{s}}{60}  + \cdots \label{kvspart}\, .
\eeqa The above is the contribution only from the normalizable modes. In order to define the heat kernel as the trace over complete functional space, we add the contribution of the 1-form boundary zero modes, which is given by
  \beqa \label{adsbdyzm}
  K_{\text{AdS}}^{\Lambda} (\bar{s}) = n^{A^\textrm{bdry}}_{\textrm{zm}} = -1\, . 
  \eeqa
  
 Therefore, at long last, from  \eqref{heatkernela1} , \eqref{kvspart} and   \eqref{adsbdyzm} and we have 
\beqa \label{heatkenelsigmaa2}
K_{\text{AdS}}^{\left( A_\mu, \sigma \right) } (\bar{s}) &=&K^{a_{1}   }_{\text{AdS}}  \left(\bar{s}\right)   + K_{\text{AdS}}^{\S+a_2} (\bar{s})+K_{\text{AdS}}^{\Lambda} (\bar{s})\, .
\eeqa 
So, using \eqref{Kads} and \eqref{heatkenelsigmaa2}, taken altogether, we have
\beqa\label{resultHKads}
 K^{\text{v.m.}}_{\text{AdS}}(\bar{s})&=& K_{\text{AdS}}^{\left( A_\mu, \sigma \right) } (\bar{s}) + K_{\text{AdS}}^{\rho }\left( \bar{s}\right)+ K_{\text{AdS}}^{c,\bar{c}} \left( \bar{s}\right) + K_{\text{AdS}}^{\lambda,\bar{\lambda}}\left( \bar{s}\right)  = - \half +\mathcal{O}\left(\bar{s} \right)  \qquad 
\eeqa 
and thus the $\bar{s}$ independent Seeley-DeWitt coefficient, 
$a^{\text{v.m.}}_{0,\text{AdS}} $  of $ K^{\text{v.m.}}_{\text{AdS}}(\bar{s}) $ is  given by 
\begin{eqnarray} \label{a0vmads}
 a^{\text{v.m.}}_{0,\text{AdS}} = -\half \, .
\end{eqnarray}
Note here, like the \ss2 case, the  ${\cal O}(1/\bar{s})$ divergences coming from trace of heat kernel of each field  cancel. However, $\mathcal{O}\left( \bar{s}\right) $ terms survive, which was not the case in \ss2.

%%%%%%%%%%%%%%%%%%%%%%%%%%%%%%%%%%%%%%%%%%%%%%%%%%%%%%%%%%%%%%%%%%%%%%%%%%%%%%%%%%%%%%%%%%%%%%%%%%%%%%%%%%%%%%%%%%%%

\subsubsection*{The chiral multiplet} \label{ads2cm}

The contribution of the chiral multiplet to the heat kernel on \ads2 is calculated as we did for the case on \ss2. The contribution is obtained as
\beqa\label{resultHKadscm}
a^{\text{c.m.}}_{0,\text{AdS}} = 2 K^{s}_{\textrm{AdS}}\left( \bar{s},\bar{m}_b \right)  +K^{f}_{\textrm{AdS}}\left( \bar{s},\bar{m}_f\right)  \, . 
\eeqa 
The masses for the scalars and the fermions are summarised in equations \eqref{MassScalar} and \eqref{MassFermion} respectively and we have calculated the contribution to the heat kernel of massive scalars and massive fermions in  \eqref{massivescalarads} and \eqref{massivefermionads} respectively and have been summarised in \eqref{massivescalarads} and \eqref{massivefermionads} and .  Using them, we get 
\beqa \label{a0cmads}
a^{\text{c.m.}}_{0,\text{AdS}} 
=  \half \left(1-r\right)  + \rho_0 L  \,  .
\eeqa

Equations \eqref{a0vmads} and \eqref{a0cmads} conclude our evaluation of Seeley-DeWitt coefficients and the contribution to the trace of the heat kernel of the vector and chiral multiplet for the theory on \ads2. They are presented in \eqref{a0ads2}. 

%%%%%%%%%%%%%%%%%%%%%%%%%%%%%%%%%%%%%%%%%%%%%%%%%%%%%%%%%%
%%%%%%%%%%%%%%%%%%%%%%%%%%%%%%%%%%%%%%%%%%%%%%%%%%%%%%%%%%%%%%%%%%%

%%%%%%%%%%%%%%%%%%%%%%%%%%%%%%%%%%%%%%%%%%%%%%%%%

\section{Conclusions} \label{sec:conclusions}

During our study of supersymmetric localization on \ads2, we have shed light on several aspects that we hope will aid a systematic application of this powerful tool in other non-compact spaces. Let us now highlight those issues we have clarified as well as several possible open problems that we leave for the future. 
\begin{itemize}
\item We have refined the construction of the equivariant superalgebra for supersymmetric field theory on~\ads2. 
  This is done by defining the quantity $\Lambda_0$ as we did in \eqref{L0ads2} such that  it removes the non-normalizable contributions to the superpartners of ghost.  Our new understanding of the equivariant algebra is that $\qeq^2$ closes to the isometry plus large gauge transformation. It is natural to expect that this structure will go through for generic gauge theories including supergravity.  The equivariant algebra will close to all the gauge transformations with large gauge parameters that include diffeomorphism, Lorentz transformation and local supersymmetry.

\item We have noticed that the supersymmetric boundary condition for localization computation requires us to use seemingly non-normalizable mode for some fields.
An important observation is that the heat kernel calculation is  carried out using a different set of boundary conditions. Nevertheless, we see that the heat kernel computation matches with the result of localization. The reason for this agreement is yet to be understood. A rigorous explanation will be provided in our upcoming paper~\cite{GonzalezLezcano:2023uar}. 

\item We have chosen a Dirichlet type (normalizable) boundary condition  for bosonic fields compatible with supersymmetry and the variational principle. A direction that can immediately be explored is to find a more systematic treatment of the most general boundary conditions that are allowed by the variational principle and supersymmetry.  It would be interesting to explore further in this direction by including more general boundary terms. 

  \item   Based on this supersymmetric boundary condition, we have identified which zero modes are part of the spectrum and how they are organized according to supersymmetry. Specifically, among those zero modes we find the superpartners of 1-form boundary zero modes whose existence we have proved. This  was crucial to capture the full scale dependence of the partition function. 
   Based on the fact that  the superpartner of boundary 1-form zero modes exist in the spectrum of normalizable modes of fermions, it is natural to expect that this situation generalizes such that, in supergravity, boundary zero modes of graviton and gravitino have their superpartners exist in the spectrum of  normalizable basis of  fermion and boson respectively~\cite{Sen:2023dps}. 
   
  \item As we have explained in the body of the paper, the outcome of applying the index method is intrinsically ambiguous. We understand that the
expansion of the index should be such that its constant part encodes the number of zero modes admitted by the boundary condition. Based on this, we single out the correct way to expand the index in terms of the equivariant parameter thus yielding the correct 1-loop determinant.
We expect this prescription to be applicable to more generic situations including supergravity in non-compact backgrounds.

\item Our result of \ads2 partition function coincides with the hemisphere partition function with Dirichlet boundary condition for chiral multiplet up to overall scaling factor and Chan-Paton factors. It may be natural that the local part of partition function is insensitive to the local Weyl scaling that maps the hemisphere to \ads2.  Making this relation more precise  may require extension of our result to non-Abelian gauge theories.

  \item Any extremal supersymmetric asymptotically flat black hole has an AdS$_2$ factor in the near horizon geometry.
 By dimensionally reducing along the internal manifold, we would obtain a theory on \ads2 containing an infinite tower of Kaluza-Klein modes. Therefore, inclusion and treatment of Kaluza-Klein modes in our theory would give us non-trivial insight into the physics of these supersymmetric black holes. 
 
 \item Since we have successfully applied the localization method in the example of \ads2, we would also like to investigate a larger class of supersymmetric observables in general  dimensional anti-de Sitter space. This may even include supergravity theories, and will provide an exact test for the AdS$_{d+1}$/CFT$_d$ correspondence including both perturbative and non perturbative effects. 
 \end{itemize}

%%%%%%%%%%%%%%%%%%%%%%%%%%%%%%%%%%%%%%%%%%%%%%%%%%%%%%%%%%%%%%%%%%%%%%%%%%%%%%%%%%%%%%%%%%%%%%%%%%%%%%%%%%%%%%%%%%%%
%%%%%%%%%%%%%%%%%%%%%%%%%%%%%%%%%%%%%%%%%%%%%%%%%%%%%%%%%%%%%%%%%%%%%%%%%%%%%%%%%%%%%%%%%%%%%%%%%%%%%%%%%%%%%%%%%%%%
%%%%%%%%%%%%%%%%%%%%%%%%%%%%%%%%%%%%%%%%%%%%%%%%%%%%%%%%%%

\section*{Acknowledgements}
This work is supported by an appointment to the JRG Program at the APCTP through the Science and Technology 
Promotion Fund and Lottery Fund of the Korean Government, by the Korean Local 
Governments - Gyeongsangbuk-do Province and Pohang City, and  by the National 
Research Foundation of Korea (NRF) grant funded by the Korea government (MSIT) (No. 2021R1F1A1048531).
We wish to specially thank Ashoke Sen for in-depth discussions that helped us improve our understanding of the result. 
We also  thank  Francesco Benini, Dongwook Ghim,  Rajesh Gupta, Christopher Herzog,  Kazuo Hosomichi, Sungjoon Kim, Bum-Hoon Lee, Sungjay Lee,  Sameer Murthy and Daisuke Yokoyama  
for many interesting and useful discussions related to the topics discussed in this paper. We also wish to specially thank Hee-Joong Chung and Rak-Kyeong Seong for their hospitality in Jeju National University and UNIST respectively during the initial stages of the project.

\appendix
\section{Gamma matrix in Euclidean $2d$}\label{gammaconvention}
In the Euclidean two dimensions, there are two choices of gamma matrix representation
\be\ba{lll}
\gamma_{a}^{\dagger}=\gamma_{a}\,,~&~\\
\gamma_{a}^{*}=\gamma_{a}^{T}=\pm C_{\pm}\gamma_{a}C_{\pm}^{-1}\,,~&~C_{\pm}^{\dagger}=C_{\pm}^{-1}\,,~&~C^{T}_{\pm}=\pm C_{\pm}\,~\Leftrightarrow~C_{\pm}^{*}C_{\pm}=\pm1\,.\\
\ea\ee 
Chirality operator is defined as
\be
\gamma_{(3)}\equiv -\i\gamma_{12}\,.
\ee
Majorana spinor are defined only using $C_{+}$, and the spinor is not compatible with Weyl condition.\\
\be
\psi^{*}_\pm= C_{+}\psi_\mp\,,
\ee
We can use Pauli matrix for the gamma matrix $\gamma_a = (\tau_1\,,\tau_2\,,\tau_3 )$, and the charge conjugation matrix  $C_\pm$ can be chosen as \footnote{ The authors of \cite{Benini:2012ui} use the $C_-$. }
\be
C_+ = \gamma_1\,,~~~~C_- = \gamma_2\,.
\ee
Symmetric property is
\be
(C_\pm)^T = \pm C_\pm\,,  ~~(C_\pm\gamma_a)^T = C_\pm \gamma_a\,,~~ (C_\pm \gamma_{ab})^T= \mp C_\pm \gamma_{ab}\,,
\ee
which is followed by, dealing with  anti-commuting spinors,
\be
\tilde{\epsilon}_{\alpha}\lambda^\alpha \equiv  \epsilon^{T\alpha} C_{\pm \alpha \beta}\lambda^\beta = \mp \tilde{\lambda}_\alpha \epsilon^\alpha\,.
\ee
\be
\tilde{\epsilon} (\gamma_a) \lambda = -  \tilde{\lambda}\gamma_a\epsilon\,,~~~~\tilde{\epsilon}\gamma_{ab} \lambda = \pm  \tilde{\lambda}\gamma_{ab}\epsilon\,,
\ee
where we define the conjugate spinor as $\tilde{\epsilon}\equiv \epsilon^T C_{\pm}$\,.

 Throughout the main body of this paper, we have used the gamma matrix representation with the charge conjugation matrix $C_-$ and simply call it $C$,%$C$. 
 i.e.
 \be
 C\equiv C_- = \gamma_2\,.
 \ee
\paragraph{Summary}
Gamma matrix representation 
\be\ba{lll}
\gamma_{a}^{*}=\gamma_{a}^{T}=- C\gamma_{a}C^{-1}\,,~&~C^{\dagger}=C^{-1}\,,~&~C^{T}=- C\,~\Leftrightarrow~C^{*}C=-1\,.\\
\ea\ee 
is followed by the properties
\be
(C\gamma_a)^T= C\gamma_a\,,~~(C\gamma_{ab})^T= C\gamma_{ab}\,.
\ee
We define multiplication of two spinors $\psi$ and $\chi$ as
\be
\psi \chi \equiv \psi^T C \chi\,.
\ee
Then, for Grassmann even $\psi $ and $\chi$, the bi-spinors follows the symmetry properties
\be
\psi \chi = - \chi \psi\,,~~~\psi \gamma_a \chi =  \chi \gamma_a\psi\,,~~~\psi \gamma_{ab} \chi =  \chi \gamma_{ab}\psi\,.
\ee
\paragraph{Fierz identity} For Grassmann even spinors,
\be\label{Fierzid}
(\psi \lambda )(\rho \epsilon)=\half (\rho \lambda)(\psi \epsilon) +\half (\rho \gamma_a \lambda)(\psi \gamma^a \epsilon)+ \half (\rho \gamma_3 \lambda)(\psi\gamma_3 \epsilon)\,.
\ee
If $\lambda$ and $\rho$ are Grassmann odd, the overall sign changes.

\section{Killing spinors} \label{sec:appKilling}

\subsection{S$^2$ Killing spinor}\label{S2KS}
 On the S$^2$ background with the size $L$ described by the following metric
\be \label{metricS2}
ds^2 = \R^2(d\psi^2 + \sin^2\psi\, d\theta^2)\,,
\ee
we have Killing spinors satisfying two conformal Killing spinor equations for the possible two sign choice $s=\pm 1$\,, 
\be
D_\mu \epsilon =  s  \frac{\i}{2L} \gamma_\mu \epsilon\,.
\ee
Here the covariant derivatives on $\epsilon$ are $D_\psi  = \partial_\psi $ and $D_\theta  = \partial_\theta  -\frac{1}{2}\cos\psi \,\gamma_{12}$ .\\
These equations are solved by the general solution
\be
\epsilon^s = \sqrt{L}\exp\left({s \frac{\i}{2}\gamma_1 \psi} \right) \exp\left( \frac{\theta}{2}\gamma_{12}\right)\epsilon_{0}\,,
\ee
where the $\epsilon_0 $ is  a complex constant spinor.

Let us choose the gamma matrix representation by the Pauli sigma matrix $  \gamma_a =\tau_a $, and let $\beta_\pm$ are the chiral and anti-chiral component of the constant spinor $\epsilon_0$ with respect to the $\tau_3$. 
\be
\epsilon = \beta_+ \varepsilon^s_+ + \beta_- \varepsilon^s_-\,,
\ee
where
\be\label{KSS2}
\ve^s_+=\sqrt{L}{\rm e}^{\i\theta/2}\begin{pmatrix}  \cos\frac{\psi}{2}\\ \i s \sin \frac{\psi}{2} \end{pmatrix}\,,\quad \quad \ve^s_{-}=\sqrt{L}{\rm e}^{-\i\theta/2}\begin{pmatrix}  \i s \sin\frac{\psi}{2}\\ \cos \frac{\psi}{2} \end{pmatrix}\,.
\ee
They satisfy
\be
(\ve^s_\pm)^\dagger  = \mp \i (\ve_\mp^s)^T \tau_2\,,\quad \quad(\ve^s_\pm)^\dagger  =  (\ve_\mp^{-s})^T \tau_1\,.
\ee
Let  us define  bispinors $(K^{ss'}_{\alpha \alpha'} )_A = \wt{\ve^s_\alpha} \tau_A \ve^{s'}_{\alpha'}$, where we denoted $\{\tau_1\,,\tau_2\,,\tau_3\,, \tau_4 \} \equiv \{\tau_1\,, \tau_2\,,\tau_3\,, 1\}$ and $\wt{\ve} \equiv \ve^T \tau_2$. Then,
\be\ba{ll}
(K^{++}_{++} )_A = L {\rm e}^{\i \theta }\{ -\i \,, \cos\psi\,, -\sin\psi \,,0\}\,, ~~~&(K^{++}_{+-} )_A = L \{ 0 \,, \i\sin\psi\,, \i \cos\psi \,,-\i \}\,,
\\
(K^{++}_{--} )_A = L{\rm e}^{-\i \theta }\{ \i \,, \cos\psi\,, -\sin\psi \,,0\}\,,
 ~~~&(K^{++}_{-+} )_A = L \{ 0 \,, \i \sin\psi\,, \i \cos\psi \,,\i \}\,, 
\\
(K^{+-}_{++} )_A = L {\rm e}^{\i \theta }\{ -\i \cos\psi \,, 1 \,, 0 \,,-\sin\psi\}\,, ~~~&(K^{+-}_{+-} )_A = L \{ -\sin\psi \,,0\,, \i\,, - \i \cos\psi  \}\,, 
\\
(K^{+-}_{--} )_A = L {\rm e}^{-\i \theta }\{ \i \cos\psi\,, 1 \,, 0 \,, \sin\psi\}\,,
 ~~~&(K^{+-}_{-+} )_A = L \{ \sin\psi \,,0\,, \i\,,   \i \cos\psi \}\,,
\\
(K^{-+}_{++} )_A = L{\rm e}^{\i \theta }\{ -\i \cos\psi \,, 1 \,, 0 \,,\sin\psi\}\,, ~~~&(K^{-+}_{+-} )_A = L  \{ \sin\psi \,,0\,, \i\,, - \i \cos\psi  \}\,, 
\\
(K^{-+}_{--} )_A = L {\rm e}^{-\i \theta }\{ \i \cos\psi\,, 1 \,, 0 \,, -\sin\psi\}\,,
 ~~~&(K^{-+}_{-+} )_A = L \{ -\sin\psi \,,0\,, \i\,,   \i \cos\psi \}\,,
\\
(K^{--}_{++} )_A = L {\rm e}^{\i \theta }\{ -\i \,, \cos\psi\,, \sin\psi \,,0\}\,, ~~~&(K^{--}_{+-} )_A = L \{ 0 \,, -\i\sin\psi\,, \i \cos\psi \,,-\i \}\,, 
\\
(K^{--}_{--} )_A = L {\rm e}^{-\i \theta }\{ \i \,, \cos\psi\,, \sin\psi \,,0\}\,, 
 ~~~&(K^{--}_{-+} )_A = L \{ 0 \,, -\i \sin\psi\,, \i \cos\psi \,,\i \}\,.
\ea\ee
In this paper, for the purpose of localization we  identify $\epsilon \equiv \ve_+^+$ and $\bar\epsilon\equiv -  \ve_-^+$. Then, 
\beqa 
\bar{\epsilon}\tau^\mu \epsilon = \Bigl(0\,,-\i\Bigr)\,,\qquad \bar\epsilon \tau_3 \epsilon = -\i L  \cos\psi\,,\qquad \bar\epsilon  \epsilon = -\i L \,.
\eeqa
These Killing spinors satisfy the following projection condition:
\be
\epsilon = P_+ \epsilon \,,\qquad \bar\epsilon = {P}_- \bar\epsilon \,,
\ee
where we define the projectors
\begin{equation}
    P_\pm \= \frac{1}{2}\left(1 \pm {\rm{e}}^{ \i \gamma_1 \psi }\gamma_3 \right)\=
    \frac{1}{2}\left(1 \pm \cos\psi \gamma_3  \pm \sin\psi \gamma_2 \right)
    \,,
\end{equation}
satisfying the projection property, $P_\pm^2 = P_\pm \,$. The projectors are conjugate to each other as
\be
(P_\pm)^\dagger \= C^{-1}({P}_\mp)^T C\,.
\ee
 Using the expression of the bispinors in \eqref{bispinorsS2}, we may re-express the projectors as
\begin{equation}
    P_\pm = \frac{1}{2}\left( 1 \pm \frac{\i}{L} (\Bar{\epsilon}\gamma_3\epsilon) \gamma_3 \pm \frac{\i}{L} (\Bar{\epsilon}\gamma_2 \epsilon) \gamma_2 \right)\,.
\end{equation}

\subsection{AdS$_2$ Killing spinor}\label{AdS3KS}
On the AdS$_2$ background with the size $L$ described by the following metric
\be \label{metricads2}
ds^2 = \R^2(d\eta^2 + \sinh^2\eta\, d\theta^2)\,,
\ee
we have Killing spinors satisfying two conformal Killing spinor equations for the possible two sign choice $s=\pm 1$ \footnote{This Killing spinor equation with  representation $\gamma_a = \tau_a$ is equivalent to choose Killing spinor equation $D_\mu \epsilon =  s  \frac{1}{2L} \gamma_\mu \epsilon$ with representation $\gamma_a = -\i \tau_a \tau_3$. }\,,
\be
D_\mu \epsilon =  s  \frac{\i}{2L} \gamma_\mu \gamma_3 \epsilon\,.
\ee
Here the covariant derivatives on $\epsilon$ are $D_\eta  = \partial_\eta $ and $D_\theta  = \partial_\theta  -\frac{1}{2}\cosh\eta \,\gamma_{12}$ .\\
These equations are solved by the general solution
\be
\epsilon^s = {\rm e}^{s\tau_1 \frac{\eta}{2}} \epsilon(\theta) \equiv  \sqrt{L} \exp\left({s \frac{\i}{2}\gamma_1\gamma_3\eta}\right) \exp\left( \frac{\theta}{2}\gamma_{12}\right)\epsilon_{0}\,,
\ee
where the $\epsilon_0 $ is  a complex constant spinor. 

Let us choose the gamma matrix representation by the Pauli sigma matrix $  \gamma_a =\tau_a $, and let $\alpha_\pm$ are the chiral and anti-chiral component of the constant spinor $\epsilon_0$ with respect to the $\tau_3$. 
\be
\epsilon = \alpha_+ \varepsilon^s_+ + \alpha_- \varepsilon^s_-\,,
\ee
where
\be\label{KSads}
\ve^s_+=\sqrt{L} {\rm e}^{\i\theta/2}\begin{pmatrix}  \cosh\frac{\eta}{2}\\ s \i \sinh \frac{\eta}{2} \end{pmatrix}\,,\qquad \ve^s_{-}=\sqrt{L}{\rm e}^{-\i\theta/2}\begin{pmatrix}  - s \i \sinh\frac{\eta}{2}\\ \cosh \frac{\eta}{2} \end{pmatrix}\,.
\ee
Two spinors are related by the complex conjugation as
\be
(\ve^s_\pm)^\dagger =  (\ve^s_{\mp})^T \tau_1\,,\quad \quad (\ve^{s}_\pm)^\dagger = \mp \i  (\ve^{-s}_\mp)^\dagger \tau_2\,.
\ee
Let  us define  bispinors $(K^{ss'}_{\alpha \alpha'} )_A = \wt{\ve^s_\alpha} \tau_A \ve^{s'}_{\alpha'}$, where we denote $\{\tau_1\,,\tau_2\,,\tau_3\,, \tau_4 \} \equiv \{\tau_1\,, \tau_2\,,\tau_3\,, 1\}$ and $\wt{\ve} \equiv \ve^T \tau_2$. Then,
\be\ba{ll}
(K^{++}_{++} )_A = L {\rm e}^{\i \theta }\{ -\i  \cosh\eta\,,1 \,, - \sinh\eta \,,0\}\,, ~~~&(K^{++}_{+-} )_A = L  \{ -  \sinh\eta\,,0\,, \i \cosh\eta \,,-\i \}\,, 
\\
(K^{++}_{--} )_A = L{\rm e}^{-\i \theta }\{ \i  \cosh\eta\,,1\,,  \sinh\eta \,,0\}\,, 
 ~~~&(K^{++}_{-+} )_A = L \{ -  \sinh\eta\,,0\,, \i \cosh\eta \,,\i \}\,,
\\
(K^{+-}_{++} )_A = L{\rm e}^{\i \theta }\{ -\i \,, \cosh\eta  \,, 0 \,,- \sinh\eta\}\,, ~~~&(K^{+-}_{+-} )_A = L \{0\,, \i\sinh\eta \,, \i\,, - \i \cosh\eta  \}\,, 
\\(K^{+-}_{--} )_A = L {\rm e}^{-\i \theta }\{ \i \,, \cosh\eta \,, 0 \,, -\sinh\eta\}\,, 
 ~~~&(K^{+-}_{-+} )_A = L \{ 0\,,-\i \sinh\eta \,, \i\,,   \i \cosh\eta \}\,,
 \\
(K^{-+}_{++} )_A = L {\rm e}^{\i \theta }\{ -\i\,,  \cosh\eta  \,, 0 \,, \sinh\eta\}\,, ~~~&(K^{-+}_{+-} )_A = L \{0\,, -\i\sinh\eta \,, \i\,, - \i \cosh\eta  \}\,, 
\\(K^{-+}_{--} )_A = L {\rm e}^{\i \theta }\{ \i\,, \cosh\eta \,, 0 \,, \sinh\eta\}\,, 
 ~~~&(K^{-+}_{-+} )_A = L \{ 0\,,\i \sinh\eta \,, \i\,,   \i \cosh\eta \}\,,
\\
(K^{--}_{++} )_A = L {\rm e}^{\i \theta }\{ -\i \cosh\eta\,,1\,,  \sinh\eta\,,0  \}\,, ~~~&(K^{--}_{+-} )_A = L \{   \sinh\eta\,, 0\,,\i \cosh\eta\,,-\i  \}\,, \\
(K^{--}_{--} )_A =L {\rm e}^{-\i \theta }\{ \i  \cosh\eta\,, 1 \,, -\sinh\eta \,,0\}\,, ~~~&(K^{--}_{-+} )_A = L \{ \sinh\eta\,, 0\,, \i \cosh\eta\,,\i  \}\,.
\ea\ee
In this paper, for the purpose of localization
we  identify $\epsilon \equiv \ve_+^+$ and $\bar\epsilon\equiv -  \ve_-^-$. Then,
\be
 \bar{\epsilon}\tau^\mu \epsilon = \Bigl(0\,,-\i \Bigr)\,,\qquad \bar\epsilon \tau_3 \epsilon = -\i L\,, \qquad \bar\epsilon  \epsilon = -\i L\cosh\eta\,.
\ee
 These Killing spinors  satisfy the following projection condition:
\be\label{projectionKSads2}
\epsilon = P_+ \epsilon \,,\qquad \bar\epsilon = \bar{P}_+ \bar\epsilon \,,
\ee
where we define the projectors
\beqa
    P_\pm &\=& \frac{1}{2}\left(1 \pm {\rm{e}}^{ \gamma_2 \eta }\gamma_3 \right)
    \=\frac{1}{2}\left(1 \pm \cosh\eta\,  \gamma_3  \pm \i \sinh\eta\,  \gamma_1 \right)
    \,,
    \\
    \bar{P}_\pm &\=& \frac{1}{2}\left(1 \mp  {\rm{e}}^{ -\gamma_2 \eta }\gamma_3  \right)
    \=\frac{1}{2}\left(1 \mp \cosh\eta\,  \gamma_3  \pm \i \sinh\eta\,  \gamma_1 \right)
    \,,\nn
\eeqa
and indeed satisfy the projection property, $P_\pm^2 = P_\pm $ and $\Bar{P}_\pm^2 = \Bar{P}_\pm \,$.
Two projectors are conjugate to each other as
\be
(P_\pm)^\dagger \= C^{-1}(\Bar{P}_\pm)^T C\,.
\ee
 Using the expression of the bispinors in \eqref{bispinorsAdS2}, we may re-express the projectors as
\begin{equation}
    P_\pm = \frac{1}{2}\left( 1 \pm \frac{\i}{L}(\Bar{\epsilon}\epsilon) \gamma_3 \mp  \frac{1}{L}(\Bar{\epsilon}\gamma_2 \epsilon) \gamma_1 \right)\,,\qquad 
      \bar{P}_\pm = \frac{1}{2}\left(1 \mp \frac{\i}{L}(\Bar{\epsilon}\epsilon) \gamma_3 \mp \frac{1}{L}(\Bar{\epsilon}\gamma_2 \epsilon) \gamma_1 \right)\,.
\end{equation}

\section{Atiyah-Bott fixed point formula} \label{app:Atiyah}
In this section, we present the Atiyah-Bott fixed point formula \cite{atiyah1966lefschetz,10.2307/1970694,10.2307/1970721}.  In order to show how the formula works, we take simple examples for de-Rham cohomology on S$^2$ and AdS$_2$.  Although the fixed point formula was formulated in compact manifold as far as we understand, we apply it also to the AdS$_2$ which is non-compact space and test it by comparing with direct computation.

Consider an elliptic complex of vector bundles $E$ on a smooth manifold $X$, which is a sequence of smooth vector bundle $E_i$ over $X$ and differential operators on the smooth sections of the bundles $d_i : \Gamma(E_i) \rightarrow \Gamma(E_{i+1})$.
For a linear map $T_i: \Gamma(E_i) \rightarrow \Gamma(E_{i})$, we define the {\it Lefschetz number } of $T$ by 
\be\label{LefschetzN}
L(T) := \sum_{i=0}^n (-1)^i {\rm Tr}_{H^i}T \,,
\ee
where  $H^i =$Kernel$(d_i)/$Image$(d_{i-1})$. 
For a map $f: X\rightarrow X$, which sends a point $x \in X$ to~$f(x)$, we can define~$f^\ast E^i$, which is a pullback of $E^i$ by $f$. Moreover the map $f$ induces a map $\gamma_i: E^i_{f(p)} \rightarrow E^i_p$, where $p \in X$. 
Let the linear map $T_i$ be a geometric endomorphism of $E^i$ associated to $(f\,,\gamma_i)$, i.e. $T := \gamma 
\circ f^\ast $. If  the fixed points on $X$ under $f$ are isolated, we have the following formula,
\be\label{ABformula}
L(T)=\sum_{\{x| f({x})=x\}} \frac{(-1)^i{\rm Tr}_{E^i}\,\gamma}{\det(1- \partial f(x)/\partial x) }\,.
\ee

\paragraph{Examples}: We take de-Rham cohomology on two dimensions as an example, 
\be
0 \xlongrightarrow[]{d} \Omega^0 \xlongrightarrow[]{d} \Omega^1 \xlongrightarrow[]{d} \Omega^2 \xlongrightarrow[]{d} 0\,,
\ee
where the cohomology group is defined as
\be
H^p  = \frac{Z^p}{B^p}= \frac{\text{closed}\,\Omega^p}{\text{exact}\,\Omega^p}\,.
\ee
\paragraph{On S$^2$}: The metric is given by
\be
ds^2 = \frac{4 dz d\bar{z}}{(1+ z\bar{z})^2}\,.
\ee
Consider the map $T= {\rm e}^{-\i t J^3}$ that is associated to the rotation of spacetime coordinate  $z \rightarrow f(z)={\rm e}^{\i t }z$.  The fixed points are given by~$z=0$ or $1/z=0$ which are the north pole or  south pole of S$^2$ respectively.
At both of the north and south pole,  the determinant factor in the denominator of \eqref{ABformula} is, with $\zeta={\rm e}^{\i t}$, 
\be\label{ABdet}
\det (1-\partial f(x)/\partial x)= (1- \partial f(z)/\partial z )(1- \partial f(\bar{z})/\partial \bar{z}) = (1-\zeta)(1-\zeta^{-1})\,.
\ee
Furthermore, for each $\Omega^p$, we have 
\be\label{ABTrace}
{\rm Tr}_{\Omega^0}\,{\rm e}^{-\i t J^3}= 1\,,~~~~{\rm Tr}_{\Omega^1}\,{\rm e}^{-\i t J^3}= \zeta+{\zeta}^{-1}\,,~~~~{\rm Tr}_{\Omega^2}\,{\rm e}^{-\i t J^3}= 1\,.
\ee
Therefore, the fixed point formula \eqref{ABformula} gives the {\it Lefschetz  number} of ${\rm e}^{-\i t J^3}$ as
\be
L({\rm e}^{-\i J^3})  = 2 \frac{1- \zeta -{\zeta}^{-1}+1}{(1-\zeta)(1-{\zeta}^{-1})}=2\,.
\ee For direct computation, we identify the cohomology group $H^p$, which are
\be
{H^0}=\text{constant}\,,~~~{H^1}=\emptyset   \,,~~~{H^2}=\text{constant}\,.
\ee
Therefore, the definition \eqref{LefschetzN} leads us to obtain
\be
L({\rm e}^{-\i J^3})=\sum_{p=0}^{2}{\rm Tr}_{H^p} (-1)^p {\rm e}^{-\i t J^3} ={\rm Tr}_{H^0}  {\rm e}^{-\i t J^3} +{\rm Tr}_{H^2}  {\rm e}^{-\i t J^3}= 2\,.
\ee
\paragraph{On AdS$^2$}: The metric is given by
\be
ds^2 = \frac{4 dz d\bar{z}}{(1- z\bar{z})^2}\,,
\ee
which has topology of a disk. 
We consider the map $T= {\rm e}^{-\i t L_0}$ that  is associated to the rotation of spacetime coordinate  $z \rightarrow f(z)={\rm e}^{\i t }z$.  The fixed point is given by~$z=0$ which is the center   of AdS$_2$.
At  the center,  the determinant factor in the denominator and the trace of $T$ over the $\Omega^p$ in the numerator  of \eqref{ABformula}  are same as S$^2$, which are \eqref{ABdet} and~\eqref{ABTrace}.   
Since there is one fixed point, the fixed point formula \eqref{ABformula} gives the {\it Lefschetz  number} of~${\rm e}^{-\i t L_0}$ as half of the one for S$^2$,
\be
L({\rm e}^{-\i t L_0})=  \frac{1- \zeta -{\zeta}^{-1}+1}{(1-\zeta)(1-{\zeta}^{-1})}=1\,.
\ee For direct computation, we identify the cohomology group $H^p$, which are
\be
{H^0}=\emptyset\,,~~~{H^1}= \{  A^{(\ell)}   \}\,,~~~{H^2}=\emptyset\,,
\ee
where $A^{(\ell)}\,,~\ell= \pm 1\,, \pm 2\,, \cdots$, is the boundary zero mode as presented in \eqref{bdyzero}. Therefore, the definition \eqref{LefschetzN} leads us to obtain
\beqa
L({\rm e}^{-\i L_0})&=&\sum_{p=0}^{2}{\rm Tr}_{H^p} (-1)^p {\rm e}^{-\i t L_0} \=-{\rm Tr}_{H^1}  {\rm e}^{-\i t L_0}
\\
& =& -\sum_{\ell =\pm1}^{\pm \infty}\zeta^\ell 
\= -\frac{\zeta}{1-\zeta} - \frac{\zeta^{-1}}{1-\zeta^{-1}} \=1\,.
\eeqa

\section{%Sketching Zeta function 
Regularization formula } \label{subsubsec:Reg}
In this section, we sketch the Zeta function regularization of the infinite products that have typically appeared in the calculation of $1$-loop determinants. For further details, see \cite{quine1993zeta} where the case of complex $x$ was proven, here we shall only focus on $\text{Re}(x) \geq -1 $, which we assume form now on just to illustrate the process of regularization. The generic building blocks forming the $1$-loop contribution are the following infinite products:
\be \label{toyZ}
\cZ_{x,L} = \prod_{n=1}^{\infty} \left(\frac{n+x}{L}\right)\,.
\ee
Since there are infinite powers of the length scale $L$ and we want to keep track of the dependence of the partition function on $L$, we device a regularization prescription that appropriately keeps track of such a dependence.   Note that since the product diverges, one cannot factorize it  into two product. i.e. 
\be\label{factorizationofProd}
 \prod_{n=1}^{\infty} \left(\frac{n+x}{L}\right)\neq \prod_{n=1}^{\infty} \left({n+x}\right)\times \prod_{n=1}^{\infty} \frac{1}{L}\,.
\ee  
To illustrate, we can consider when $x$ is a positive integer. Then we can rewrite the $\cZ_{x,L}$ as  
\be\label{integerq}
 \prod_{n=1}^{\infty} \left(\frac{n+x}{L}\right) = \frac{L^x}{x!} \prod_{n=1}^{\infty} \left(\frac{n}{L}\right)\,,
\ee
where  appearance of the factor $L^x$ proves the \eqref{factorizationofProd}. The rewriting \eqref{integerq} seems to suggest the regularized function of the quantity $\cZ_{x,L}$  by promoting the $r$ to non-integer number. To justify this suggestion and find the finite function,  we use the regularization process as follows.

\paragraph{Derivative and integration} \par 
 To find the regularized function of $L$ and $x$, we take the logarithm of the quantity \eqref{toyZ}
 \be
 \log \cZ_{x,L} =  \sum_{n=1}^\infty \log \frac{n+x}{L} \,,
  \ee
and take derivative with respect to $x$ twice. Let us assume that the $ L$ dependence in $\log \cZ_{x,L}$  appears up to linear order in $x$ \footnote{ The derivative and the infinite summation does not commute as the summation of the series diverges. This assumption may imply that $\partial_x \sum_n f_{(n)}(x,L) = \sum_n \partial_x  f_{(n)}(x,L) + g(L)$, where commuting the derivative and the summation generates some function of $L$, $g(L)$.   }. 
 This seems to be a valid assumption by looking at the \eqref{integerq}, and in fact is justified later in \eqref{logLterm}.  Then, we can write  the two derivatives  $L$ independently as
\be
\partial_x^2 \log\cZ_{x,L} = -\sum_{n=1}^{\infty} \frac{1}{(n+x)^2} \=- \psi^{(1)}(x) \,,
\ee
where $\psi^{(1)}(x)$ is the well known Polygamma function, and we can recover $\log \cZ_{x,L}$ by integration twice
\be\label{integrationconsts}
\log \cZ_{x,L} \=  c_0(L) + c_1(L) x - \log\Gamma(x+1) \,.
\ee
Here we have two integration constants, $c_0(L)$ and $c_1(L)$, as functions of $L$. 

\paragraph{Heat kernel regularization} To find the integration constants $c_0(L)$ and $c_1(L)$, we use the heat kernel technique. We take the integral representation of log and use heat kernel regularization
\beqa
\log \cZ_{x,L} &=& \sum_{n=1}^\infty \log \frac{n+x}{L} \= -\sum_{n=1}^\infty \lim_{\epsilon \rightarrow 0} \int_\epsilon^\infty \frac{{\rm d}s}{s} {\rm e}^{-s(n+x)/L} \nn
\\
& = & - \lim_{\epsilon \rightarrow 0}\int_{\epsilon/L}^\infty \frac{{\rm d}\bar{s}}{\bar{s}} \sum_{n=1}^\infty {\rm e}^{-\bar{s}(n+x)} \= -\lim_{\epsilon \rightarrow 0} \int_{\epsilon/L}^\infty \frac{{\rm d}\bar{s}}{\bar{s}}  {\rm e}^{-\bar{s}x} \frac{ {\rm e}^{-\bar{s}}}{1-{\rm e}^{-\bar{s}}} 
\\
&=& \lim_{\epsilon \rightarrow 0}\int^\infty_{\epsilon/L} \frac{{\rm d}\bar{s}}{\bar{s}}\left( - \frac{1}{\bar{s}} + \left(\frac{1}{2}+x\right)+ O({\bar{s}})
\right)\,.\nn
\eeqa
Here we have introduced the UV cutoff $\epsilon$ that is  taken to be zero later. 
In the small $\bar{s}$ expansion, the integration gives
\be\label{logLterm}
\log \cZ_{x,L}=-\frac{L}{\epsilon}+ \left(\frac{1}{2} +x\right)\log \frac{L}{\epsilon} + \cdots\,,
\ee
and we throw the divergence away when $\epsilon \rightarrow 0$.
Here, the $``\cdots"$ part is from the $O(\bar{s})$ terms in the integrand and is guaranteed to be finite  as the series is suppressed by ${\rm e}^{-\bar{s}}$. Thus, it does not depend on $L$ as~$\epsilon \rightarrow 0$. The only finite term that has $L$ dependence is the logarithmic one. This finds the integration constants in \eqref{integrationconsts} as $c_0(L)= \half \log L $ and~$c_1(L)=  \log L$. Therefore, we obtain
\be \label{eq:regen}
\cZ_{x,L} \Big{|}_{\text{reg}}= \prod_{n=1}^{\infty} \left(\frac{n+x}{L}\right)=  \frac{L^{\frac{1}{2}+x}}{\Gamma(1+x)}\,.
\ee

%%%%%%%%%%%%%%%%%%%%%%%%

%%%%%%%%%%%%%%%%%%%%%%%%%

\section{The basis functions on \ss2 and \ads2 } \label{basisfunctions }
This section mostly follows the analysis in the appendix A of  \cite{Sen:2012kpz}.
\subsection{ \ss2} 
We have defined the \ss2 metric in \eqref{spheremetric}. The following is based on that metric. 

\subsubsection*{Scalar Modes} \label{s2scalarmodes}
The delta function normalized eigenfunctions of $-\nabla^2_{\textrm{S$^2$}}$ are given by the spherical harmonics $Y_{lm}(\psi,\theta)$, which are defined as
\beqa \label{eq:S2eigenfunctions}
Y_{lm}(\psi,\theta) = (-1)^m \sqrt{\dfrac{2l+1}{4 \pi} \dfrac{(l-m)!}{(l+m)!}} ~{\rm e}^{\i m \theta} P_l^m \left( \cos \psi \right)  \, , \quad 0\leq l < \infty\, , \quad - l \leq m \leq l\,, \qquad
\eeqa where the associated Legendre Polynomials are defined as: 
\beqa
P_l^m \left( x \right) = \dfrac{\left(-1 \right)^m }{2^l l!} \left( 1-x^2\right)^{\tfrac{m}{2}}  \dfrac{d^{l+m}}{d x^{l+m}}  \left(x^2 - 1 \right)^l \, . 
\eeqa
The spherical harmonics satisfy the eigenvalue equation given by
\be
-\nabla^2_{\textrm{S$^2$}} Y_{lm}(\psi,\theta)= \dfrac{l\left(l+1\right)}{\R^2} Y_{lm}(\psi,\theta) \, . \label{scalareigenvaluess2}
\ee
Note that the complex conjugate of the eigenfunctions is
\be
Y^*_{l,m}(\psi,\theta) =Y_{l,-m}(\psi,\theta) \,.
\ee

\subsubsection*{Vector Modes}
The normalized basis of vector fields on \ss2 is taken as
\be
\dfrac{1}{L\sqrt{ \kappa_l}}\partial_\mu Y_{lm} \,,~~~~~~~ \dfrac{1}{L\sqrt{ \kappa_l}} \varepsilon_{\mu\nu}\partial^\nu Y_{lm} \qquad \text{for}~l \geq 1\, , 
\ee
where $\kappa_l$ has been defined by the relation in  \eqref{scalareigenvaluess2}. They are eigenmodes of $- \nabla^2_{\textrm{S$^2$}}$ with eigenvalues given by $\left( \dfrac{l\left( l+1\right)-1 }{\R^2}\right) $. Note that the $l=0$ mode does not exist for the vectors as $Y_{0,0}=$ constant and the derivative vanishes.

\subsubsection*{Spinor Modes} 
The Dirac operator on \ss2 is given by
\be
\slashed{D}_{\textrm{S$^2$}}= \R^{-1}\left[\tau_1\partial_{\psi}+\tau_2\frac{1}{\sin \psi}\partial_{\theta}+\half \tau_1 \cot\psi \right]\,,
\ee
and it has the following eigenmodes, 
\beqa\label{SpinorEigenmode}
\chi_{l,m}^{\pm} &=& \dfrac{1}{\sqrt{4\pi L^2}} \dfrac{\sqrt{\left( l-m\right)!\left(l+m+1 \right)!  }}{l!} {\rm e}^{-\i \left( m+\tfrac{1}{2}\right) \theta} 
\begin{pmatrix}
	\i \sin^{m+1} \tfrac{\psi}{2} \cos^{m} \tfrac{\psi}{2} P^{\left(m+1,m\right) }_{l-m}\left(  \cos \psi \right)  \\ \pm \sin^{m} \tfrac{\psi}{2} \cos^{m+1} \tfrac{\psi}{2} P^{\left(m,m+1\right) }_{l-m}\left(  \cos \psi \right) 
\end{pmatrix} \, , \nonumber \\
\eta_{l,m}^{\pm} &=& \dfrac{1}{\sqrt{4\pi L^2}} \dfrac{\sqrt{\left( l-m\right)!\left(l+m+1 \right)!  }}{l!} {\rm e}^{\i \left( m+\tfrac{1}{2}\right) \theta} 
\begin{pmatrix}
	\sin^{m} \tfrac{\psi}{2} \cos^{m+1} \tfrac{\psi}{2} P^{\left(m,m+1\right) }_{l-m}\left(  \cos \psi \right)  \\ \pm \i\sin^{m+1} \tfrac{\psi}{2} \cos^{m} \tfrac{\psi}{2} P^{\left(m+1,m\right) }_{l-m}\left(  \cos \psi \right) 
\end{pmatrix} \, ,\nonumber \\ 
&&\qquad \forall~ l,m \in \Z \, , \quad 0 \leq l \, , \quad 0\leq m \leq l \, ,
\eeqa 
where $P_{a}^{(b,c)}(x)$ is Jacobi Polynomial. They satisfy the eigenvalue equation 
\beqa \label{diraceigenvalueons}
\i\slashed{D}_{\textrm{S$^2$}}~\chi^{\pm}_{l,m}= \mp  \dfrac{ l+1 }{L}\, \chi^{\pm}_{l,m}\,, \qquad 
\i\slashed{D}_{\textrm{S$^2$}}~\eta^{\pm}_{l,m}= \mp  \dfrac{ l+1 }{L}\, \eta^{\pm}_{l,m}\, .
\eeqa
Note that some of eigenmodes \eqref{SpinorEigenmode} are related to the Killing spinors \eqref{KSS2} by
\be
\chi^\pm_{0,0}\= \pm \varepsilon^\pm_-\,,\qquad \eta^\pm_{0,0}\=\varepsilon^\pm_+\,. 
\ee
\subsection{AdS$_2$}\label{appendix:AdS2}
We have defined the \ads2 metric in \eqref{adsmetric}. The following is based on that metric. 
\subsubsection*{Scalar Modes} \label{adsscalarmodes}
%\paragraph{Eigenfunction on  AdS$_2$}: 
The delta function normalized eigenfunctions of $-\nabla^2_{\textrm{AdS}}$ are given by
\beqa
\label{AdS2eigenfunction}
f_{\lambda, k}(\eta,\theta)&=&\frac{1}{\sqrt{2\pi }}\frac{1}{2^{| k|}(| k|)!}\left| \frac{\Gamma(\i\lambda +\frac{1}{2}+| k|)}{\Gamma(\i\lambda)} \right| {\rm e}^{\i k \theta}\sinh^{| k|}\eta \\
~~~~~~~~~~~~~~~&&\,\times {}_2F_1\left(\i\lambda+\frac{1}{2}+| k|, -\i \lambda +\frac{1}{2} +| k| ; | k| +1; -\sinh^2\frac{\eta}{2} \right)\,,~~~~~ k \in \mathbb{Z}\,,~~~0  \le \lambda<\infty\,.\nn
\eeqa
where ${}_2F_1(\alpha, \beta; \gamma; z)$ is the Hypergeometric function and
\be
-\nabla^2_{\textrm{AdS}} f_{\lambda, k}(\eta,\theta)=\dfrac{1}{\R^{2}} \left(\frac{1}{4} +\lambda^2\right) f_{\lambda, k}(\eta,\theta)\,. \label{scalareigenvaluesads2}
\ee
Note that the complex conjugate of the eigenfunctions is
\be
f^\ast_{\lambda ,  k}(\eta, \theta) =f_{\lambda , - k}(\eta, \theta) \,.
\ee
Also 
\be\label{lambda=0}
f_{\lambda,  k}(\eta, \theta)=0 \quad \text{at}~ \lambda=0\, .
\ee
\paragraph{At  the origin} As $\eta =0 $
\beqa
f_{\lambda ,  k}(\eta=0, \theta) &=& \delta_{k,0}\frac{1}{\sqrt{2\pi }}\left| \frac{\Gamma(\i\lambda +\frac{1}{2})}{\Gamma(\i\lambda)} \right| = \delta_{k,0} \left| \sqrt{  \frac{\lambda}{2\pi }\tanh \pi \lambda }\right|\,.  \label{fAtOrigin}
\eeqa
\subsubsection*{Vector Modes}
The normalized basis of vector fields on AdS$_2$ is taken as
\be \label{vectorlaplacianads}
\frac{1}{ \sqrt{\tfrac{1}{4} + \l^2} }\partial_\mu {f}_{\lambda,k} \,,\qquad\frac{1}{ \sqrt{\tfrac{1}{4} + \l^2} } \varepsilon_{\mu\nu}\partial^\nu {f}_{\lambda,k} \, .
\ee
This basis has the eigenvalue  $\R^{-2}(\frac{5}{4}+\lambda^2 )$ of $-\nabla^2_{\textrm{AdS}}$ and  $f_{\lambda, k} $ is the scalar modes in \eqref{AdS2eigenfunction}. There are also additional square integrable modes 
\beqa \label{bdyzero}
A^{(\ell)}=d\Lambda^{(\ell)}_{\text{bdry}}\, ,\qquad  \Lambda^{(\ell)}_{\text{bdry}} = \dfrac{1}{\sqrt{2\pi |\ell|}} \Biggl(\dfrac{\sinh \eta }{1+\cosh \eta }\Biggr)^{|\ell|} {\rm e}^{\i \ell \theta} \, , \qquad \ell = \pm 1\,,\pm 2\,,\, \cdots  \,.
\eeqa 
To count the number of the boundary zero modes, we see
\beqa \label{vectorbdymode}
 n^{A^{bdry}}_{\text{bdry}}&\equiv&  \sum_{\ell \neq 0} \int_{\textrm{AdS}} {\rm d}^2x \sqrt{g}%\textrm{dvol} ~
\, \partial_\mu \Lambda^{(\ell)}_{\text{bdry}}(x)^\ast \partial^\mu \Lambda^{(\ell)}_{\text{bdry}}(x)
 \nonumber \\
 &
=& \sum_{\ell =\pm 1} \int_{\textrm{AdS}} {\rm d}^2x \sqrt{g}%\textrm{dvol}
~ \partial_\mu \Lambda^{(\ell)}_{\text{bdry}}(0)^\ast \partial^\mu \Lambda^{(\ell)}_{\text{bdry}}(0)  \=        -1 \,  , 
\eeqa
where in the second line, we use the homogeneity of spacetime to set $x=0$, and obtain the regularized volume of the \ads2.
Therefore, a vector field $A_\mu$ has the mode expansion as
\be\label{1-formExpansion}
A_\mu =\frac{1}{\sqrt{\quater +\lambda^2}} \int_0^\infty {\rm d}\lambda\sum_{k\in \mathbb{Z}} (a_{1\lambda,k} \partial_\mu {f}_{\lambda,k}  +a_{2\lambda,k} \varepsilon_{\mu\nu}\partial^\mu {f}_{\lambda,k}) +\sum_{\ell=\pm 1}^{\pm \infty} \alpha_{(\ell)} \partial_\mu \Lambda^{(\ell)}_{\text{bdry}}\,.
\ee
The Dirac operator on \ads2 is given by
\be
\slashed{D}_{\textrm{AdS}}= \R^{-1}\left[\tau_1\partial_{\eta}+\tau_2\frac{1}{\sinh\eta}\partial_{\theta}+\half \tau_1 \coth\eta \right]\,,
\ee
This basis has the following eigenmodes

\paragraph{Eigenstates of $\slashed{ D}_{\textrm{AdS}}$}:
\be\ba{ll}\label{EigenAdS2}
\chi_{k,\lambda}^{\pm}= &\dfrac{1}{\sqrt{4\pi }} \dfrac{1}{k !}\left( \frac{\Gamma(1+k+\i\lambda) \Gamma(1+k -\i \lambda)}{\Gamma\left(\tfrac{1}{2} +\i\lambda\right)\Gamma\left(\tfrac{1}{2}-\i\lambda\right)}\right)^{\frac{1}{2}} {\rm e}^{-\i(k+\frac{1}{2})\theta} \\
&
~~~\begin{pmatrix}\i \dfrac{\lambda}{k+1}  \cosh^k\dfrac{\eta}{2}\sinh^{k+1}\frac{\eta}{2}\,\,{}_2F_1(k+1+\i\lambda, k+1-\i\lambda; k+2;-\sinh^2\frac{\eta}{2})\\ \pm \cosh^{k+1}\frac{\eta}{2}\sinh^{k}\frac{\eta}{2}\,\,{}_2F_1(k+1+\i\lambda, k+1-\i\lambda; k+1; -\sinh^2\frac{\eta}{2})\end{pmatrix}\,,
\\
\eta^{\pm}_{k,\l}=&\dfrac{1}{\sqrt{4\pi }}  \dfrac{1}{k !}\left( \frac{\Gamma(1+k+\i\lambda) \Gamma(1+k -\i \lambda)}{\Gamma(\frac{1}{2} +\i\lambda)\Gamma(\frac{1}{2}-\i\lambda)}\right)^{\frac{1}{2}} \re^{\i(k+\frac{1}{2})\theta} \\
&
~~~\begin{pmatrix}   \cosh^{k+1}\frac{\eta}{2}\sinh^{k}\frac{\eta}{2}\,~~~~{}_2F_1(k+1+\i\lambda, k+1-\i\lambda; k+1;-\sinh^2\frac{\eta}{2})\\ 
 \pm\i \dfrac{\lambda}{k+1} \cosh^{k}\frac{\eta}{2}\sinh^{k+1}\frac{\eta}{2}\,\,{}_2F_1(k+1+\i\lambda, k+1-\i\lambda; k+2; -\sinh^2\frac{\eta}{2})\end{pmatrix}\,,\\
&~~~~~~~~~~~~k\in \mathbb{Z}\,,~~0   \le k<\infty\,, ~~ 0  \leq\lambda<\infty\,,
\ea\ee
satisfying
\be \label{diraceqn}
\i \slashed{D}_{\textrm{AdS}}\chi^{\pm}_{k,\l}= \mp \dfrac{\lambda }{\R}\, \chi^{\pm}_{k,\l}\,,\qquad 
\i \slashed{D}_{\textrm{AdS}}\eta^{\pm}_{k,\l}= \mp \dfrac{\lambda }{\R} \eta^{\pm}_{k,\lambda}\,.
\ee
Reality property
\be
(\chi^\pm_{k,\lambda})^\dagger = \pm \i (\eta^\pm_{k,\lambda})^T \tau_2\,, \qquad (\eta^\pm_{k,\lambda})^\dagger = \mp \i (\chi^\pm_{k,\lambda})^T \tau_2\,.
\ee
 At $\lambda=0$,
\be\ba{ll}\label{Fermionatorigin}
\chi_{k,0}^{\pm}=\! \frac{1}{2\pi}  {\rm e}^{-\i(k+\frac{1}{2})\theta} \!
\begin{pmatrix}0 \\ \pm \cosh^{\!-k-1}\!\frac{\eta}{2}\sinh^{k}\frac{\eta}{2}\end{pmatrix}\,,&
\eta^{\pm}_{k,0}=  \!\frac{1}{2\pi }  {\rm e}^{\i(k+\frac{1}{2})\theta} \!
\begin{pmatrix}   \cosh^{\!-k-1}\!\frac{\eta}{2}\sinh^{k}\frac{\eta}{2}\\ 
0\end{pmatrix}\,,
\ea\ee
which are non-zero differently from the scalar basis function $f_{\lambda\,,k}$.
\paragraph{At  the origin} At $\eta =0 $,
\be\ba{l}
\chi^\pm_{\lambda ,  k}(\eta=0, \theta) = \eta^\pm_{\lambda ,  k}(\eta=0, \theta)= 0 ~~~\mbox{for} ~k\neq 0\,,  \\
\chi^\pm_{\lambda ,  k}(\eta=0, \theta) =\frac{1}{\sqrt{4\pi }} \left( \frac{\Gamma(1+\i\lambda) \Gamma(1-\i \lambda)}{\Gamma(\frac{1}{2} +\i\lambda)\Gamma(\frac{1}{2}-\i\lambda)}\right)^{\frac{1}{2}} {\rm e}^{-\i\frac{1}{2}\theta} \Biggl( \ba{c}0 \\ \pm 1 \ea\Biggr)={\rm e}^{-\i\frac{1}{2}\theta} \left(\frac{\lambda \coth \pi \lambda }{4 \pi }\right)^{\frac{1}{2}} \Biggl( \ba{c}0 \\ \pm 1 \ea\Biggr)~~~\mbox{for} ~k= 0\,,  \\
\eta^\pm_{\lambda ,  k}(\eta=0, \theta) =\frac{1}{\sqrt{4\pi }} \left( \frac{\Gamma(1+\i\lambda) \Gamma(1-\i \lambda)}{\Gamma(\frac{1}{2} +\i\lambda)\Gamma(\frac{1}{2}-\i\lambda)}\right)^{\frac{1}{2}} {\rm e}^{\i\frac{1}{2}\theta} \Biggl( \ba{c}  1 \\ 0 \ea\Biggr)={\rm e}^{\i\frac{1}{2}\theta} \left(\frac{\lambda \coth \pi \lambda }{4 \pi }\right)^{\frac{1}{2}} \Biggl( \ba{c} 1  \\ 0 \ea\Biggr)~~~\mbox{for} ~k= 0\,.  
\ea\ee
\section{Heat kernel calculations} \label{appendixheatkernel}
Here, we compute the  trace of the heat kernel  \eqref{Kexpansion} for the massive scalar and Dirac fermions both on \ss2 and \ads2. 

\subsection{Massive scalar} \label{scalarheatkernelappendix}

Here, we compute the trace of the heat kernel associated with the bosonic operator 
\be \label{bosonkineticoperator}
\cO_b = - \nabla^2 +\frac{\bar{m}_b^2}{L^2} \, . 
\ee 
%\be
%K(\bar{s})= \Tr \,{\rm e}^{-\bar{s}\cO_b}
%\ee

\subsubsection*{\ss2 }

 We know from \cite{Sen:2012kpz} that the contribution of the massless scalar field to the trace of the heat kernel is given by 
 \beqa
K_{\text{S}^2}^{sc} \left( \bar{s}\right)  
&=& \dfrac{1}{\bar{s}} + \dfrac{1}{3} + \dfrac{\bar{s}}{15} + \cdots \, . \label{heatkernelscalarfinalresults}
\eeqa
From this we can obtain the heat kernel expansion for massive scalar. 
Since the kinetic operator is shifted as given by  \eqref{bosonkineticoperator},  the eigenvalue is shifted by a constant, viz., $\kappa_n \longrightarrow \kappa^{new}_n = \kappa_n + \frac{\bar{m}_b^2}{L^2} $, where $\kappa_n$ is the eigenvalue of the $-\nabla^2 $ operator. The scalar eigenfunctions of the Laplacian operator $\lbrace f_n\left( x \right)\rbrace$ are spherical harmonics \ref{s2scalarmodes}. Thus, the trace of the heat kernel is obtained as
 \beqa
K_{\text{S}^2}^{sc} \left( \bar{s},\bar{m}_b\right)  &=& \sum_n \frac{1}{L^2}\int_{\text{S}^2} \textrm{d}^2 x \sqrt{g}~ f^*_n (x) f_n (x) \textrm{e}^{-\bar{s} \kappa^{new}_n} = \exp\left(-\bar{s} \bar{m}_b^2\right)\left(\dfrac{1}{\bar{s}} + \dfrac{1}{3} + \dfrac{\bar{s}}{15} + \cdots\right) \nonumber \\
   &=& \dfrac{1}{\bar{s}} + \left( \dfrac{1 }{3} - \bar{m}_b^2 \right) + \dfrac{2-10 \bar{m}_b^2+ 15 \bar{m}_b^4 }{30} \bar{s} + 
 \cdots \,.
 \label{massivescalars} \eeqa 
 This naturally reduces to  \eqref{heatkernelscalarfinalresults} when $\bar{m}_b= 0$.
 
\subsubsection*{\ads2 \label{heatkernelscalarads}}

We know from \cite{Sen:2012kpz} that the contribution of the massless scalar field to the trace of the heat kernel is given by 
\beqa
K_{\text{AdS}}^{sc} \left( \bar{s}\right) &=&  -  \dfrac{1}{2\bar{s}} + \dfrac{1}{6} - \dfrac{\bar{s}}{30}+ \cdots \, .  \label{heatkernelscalarfinalresultads} \eeqa Let us now introduce mass terms for the bosonic fields. The analysis for the massive bosonic fields on \ads2 is similar to the one on described in the \ss2 case.  The scalar eigenfunctions of the Laplacian operator $\lbrace f_n\left( x \right)\rbrace$ are given in \ref{AdS2eigenfunction}.  Thus, the trace of the heat kernel is obtained as
\beqa
K_{\text{AdS}}^{sc} \left( \bar{s},\bar{m}_b\right) &=& \sum_n \frac{1}{L^2}\int_{\text{AdS}} \textrm{d}^2 x \sqrt{g}~ f^*_n (x) f_n (x) \textrm{e}^{-\bar{s} \kappa^{new}_n} = \textrm{e}^{ - \bar{s} \,\bar{m}^2_b } \Bigg( -  \dfrac{1}{2\bar{s}} + \dfrac{1}{6} - \dfrac{\bar{s}}{30} + \cdots  \Bigg) \nonumber \\ 
&=& -  \dfrac{1}{2\bar{s}} +\left(  \dfrac{1}{6} + \dfrac{\bar{m}^2_b}{2} \right) - \dfrac{\left(2 + 10 \bar{m}^2_{b}  +15 \bar{m}^4_{b} \right) \bar{s}}{60} + \cdots \, . \label{massivescalarads}
\eeqa This naturally reduces to  \eqref{heatkernelscalarfinalresultads} when $\bar{m}_b = 0$.

\subsection{Massive fermion}  \label{fermionheatkernelappendix}
\subsubsection*{\ss2 } 
In this section we derive the trace of the heat kernel of the fermionic kinetic operator 
\beqa
\mathcal{O}_f^2 = \left( \i \slashed{D}_{\textrm{S$^2$}} + \dfrac{\bar{m}_D}{\R} \right)^2 \, .  
\eeqa Recall that this Dirac mass term appears from \eqref{MassFermion} by setting $G=0$. The eigenvalues of our basis states $\lbrace \chi^\pm_{l,m}, \eta^\pm_{l,m} \rbrace$ are as given in \eqref{diraceigenvalueons}. Then,  the trace of the heat kernel is obtained as
\beqa
&&K_{\textrm{S}^2}^f  (\bar{s},\bar{m}_D) =   - \frac{1}{L^2} \int_{\textrm{S$^2$}} \textrm{d}^2 x \sqrt{g}~ \sum_{m} \sum^\infty_{l=0} ~ \bigg[ ||\chi_{l,m}^{+}||^2 \exp\left(- \bar s \, \k^+_{l+1}  \right)   +||\chi_{l,m}^{-}||^2 \exp\left(- \bar s \,\kappa^-_{l+1}  \right)  \nonumber \\ 
&& 
\qquad \qquad \qquad \qquad\qquad \qquad  +  ||\eta_{l,m}^{+}||^2 \exp\left(- \bar s \,\kappa^+_{l+1}) \right) + ||\eta_{l,m}^{-}||^2 \exp\left(- \bar s \,\kappa^-_{l+1}  \right)   \biggr] \,,  
 \eeqa where $\kappa^{\pm}_{l} = \left( l \mp \bar{m}_D\right)^2$, i.e., they are the eigenvalues shifted by the appropriate masses. This integral can be performed using the techniques illustrated in \cite{Sen:2012kpz}.  The above integral reduces to
 \beqa
K_{\textrm{S}^2}^f  (\bar{s},\bar{m}_D) = 2 \i \int^{\infty}_{0} d\l ~ \l \cot \pi \l \left[ \exp\left(-\bar{s} (\bar{m}_D + \l )^2 \right) +   \exp\left(-\bar{s} (\bar{m}_D - \l )^2 \right)\right] \, \, . \qquad
\eeqa
 We will use the series expansion of cotangent  functions. Splitting the integral into two parts, we have 
\beqa
I_1 &&= -2 \int^{\infty}_{0} d\l ~ \l  \left[ \exp\left(-\bar{s} (\bar{m}_D +\l )^2\right)   + \exp\left(-\bar{s} (\bar{m}_D - \l)^2 \right)\right] \, \, .  \nonumber \\ 
&& = - \dfrac{2}{\bar{s}} - 2 \bar{m}^2_D  + \dfrac{1}{3} \bar{m}^4_D ~\bar{s} + \cdots\,, \nonumber \\
I_2 &&= -4  \sum_{k=1}^{\infty} \int^{\infty}_{0} d\l ~ \l \sum^\infty_{p=0}\dfrac{\left(-1 \right)^p \bar{s}^p }{p!}\left[(\bar{m}_D + \l )^{2p}  + (\bar{m}_D - \l )^{2p} \right] \textrm{e} ^{-2 \pi \i k \l} \qquad \nonumber \\
&& = \dfrac{1}{3} - \dfrac{1}{30}\left(10\bar{m}^2_D-1\right)\bar{s}  + \cdots\,. \nonumber
\eeqa
Adding $I_1$ and $I_2$, we have
\beqa \label{massivefermions2}
K_{\textrm{S}^2}^f  \left( \bar{s},\bar{m}_D\right) &&= -\dfrac{2}{\bar{s}} + \dfrac{1}{3}- 2 \bar{m}^2_D+ \left(\dfrac{1}{30} - \dfrac{\bar{m}^2_D}{3}  + \dfrac{\bar{m}^4_D}{3} \right) \bar{s} + \cdots\,.
\eeqa As a consistency check, we see that  \eqref{massivefermions2} reduces to known results in the massless limit. 

\subsubsection*{\ads2 } 
In this section we derive the trace of the heat kernel of the fermionic kinetic operator \begin{align}\mathcal{O}_f^2 = \left( \i \slashed{D}_{\textrm{AdS}} + \dfrac{\bar{m}_c}{\R} \gamma_3 \right)^2 \, .
\end{align}  
Recall that the mass term appears from \eqref{MassFermion} by setting $H=0$. The eigenvalues of our basis states $\lbrace \chi^\pm_{k,\l}, \eta^\pm_{k,\l} \rbrace$  as given in \eqref{EigenAdS2}. Furthermore, the chirality matrix acts on them as follows
\beqa
\gamma_3 \chi_{k,\l}^\pm = \chi_{k,\l}^\mp \, , \qquad \gamma_3 \eta_{k,\l}^\pm = \eta_{k,\l}^\mp \, . \qquad 
\eeqa
Then, the trace of the heat kernel is obtained as
\beqa \label{generalfermionheatkernel}
&& K_{\text{AdS}}^f  (\bar{s},\bar{m}_c) =   - \frac{1}{L^2}\int_{\text{AdS}} \textrm{d}^2 x \sqrt{g}~  \sum^\infty_{k=0}  \int_0^\infty d\l~ \bigg[ ||\chi_{k,\l}^+||^2 \exp\left(- \bar s  \k  \right) +|| \chi_{k,\l}^-||^2 \exp\left(- \bar s \k  \right)   \nonumber \\ 
& &\qquad \qquad \qquad \qquad \qquad \qquad \quad+  ||\eta_{k,\l}^+||^2 \exp\left(- \bar s \k \right) + ||\eta_{k,\l}^-||^2 \exp\left(- \bar s \k  \right)   \bigg] \, ,
\eeqa where $\kappa =  \l^2 + \bar{m}^2_c$. Now, using the homogeneity of \ads2 spacetime, we conclude that the above integral receives contribution only from $\eta = 0$. At $\eta = 0$, $||\chi_{k,\l}^+||^2 = ||\eta_{k,\l}^+||^2 =  ||\chi_{k,\l}^-||^2 = ||\eta_{k,\l}^-||^2= \left(\l \coth \pi \l / 4\pi\right)  ~ \delta_{k,0} $ .   The integration then can be performed directly and  we have
\beqa \label{massivefermionads}
K_{\text{AdS}}^f  \left( \bar{s},\bar{m}_c\right) &=& \dfrac{1}{\bar{s}} + \left(\dfrac{1}{6}  - \bar{m}_c^2  \right) 
  -   \left(\dfrac{1}{60} +\dfrac{\bar{m}_c^2}{6}   - \dfrac{\bar{m}_c^4}{2}  \right)\bar{s}  + \cdots \, . 
\eeqa 
As a consistency check, we see that  \eqref{massivefermionads} reduces to  known results in the massless limit.

\bibliographystyle{JHEP}
\bibliography{BibFile.bib}

\end{document}